\documentclass{article}
\usepackage[utf8]{inputenc}
\usepackage[sort&compress,numbers]{natbib}
\usepackage{graphicx}
\usepackage{color}
\usepackage{authblk}
\usepackage{authblk}
\usepackage{amsmath,amssymb}
\usepackage{url}
\usepackage{upgreek} 
\usepackage{currvita} 
\usepackage[T1]{fontenc}
\usepackage{aas_macros}
\usepackage{longtable}
\usepackage[colorlinks]{hyperref}
\usepackage{xspace}

\newcommand{\gr}{$\gamma$-ray}

\newcommand{\bx}{\mathbf{x}}
\newcommand{\bz}{\mathbf{z}}
\newcommand{\btheta}{\boldsymbol{\theta}}

\newcommand{\fex}{\textit{e.g.},\xspace}
\newcommand{\AS}{sec:astrostatistics}
\newcommand{\PS}{sec:particlesfromstars}

\newcommand{\TM}{sec:travelingmessengers}
\newcommand{\EU}{sec:earlyuniverse}
\newcommand{\NA}{sec:nuclearastro}
\begin{document}
\title{}
\begin{titlepage}
   \begin{center}
       \vspace*{1cm}
\huge
       \textbf{EuCAPT White Paper}
       
\huge
       \vspace{0.5cm}
       \textbf{Opportunities and Challenges for Theoretical Astroparticle Physics in the Next Decade}
            
       \vspace{1.5cm}
       \includegraphics[width = 80mm]{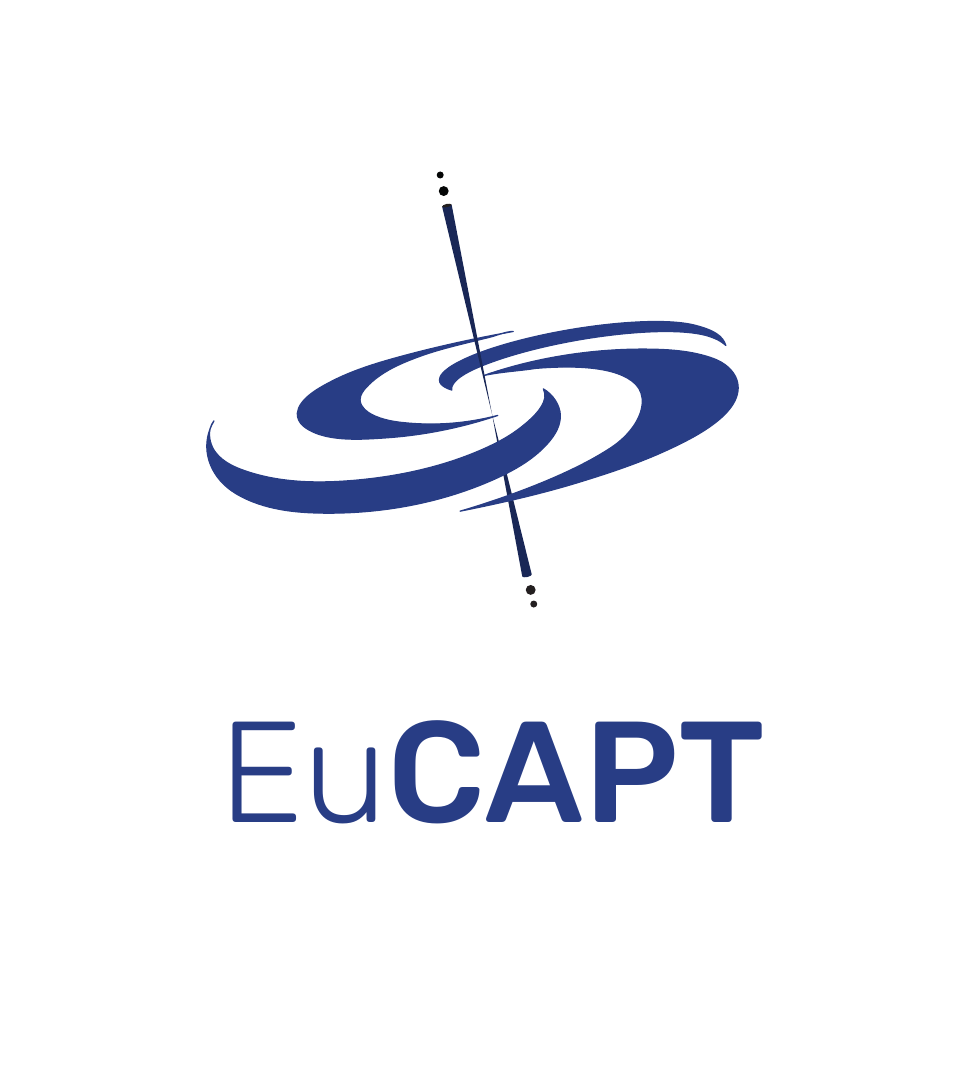}
   \end{center}
   
 \large
 \centerline{\textbf{Abstract} }
 \vskip 0.2cm
 \noindent
 Astroparticle physics is undergoing a profound  transformation, due to a series of extraordinary new results, such as the discovery of high-energy cosmic neutrinos with IceCube, the direct detection of gravitational waves with LIGO and Virgo, and many others. This white paper is the result of a collaborative effort that involved hundreds of theoretical astroparticle physicists and cosmologists, under the coordination of the European Consortium for Astroparticle Theory (EuCAPT). Addressed to the whole astroparticle physics community, it explores upcoming theoretical opportunities and challenges for our field of research,  with particular emphasis  on the possible synergies among different subfields, and the prospects for solving the most fundamental open questions with multi-messenger observations.

\end{titlepage}




\author[1]{R.~{Alves~Batista}}
\affil[1]{Instituto de Física Teórica UAM/CSIC, U. Autónoma de Madrid, C/ Nicolás Cabrera 13-15, Cantoblanco, 28049, Madrid, Spain}
\author[2]{M.~A.~Amin}
\affil[2]{Physics and Astronomy Department - Rice University,
6100 Main MS-61, Houston, TX 77005-1827, U.S.}
\author[3]{G.~Barenboim }
\affil[3]{Departament de F\'{\i}sica Te\`orica and IFIC, Universitat de 
Val\`encia-CSIC, E-46100, Burjassot, Spain}
\author[4]{N.~Bartolo}
\affil[4]{Dipartimento di Fisica e Astronomia Galileo Galilei - Università degli Studi di Padova,
via F. Marzolo, I-35131 Padova, Italy}
\author[5,122]{\textbf{D.~Baumann}}
\affil[5]{Gravitation Astroparticle Physics Amsterdam (GRAPPA),
Institute for Theoretical Physics Amsterdam and Delta Institute for Theoretical Physics, University of Amsterdam, Science Park 904, 1098 XH Amsterdam, The Netherlands}
\author[6,7]{A.~Bauswein}
\affil[6]{GSI Helmholtzzentrum für Schwerionenforschung, D-64291 Darmstadt, Germany}
\affil[7]{Helmholtz Research Academy Hesse for FAIR (HFHF), GSI Helmholtz Center for Heavy Ion Research, Campus Darmstadt, Germany}
\author[8]{E.~Bellini}
\affil[8]{D\`epartement de Physique Th\`eorique, Universit\`e de Gen\`eve, 24 quai Ernest Ansermet, 1211 Gen\`eve 4, Switzerland}
\author[9,10]{D.~Benisty}
\affil[9]{DAMTP, Centre for Mathematical Sciences, University of
Cambridge, Wilberforce Road, Cambridge CB3 0WA, United Kingdom}
\affil[10]{Kavli Institute of Cosmology (KICC), University of Cambridge, Madingley Road, Cambridge, CB3 0HA, UK}
\author[5]{\textbf{G. Bertone (Editor)}}
\author[11,12]{P.~Blasi}
\affil[11]{Gran Sasso Science Institute, Via Michele Iacobucci 2, 67100 L’Aquila, Italy}
\affil[12]{INFN, Laboratori Nazionali del Gran Sasso, 67100 Assergi (AQ), Italy}
% ...


\date{\vspace{-5ex}}


\maketitle
\newpage

\newpage
\setcounter{tocdepth}{2}
{
  \hypersetup{linkcolor=blue}
  \tableofcontents
}
  
\clearpage
\hypersetup{citecolor=blue}
\section{Introduction}
\paragraph{Astroparticle physics.} Astroparticle physics is a fascinating field of research at the intersection among astronomy, cosmology and particle physics. It is concerned with the study of particles of astronomical origin, and more in general with the study of the origin, structure and evolution of the universe with the tools of theoretical particle physics. As such, it explores the connection between  the micro- and the macro-cosm. 

This field of research is today undergoing a profound  transformation, due a series of extraordinary new results, such as the discovery of high-energy cosmic neutrinos with IceCube, the direct detection of gravitational waves with LIGO and Virgo, and many others. Yet formidable challenges remain open:  understanding the nature of dark matter and dark energy, testing gravity and the dynamics of compact objects, elucidating the origin of cosmic rays, understanding the matter-antimatter asymmetry problem, and so on. Addressing such fundamental questions  requires a diverse community of scientists active on all sides, and a strong synergy among experimentalists, observers, and theorists. 

\paragraph{EuCAPT.} To respond to such challenges, 
the European Consortium for Astroparticle Theory (EuCAPT) has been recently established to bring together the European community of theoretical astroparticle physicists and cosmologists. The goals of EuCAPT are to increase the exchange of ideas and knowledge;
to coordinate scientific and training activities;
to help scientists attract adequate resources for their projects; and
to promote a stimulating, fair and open environment in which young scientists can thrive.

\paragraph{White Paper.} This White Paper (WP) was initiated by the Steering Committee of EuCAPT, with the aim to identify the opportunities in our field for the next decade, and to strengthen the coordination of Astroparticle Theory in Europe. It is addressed to the whole astroparticle community, and rather than attempting an impossible review of the current status of the whole field of research, it focuses on the upcoming theoretical opportunities and challenges,  with particular emphasis  on the possible synergies among  different subfields.

The material is organised around 10 themes of research, each of them coordinated by 2 or 3 scientists:
 \begin{itemize}
\item {\bf Early universe}: Daniel Baumann and Laura Covi;
 
\item {\bf Dynamical spacetimes}: Rafael Porto and Philipp Moesta;
 
\item {\bf Nuclear Astrophysics}: Tetyana Galatyuk and Tanja Hinderer;
 
\item {\bf Cosmic accelerators}: Sera Markoff, James Matthews, and Enrico Ramirez Ruiz;
 
\item {\bf Traveling Messengers}: Daniele Gaggero and Kumiko Kotera;
 
\item {\bf Neutrino Properties}: Thomas Schwetz and Olga Mena;

\item {\bf Particles from stars}: Aldo Serenelli and Irene Tamborra;

\item {\bf Dark Matter}: Francesca Calore, David J. E. Marsh, and Christian Byrnes;
 
\item {\bf Dark Energy}: Alessandra Silvestri, and Julien Lesgourgues;
 
\item {\bf Astrostatistics}: Christoph Weniger and Roberto Trotta.

\end{itemize}

\noindent
Needless to say, the definition of these research themes is somewhat arbitrary, as there are many connections and no real clear boundaries among many of them. Yet we believe this structure offers a perspective which is more natural for the practicing theoretical astroparticle physicists and cosmologists, and complementary to that offered by e.g. the European Astroparticle Physics Strategy document, which is instead organised around 'messengers'. 

 The final document includes contributions from 135 scientists, who participated in the brainstorming sessions  at the first EuCAPT annual Symposium held at CERN from May 5 to 7, 2021, and provided feedback via a dedicated channel on the CERN Mattermost community platform. The final WP has been endorsed by about 400 members of the community.    
 

\section{Early Universe}
\label{sec:earlyuniverse}
{\bf Coordinators:} Daniel Baumann and Laura Covi.\\[2pt]
 {\bf Contributors:} Mustafa Amin, Nicola Bartolo, Chris Byrnes, Alex Cole, Daniel Cutting, Guillem Dom\`enech, Marco Drewes, Matteo Fasiello, Ricardo Ferreira, Julia Harz, Peter Millington, Claudio Paganini, Lucas Pinol, Rafael Porto, Sebastien Renaux-Petel, Angelo Ricciardone, Kai Schmitz, Pedro Schwaller, Evangelos Sfakianakis,  Daniele Teresi, Anna Tokareva and Sebastian Zell.
 \\

 \noindent
Progress in cosmology has been nothing short of astounding.
Observations of the cosmic microwave background, the clustering of galaxies and the luminosity distances of supernovae have culminated in a simple standard model of cosmology---the $\Lambda$CDM model. However, while the model is phenomenologically very successful, it raises many important questions, especially about the physics of the early universe: What happened in the first second of the universe's existence? What matter filled the universe during that time and what physical laws governed its interactions? What created the initial fluctuations that ultimately grew into all of the structures we see around us?  What created the baryon asymmetry of the universe? 
What is the nature of dark matter, where did it come from
or how was it generated?
In this chapter, we will discuss the theoretical challenges that need to be addressed to shed light on these important open problems.

\subsection{Primordial Correlations}

The fundamental observables in cosmology are spatial correlation functions. By measuring these correlations at different times in the universe's history, we learn both about the evolution of the universe and its initial conditions. Under relatively mild assumptions, the observed correlations can be traced back to a spatial surface at the origin of the hot Big Bang. The primordial correlations on this surface provide the initial conditions for the hot Big Bang.
To explain the large-scale correlations that are observed in the cosmic microwave background, however, requires that the hot Big Bang was in fact not the beginning of time, but instead the end of an earlier high-energy phase. 
A key challenge for modern cosmology is to uncover what precisely happened before the universe reached a state of thermal equilibrium and how the physics during that area gave rise to the primordial fluctuations. The difficulty lies in extracting clues about the physics of the primordial universe from measurements made in the late universe. On the theoretical side, this requires very accurate predictions for cosmological correlation functions, that can then be matched against the data.

\subsubsection{The Physics of Inflation} 

Although there is intriguing observational evidence that something like inflation occurred in the early universe, it must be emphasized that inflation is not yet a fact---at the same level that, for example, Big Bang nucleosynthesis is a fact. Both theoretical and observational advances are needed to either confirm or falsify the inflationary paradigm.

\vskip 4pt
{\bf UV sensitivity:} At a theoretical level, the challenge is that the microphysics during the inflationary era is highly uncertain, but at the same time the inflationary dynamics is sensitive to the details of the UV completion~\cite{Baumann:2014nda}.
There are two aspects to the problem: 1) Including high-energy degrees of freedom can spoil an otherwise successful inflationary model. To have confidence in the theory, it therefore either has to be demonstrated that the inflationary model is immune to the most general set of UV corrections, or the model must be embedded in a UV-complete framework such as string theory. 2) Less dramatically, the UV completion may not destroy inflation, but lead to interesting effects on cosmological observables, which should be included in the effective description of inflation~\cite{Cheung:2007st}. For example, as has been seen explicitly in a variety of cases, integrating out, or dynamically producing heavy fields, can affect the tilt of the scalar power spectrum, the tensor-to-scalar ratio, and the (non)-Gaussianity of the fluctuations.
It is also common in UV-complete theories to have several fields that are light enough to play a dynamical role during inflation, which should be described by a multi-field theory~\cite{Senatore:2010wk}, again affecting all observables (see e.g.~\cite{Chen:2009zp,Wands:2010af,
Achucarro:2010da,Renaux-Petel:2015mga,Slosar:2019gvt}).
However, a systematic charaterization of the effects of the UV completion of inflation on cosmological observables is still lacking. Recent developments in string theory and black hole physics even suggest that quantum gravity effects could be larger and more constraining than previously thought \cite{bhinfo, bhinfo-cosmo, bhinfo-swampland}, but the implications of these results for inflation are still unknown.

\vskip 4pt
{\bf Tensor modes:} B-mode measurements continue to progress steadily via a multi-faceted observational campaign.  
The next generation of CMB polarization experiments (such as the BICEP Array~\cite{Hui:2018cvg}, CMB-S4~\cite{CMB-S4:2016ple}, LiteBIRD~\cite{LiteBIRD:2020khw} and the Simons Observatory~\cite{SimonsObservatory:2018koc}) will be sensitive to primordial tensor perturbations characterized by a tensor-to-scalar ratio as small as $r \sim 10^{-3}$. 
Although string theory predicts the detectability  of the tensor-to-scalar-ratio in some models, there is no universally guaranteed signal. Nevertheless, the  observations will discriminate among qualitatively distinct classes of inflationary mechanisms.  
A B-mode detection would demonstrate that inflation occurred at a very high energy scales, making effects of the UV completion observationally more accessible. This would provide us with the luxury problem of looking for additional correlated signatures that would teach us more about the inflationary mechanism, e.g.~parity violations in the CMB \cite{Sorbo:2011rz}.
In addition to B-mode measurements, the direct detection of stochastic gravitational wave backgrounds by ground- and space-based interferometers, throughout the frequency range, has the potential to reveal the otherwise inaccessible inflationary dynamics on small scales. UV embeddings of inflation indeed motivate scenarios in which GWs are generated during inflation by nontrivial dynamics, involving for instance gauge or spinning fields~\cite{Cook:2011hg,Franciolini:2017ktv,Iacconi:2019vgc}, or after inflation by primordial scalar fluctuations of large amplitude~\cite{Acquaviva:2002ud,Ananda:2006af,Baumann:2007zm}. The frequency profile, chirality, anisotropies and cross-correlations with other probes 
offer precious insight into the mechanisms sourcing these GWs (see e.g.~\cite{Maleknejad:2012fw,Dimastrogiovanni:2019bfl,Adshead:2020bji,Fumagalli:2020nvq}), and their systematic exploration and characterization is an important open problem.

\vskip 4pt
{\bf Non-Gaussianity:} A lot could be learned about the dynamics of inflation if we could measure primordial higher-point correlations (non-Gaussianity) (see~\cite{Bartolo:2004if,Chen:2010xka,Wang:2013zva,Renaux-Petel:2015bja,Meerburg:2019qqi} for reviews). In particular, their shapes and scale-dependences would teach us about the field content and the interactions during inflation, while current observations only probe the free theory. 
A particular role is played by soft limits of correlators, most notably the squeezed limit of the three-point function, that have been identified as robust probes of the masses and spins of the fields active during inflation (see e.g.~\cite{Maldacena:2002vr,Creminelli:2004yq,Chen:2009zp,Baumann:2011nk,Noumi:2012vr,Assassi:2012zq,Arkani-Hamed:2015bza,Kehagias:2017cym}).
Non-Gaussianity can also provide the opportunity to diagnose whether the origin of the primordial fluctuations was quantum or classical in nature~\cite{Green:2020whw}. 
Many interesting examples of primordial non-Gaussianity have been constructed in specific models of inflation.
While these provide interesting targets for future observations (see e.g.~\cite{Dore:2014cca}) a systematic exploration of the allowed space of non-Gaussian correlators is an important open problem. Recently, a new bootstrap perspective was developed to carve out the space of inflationary correlations that are consistent with basic physical principles, such as causality, unitarity and locality (see e.g.~\cite{Arkani-Hamed:2018kmz,Goodhew:2020hob,Sleight:2019hfp, Arkani-Hamed:2017fdk}). However, a complete characterization of all consistency requirements for the inflationary correlators is still an outstanding problem. 
Eventually, beyond $n$-point functions, it is an interesting challenge to characterize the full statistical properties of primordial fluctuations in a nonperturbative framework. This is both of phenomenological interest, as the tails of the distribution are relevant to reliably predict the abundance of primordial black holes \cite{Franciolini:2018vbk}, and of conceptual interest, to better understand the infrared structures of correlators during inflation (see e.g.~\cite{Riotto:2008mv,Chen:2018uul,Ezquiaga:2019ftu,Gorbenko:2019rza,Cohen:2020php,Pinol:2020cdp}).

\subsubsection{Reheating and Preheating}

The end of inflation and the subsequent ``reheating" is essential for our understanding of how the hot Big Bang began, and how the physics of inflation is connected to the lower energy physics of the Standard Model (possibly via intermediaries). The end of inflation can be dynamically rich, and include perturbative and nonperturbative physics of energy transfer from the inflaton to daughter fields (for recent reviews, see \cite{Amin:2014eta,Lozanov:2019jxc}).

\vskip 4pt
With accelerated expansion having ended, there is no `cosmic amplifier' to make the (mostly) microscopic physics of reheating easily accessible on contemporary cosmological length scales. Moreover, the potentially high-energy scale associated with the end of inflation, as well as subsequent thermalization, can further hide details of this era from our low-energy probes.  Nevertheless, the dynamics during this period can generate relics such as isocurvature perturbations, stochastic gravitational waves (see \cite{Caprini:2018mtu} for a recent review), non-Gaussianities (e.g.~\cite{Chambers:2008gu,Bond:2009xx}), dark matter/radiation \cite{Mambrini:2013iaa,Baumann:2015rya,Daido:2017tbr,Hooper:2018buz,Bernal:2018hjm,Garcia:2018wtq,Green:2019glg}, primordial black holes \cite{Green:2000he,Carr:2020gox}, topological and non-topological solitons \cite{Amin:2011hj,Gleiser:2011xj,Amin:2019ums} and other small-scale structures (e.g.~\cite{Jedamzik:2010dq,Musoke:2019ima}), matter/antimatter asymmetry \cite{Hertzberg:2013mba,Cline:2019fxx}), and primordial magnetic fields (e.g.~\cite{Enquist:1998,Bassett:2000aw,Adshead:2016iae,Patel:2019isj}), providing incisive probes of this period. 
The expansion history, as well as local dynamics after the end of inflation, also impacts the inference of inflationary 
observables such as $n_s$, $r$ and $f_{\rm NL}$ (e.g.~\cite{Martin:2014vha, Lozanov:2016hid, Allahverdi:2020bys}). 

\subsection{Relics from the Hot Big Bang}

The high energies and high densities of the early universe provide access to new regimes of physics beyond the Standard Model (BSM). 
New massive particles can be produced during inflation~\cite{Arkani-Hamed:2015bza} and in the early phases of the hot Big Bang, whose traces we can look for in the late universe.
Moreover, the high densities in the early universe allow particles to be created that are only very weakly interacting with ordinary matter probing the weak-coupling frontier of BSM physics. Finally, various phase transitions are expected to have occurred in the early universe that can probe the physics of the SM and beyond.

\subsubsection{Thermal Relics}

Any particle with a substantial coupling to the SM or within a hidden sector (e.g.~$ g > 10^{-7} $ for a Yukawa/gauge coupling, 
and mass below the reheating temperature) reaches thermal equilibrium at some early epoch and if sufficiently stable survives as a thermal relic until the present day. 

\vskip 4pt
\textbf{WIMPs and similar relics:} Relics that decoupled as non-relativistic particles, such
as Weakly Interacting Massive Particles (WIMPs) \cite{Roszkowski:2017nbc, Arcadi:2017kky},
are prime candidates for cold dark matter (CDM) (see also Section~\ref{sec:DM}).
 The precise computation of their relic density after decoupling, governed by the annihilation interactions in the standard case,
 is one of the main challenges of DM phenomenology.
This theoretical determination is becoming more and more refined and reaching very high precision, thanks to the inclusion of higher-order radiative corrections \cite{Chatterjee:2012hkk, Colucci:2018qml, Schmiemann:2019czm, Beneke:2019vhz}, nonperturbative effects like the Sommerfeld enhancement \cite{Cirelli:2007xd, Hryczuk:2010zi, Beneke:2019vhz, Beneke:2020vff}, as well as co-annihilations with nearly degenerate particles \cite{Edsjo:1997bg, Ellis:2015vaa, DAgnolo:2018wcn}. 
Moreover, other types of interactions beyond annihilations can strongly affect decoupling---e.g.~conversions \cite{DAgnolo:2017dbv, Garny:2017rxs}, semiannihilations \cite{DEramo:2010keq, Kamada:2017gfc}, elastic scattering \cite{Binder:2017rgn}, two-to-three scatterings \cite{Hochberg:2014dra, Choi:2015bya}, as well as bound state formation \cite{vonHarling:2014kha, Cirelli:2016rnw, Harz:2018csl}---and they are the object of intense study.
Even the unitarity bound for thermal relics can be modified by
new effects and extended to higher masses \cite{vonHarling:2014kha, Bramante:2017obj, Kramer:2020sbb}.
All these new possibility in DM production have led to broadening the concept of WIMPs to include new characteristics and  new possibilities like multi-component DM \cite{DEramo:2010keq, Kamada:2017gfc}, as well as modifying the expected signals in direct and indirect detection.

On the other side, the production of feebly interacting or very weakly interacting particles from the thermal bath has been and is being explored further, as it is a universal feature of hidden sectors and of particles with weak couplings to the thermal bath.
Indeed the production through the FIMP \cite{Hall:2009bx, Bernal:2017kxu} or the SuperWIMP \cite{Feng:2003uy} mechanisms can be realised in many SM extensions and can lead to new signatures at colliders and in indirect detection \cite{Arcadi:2014tsa,Co:2015pka, Belanger:2018sti}.

\vskip 4pt
\textbf{Light relics:}
If the decoupling from the thermal bath happens when the particles are relativistic, they need to be light in order not to overclose the universe and so count as relativistic degrees of freedom for most of the cosmological history and are too hot to provide the full DM density. Typical relics of this type are the SM neutrinos giving rise to the cosmic neutrino background.
Light thermal relics, which remain relativistic until CMB decoupling, are strongly constrained by cosmological observations.
Indeed, the next generation of CMB experiments will improve constraints on the radiation density of the early universe by an order of magnitude~\cite{CMB-S4:2016ple}. This is one of the largest improvements in the constraints on cosmological parameters, providing an opportunity to constrain not only thermal relics, usually parametrised as additional effective neutrino species 
$ \Delta N_{\rm eff} $, see Section \ref{sec:neutrinoMass},
but also non-thermal or diluted light relics in any SM extension. Examples of the latter type are sterile neutrinos \cite{Abazajian:2017tcc, Boyarsky:2018tvu} or gravitinos \cite{Fujii:2002fv}, which can reach substantial energy densities and in some cases play the role of a non-negligible warm dark matter component. 
Future large-scale structure surveys like Euclid will be able to provide a better handle on these additional warm DM contribution, apart for possibly measuring the sum of the SM neutrino masses \cite{Amendola:2016saw}.

\subsubsection{Baryogenesis and Leptogenesis}

The baryon-to-photon ratio inferred from CMB data~\cite{Planck:2018vyg} indicates
that matter dominated over antimatter in the early universe at around $1$ part in
$10^{10}$. The necessary conditions for generating this asymmetry are baryon number
(B), charge (C) and charge parity (CP) violation, and the absence of thermodynamic
equilibrium \cite{Sakharov:1967dj}. Still the exact mechanism responsible for the baryon asymmetry of the universe has not yet been identified, nor the new physics and its particular scale. Two archetypal scenarios are:~{\it leptogenesis} \cite{Fukugita:1986hr,Fong:2012buy}
(see~\cite{Dev:2017trv, Drewes:2017zyw, Dev:2017wwc, Biondini:2017rpb, 
Garbrecht:2018mrp, Bodeker:2020ghk} for recent reviews
and the discussion in Section \ref{sec:neutrinoLG}), wherein
out-of-equilibrium decays of new states connected to the neutrino mass generation
produce a net lepton number (L) that is converted into a net baryon number by sphaleron 
processes \cite{Kuzmin:1985mm},
and {\it electroweak (EW) baryogenesis}~\cite{Kuzmin:1985mm, Riotto:1999yt}, 
wherein the baryon asymmetry is generated during a strong first-order EW phase transition 
(EWPT).\footnote{While these two scenarios are the best studied, there are
other promising ways to produce the baryon asymmetry which
exploit and explore other theoretical model characteristics, 
such as the non trivial dynamics of scalar fields as in 
 Affleck-Dine \cite{Bramante:2017obj, Bettoni:2018utf} and spontaneous baryogenesis \cite{Co:2019wyp,Domcke:2020kcp},
or the cogenesis with dark matter like 
in asymmetric DM \cite{Petraki:2013wwa, Zurek:2013wia}. It remains important to explore 
the wide range of possible baryogenesis scenarios and their observational consequences.} These two examples illustrate the open questions in baryogenesis
scenarios:

\vskip 4pt
\textbf{CP violation:} Usually additional sources of CP violation beyond the SM 
are needed; the observation of CP violation in electric dipole 
moments~\cite{Morrissey:2012db, Fuyuto:2015ida}, neutrino 
oscillations~\cite{Hagedorn:2017wjy,Chun:2017spz} or collider
searches~\cite{Chun:2017spz} could in some cases pin down the mechanism and parameter 
space of different baryogenesis models, even if in general different combinations of
phases could be important at high/low energy.

\vskip 4pt
\textbf{B-violating signatures as a guide:} $B-L$ is an accidental global symmetry
in the SM. If $B$- and $L$-violating interactions exist linked to baryogenesis,
the mechanism behind baryogenesis can be tested by the discovery of such interactions 
in the interplay of nucleon decays~\cite{Heeck:2019kgr}, future
neutron-antineutron oscillation experiments~\cite{Proceedings:2020nzz, Fridell:2021gag} or meson oscillations~\cite{Elor:2018twp}.

\vskip 4pt
\textbf{Neutrinos and LNV signatures:} The observation of $L$-violation (LNV) at low
energies can give insights into the mechanisms of leptogenesis and neutrino mass
generation, hinting at a Majorana neutrino nature. If observed at collider or
neutrinoless double $\beta$-decay experiments, it could falsify high-scale
leptogenesis; when combined with neutrino oscillation data and other collider
observables, it could pin down the underlying model in low-scale
scenarios~\cite{Chun:2017spz}. It is not yet clear if the implications of observing specific LNV processes 
can be understood in a model-independent way and what is the best way to combine different data sets to test low-scale leptogenesis.

\vskip 4pt
\textbf{Accelerator probes:} EW baryogenesis and leptogenesis require new particles
that contribute to the Higgs potential or neutrino masses, respectively, and there
is a rich program optimising the discovery potential for these particles at existing
and future facilities via new searches or additional detectors~\cite{Curtin:2014jma,
Deppisch:2015qwa, Huang:2015izx, Chun:2017spz, Beniwal:2017eik, Cai:2017mow, Agrawal:2021dbo}.

\vskip 4pt
\textbf{Gravitational waves:} The $125$ GeV Higgs mass is
heavy enough that the SM EWPT is a crossover, and EW baryogenesis can be realised
only in extensions of the SM~\cite{Morrissey:2012db}.  The required strong
first-order EWPT could produce GWs in the mHz range targeted by
LISA~\cite{Caprini:2015zlo}.  One naively expects EW baryogenesis to require
subsonic wall velocities, in tension with the relativistic velocities needed for
detectable GWs. Can EW baryogenesis be achieved for fast bubbles and how do we
calculate the asymmetry in this case (see e.g.~\cite{Cline:2020jre,Dorsch:2021ubz} and refs.~therein)?

\subsubsection{Phase Transitions}

The thermal universe is a unique playground to study new phases of matter that arise at high temperatures, for example, the quark-gluon plasma and the unbroken electroweak (EW) 
symmetry, within the SM, but also possibly new symmetries in extensions of the SM, such as grand unified theories or string theory.
As extended symmetry breaking often leads to the formation of topological defects, like strings or monopoles, strong constraints can be placed on these new models from cosmology \cite{Lopez-Eiguren:2017dmc, Auclair:2019wcv,
Ellis:2020ena, Blasi:2020mfx, Buchmuller:2020lbh}.
Phase transitions after inflation can lead to new important signatures, especially in case of a strong first-order phase transition, with strong deviations from thermal equilibrium 
and homogeneity. 

\vskip 4pt
\textbf{EW phase transition:}
The EW phase transition has been the object of intense study, also as a possible source of baryogenesis (see above). Unfortunately, the SM has not enough CP violation \cite{Gavela:1993ts}, nor realises a strong first-order
phase transition for Higgs masses above $80$ GeV \cite{Kajantie:1996mn}. 
Nevertheless, the EW phase transition can be changed by modifying the EW sector and in particular the Higgs potential, either by adding non-renormalizable operators \cite{Grojean:2004xa}, by extending the Higgs sector \cite{Espinosa:2011ax} or by adding new degrees 
of freedom with a substantial coupling to the Higgs, like 
the stop sector in supersymmetry \cite{Chung:2012vg}. All these modifications influence the Higgs couplings and often
lead to observable effects at future colliders and the next years will be crucial in order to test the nature of the EW phase transition both in cosmology and in collider physics \cite{Curtin:2014jma,Huang:2015izx, Beniwal:2017eik}.

\vskip 4pt
\textbf{BSM phase transitions:} Many extensions of the SM predict additional symmetries and therefore phase transitions that lead to the surviving SM gauge group. In grand unified theories, the scale of the symmetry breaking is often higher than the inflationary energy scale, even if one can have models combining inflation and symmetry breaking, like in hybrid inflation \cite{Senoguz:2003zw}.
On the other hand, part of the symmetry breaking could happen after inflation, even as a first order phase transition, and give more directly observable signatures \cite{Cutting:2018tjt}. Another class of important symmetries are global anomalous symmetries like the Peccei-Quinn $U(1)$ symmetry, which after breaking give rise to the axion or axion-like particles as DM candidates \cite{Gorghetto:2020qws}
or source of the baryon asymmetry \cite{Co:2019wyp, Domcke:2020kcp} 
or helping stabilising the hierarchy, and these fields could
also generate observable GW signals~\cite{Gorghetto:2021fsn,Ratzinger:2020koh, Banerjee:2021oeu}.

\vskip 4pt
{\bf Gravitational waves:} It is well known that cosmological first-order phase transitions give rise to a stochastic gravitational wave background~\cite{Witten:1984rs,Hogan:1986qda}. Now that the detection of GWs has become a reality, it has become more urgent to understand what we could learn from a detection of a stochastic GW background about the early universe.\footnote{Pulsar timing arrays (PTAs)~\cite{Burke-Spolaor:2018bvk,Taylor:2021yjx}, specifically, EPTA~\cite{Kramer:2013kea,Lentati:2015qwp}, NANOGrav~\cite{McLaughlin:2013ira,NANOGrav:2020gpb}, PPTA~\cite{Manchester:2012za,Shannon:2015ect}, and IPTA~\cite{Hobbs:2009yy,Verbiest:2016vem} are currently on the brink of discovering a stochastic GW background (SGWB).
In 2020, NANOGrav was the first PTA collaboration to present strong evidence for a new stochastic process affecting its 12.5-year data~\cite{NANOGrav:2020bcs}. 
Joint PTA analyses based on larger data sets in the next years, eventually leading up to PTA observations with FAST~\cite{Hobbs:2014tqa} and SKA~\cite{Janssen:2014dka,Weltman:2018zrl}, will help clarify whether this process really corresponds to a SGWB signal and shed more light on its origin~\cite{NANOGrav:2020spf}.
Possible explanations include the mergers of supermassive black-hole binaries~\cite{Middleton:2020asl} on the astrophysical side as well as an abundance of BSM scenarios on the cosmological side, including but not limited to cosmic strings~\cite{Ellis:2020ena,Blasi:2020mfx,Buchmuller:2020lbh}, scalar-induced GWs (SIGWs) generated at second order of perturbation theory in conjunction with the production of primordial black holes~\cite{Vaskonen:2020lbd,DeLuca:2020agl,Kohri:2020qqd,Domenech:2020ers}, cosmological phase transitions~\cite{Nakai:2020oit,Addazi:2020zcj,Ratzinger:2020koh,Neronov:2020qrl}, and axions~\cite{Ratzinger:2020koh,Banerjee:2021oeu}.} The frequency range of the potential signal is set by the Hubble radius at the time of the PT, and it is intriguing that a PT happening at temperatures around the weak scale or TeV scale would give a signal in the most sensitive region of the upcoming LISA experiment~\cite{LISA:2017pwj}. Higher PT scales could be probed by LIGO and follow-up experiments such as the Einstein Telescope and Cosmic Explorer, while PTA experiments are sensitive to PTs at later times that could potentially happen in a hidden sector~\cite{Schwaller:2015tja,Breitbach:2018ddu}.

\subsection{Outlook and Open Problems}

\begin{figure}[t!]
   \centering
\includegraphics[width=0.6\textwidth]{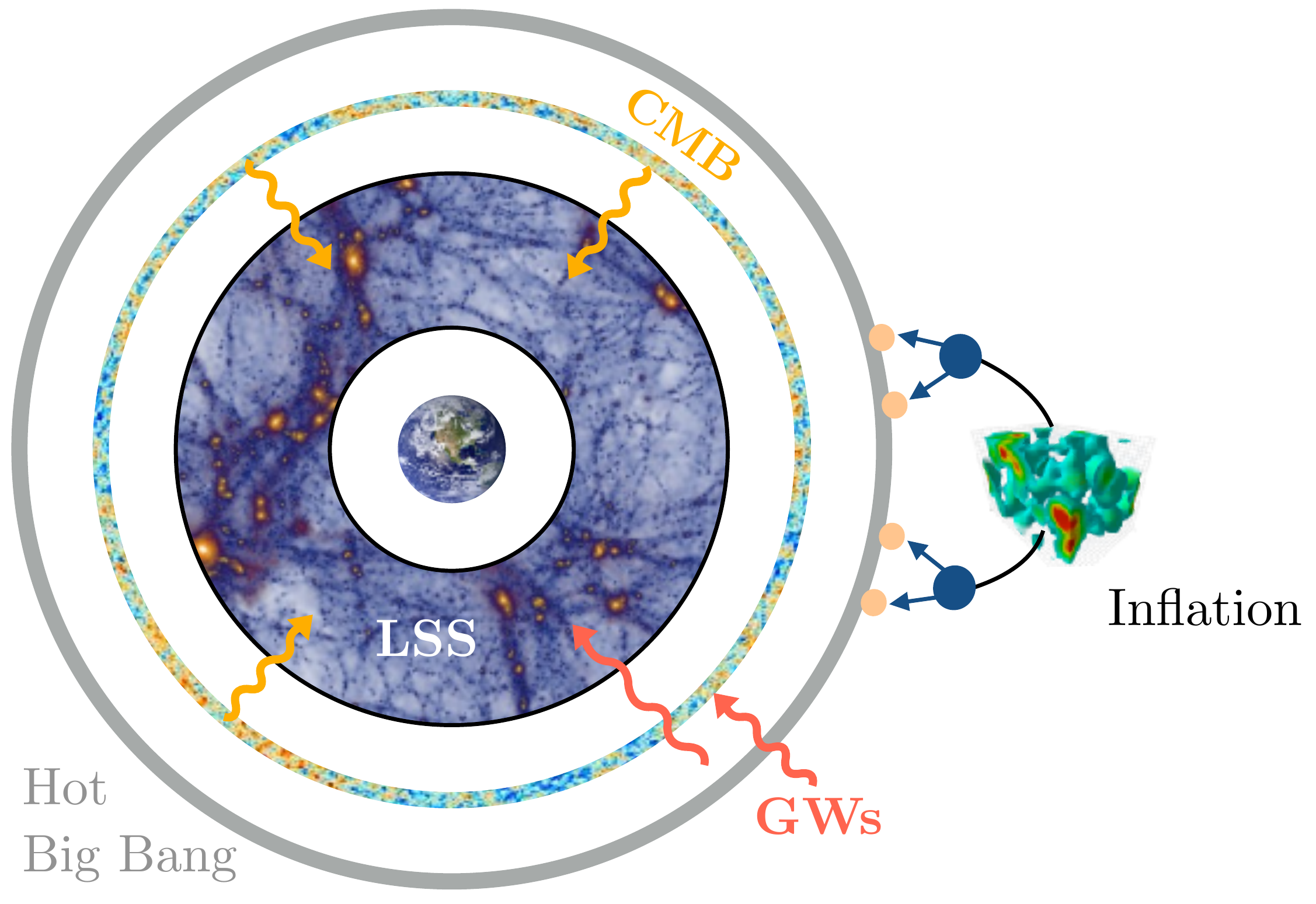}
   \caption{Traces of early universe physics can be looked for through multiple observational windows.}
   \label{fig:Probes}
\end{figure}

We are entering an age of ``multi-messenger cosmology", with multiple cosmological probes (CMB, LSS and GWs)  poised to provide new tests of the physics of the early universe (see Fig.~\ref{fig:Probes}). In order for these observations to fulfill their true potential, however, theoretical advances will be crucially needed. The following is a list of some of the most important open problems for each of the topics described above. 

\begin{itemize}
    \item {\bf Inflation:} Despite being a very successful phenomenological model, inflation is not yet a complete theory.
    In particular, the microscopic origin of the inflationary expansion is still unknown. This challenge can be addressed through two complementary approaches. On the one hand, it remains important to construct explicit models of inflation and study their observational predictions, including the effects of UV completion. On the other hand, we can carve out the space of consistent inflationary correlations starting from basic physical principles such as locality, causality and unitarity. In this way, we can hope to provide a systematic classification of the inflationary predictions. 
At a more phenomenological level, future galaxy surveys will provide interesting new constraints on non-Gaussian correlations. In order for these observations to fulfill their true potential, however, the non-Gaussianity associated to nonlinear gravitational evolution and galaxy biasing must be characterized very accurately, so that the primordial signals can be extracted reliably. This will require both advances in numerical simulations, as well as improvements in the  theory description of large-scale structure.
    
    \item  {\bf Reheating:} Some of the theoretical challenges related to the reheating era include:  (1) Delineating the model dependent vs.~relatively universal 
predictions, both from a model-building perspective and those resulting from nonlinear phenomena (e.g.~\cite{Amin:2015ftc,Ozsoy:2015rna,Figueroa:2016wxr,Amin:2017wvc,Antusch:2020iyq}). 
(2) Numerically simulating the nonperturbative physics of this period with increasingly more `realistic' field content -- scalars, fermions, Abelian and non-Abelian fields (see~\cite{Felder:2000hq,Frolov:2008hy,Easther:2010qz,Huang:2011gf,Giblin:2019nuv,Lozanov:2019jff,Figueroa:2021yhd}) and detailed accounting of metastable/solitonic structures, full quantum and gravitational effects. (3) Performing the numerical simulations 
long enough to reach full (local) thermalisation and establish thermalisation beyond the simple linear response or Boltzmann equation approximations \cite{Harigaya:2019tzu, Drees:2021lbm}, investigating which type of deviations from thermal equilibrium could be expected to survive, 
e.g. for hidden sectors or more weakly interacting particles.
(4) A more careful investigation of the inverse problem: what information from cosmic relics is essential to give us 
hints about inflaton self-couplings and the couplings of the inflaton to its daughter fields? 

\item {\bf Thermal relics:}
The theoretical precision in the computation of the dark matter energy density has to match the observational precision and therefore one of the main goals is to provide state-of-the-art computations and numerical codes for different WIMP-like models 
and explore new production mechanisms~\cite{GAMBITDarkMatterWorkgroup:2017fax, Belanger:2018ccd, Bringmann:2018lay, Ambrogi:2018jqj, Binder:2021bmg}, 
possibly even beyond the traditional WIMP 
mass window \cite{DAgnolo:2018wcn, Kramer:2020sbb}.
The present description of the number density evolution via a classical Boltzmann equation breaks down in particular cases, e.g.~in the presence of DM bound states, calling for the use of thermal field theory and the inclusion of thermal effects \cite{Biondini:2018pwp, Binder:2018znk, Binder:2019erp}. 
One of the basic questions is if it is possible to identify the dark matter as a thermal relic, via the precise measurement of 
the annihilation cross-section in indirect detection or other observations. Of course, modifications to the cosmological model and additional dark matter components can introduce degeneracies in the comparison of theory and observations that we have to be addressed.
    
\item {\bf Baryogenesis:}   The generation of the baryon asymmetry is a complex process, including inevitably CP violation and quantum effects,  and systematic first-principles approaches based on non-equilibrium quantum field theory are not always possible in full generality (see e.g.~\cite{Dev:2017trv, Dev:2017wwc, Biondini:2017rpb, Garbrecht:2018mrp}). An important issue is to either determine simplifying assumptions beyond a model-by-model basis that make these approaches tractable or to identify criteria that allow us to decide when the classic Boltzmann treatment is sufficient to reach a reasonable prediction.
    
    \item {\bf Phase transitions:} 
    Even given a concrete model, predicting the GW signal from PTs is a highly non-trivial task that requires elaborate analytical approximations and numerical simulations, since the PT is a highly non-equilibrium process. Substantial progress has been made in the past decade, with advances being made in understanding both the size and form of the GW signal in many realistic scenarios, summarised in recent reviews e.g.~\cite{Caprini:2015zlo,Caprini:2019egz,Hindmarsh:2020hop}. However, many key aspects of the determination of the GW signal require better understanding for precise GW predictions. 
     In particular, better predictions of the GW signal from sound waves in strong and very strong PTs are required~\cite{Cutting:2019zws,Jinno:2019jhi}, as well as studies of the subsequent decay of sound waves and formation of turbulence~\cite{Pen:2015qta,Ellis:2020awk,Guo:2020grp}. The production of GWs from (magneto)hydrodynamic turbulence needs to be understood~\cite{Niksa:2018ofa,RoperPol:2019wvy,Brandenburg:2021bvg} and the possible generation of magnetic fields during a PT will be important~\cite{Zhang:2019vsb,Yang:2021uid}. Recent works with nonperturbative methods call into question the validity of the perturbative expansion at finite temperature~\cite{Kainulainen:2019kyp,Tenkanen:2019aij,Croon:2020cgk,Gould:2021oba}. Predictions of PT parameters and the GW spectrum in strongly coupled theories is an open problem, with holographic methods making recent progress~\cite{Megias:2020vek,Ares:2020lbt,Bigazzi:2020phm}. The bubble wall velocity is a challenging to compute but crucially important parameter for predictions for both baryogenesis and GW production~\cite{Bodeker:2009qy,Bodeker:2017cim,Hoche:2020ysm,Azatov:2020ufh,Bigazzi:2021fmq,Bea:2021zsu}. In the event of a runaway bubble wall, the GW spectrum will be modified and its precise form is under investigation~\cite{Lewicki:2019gmv,Lewicki:2020azd,Cutting:2020nla,Gould:2021dpm}. Finally, on the observation of a GW signal we will need to be able to reconstruct the PT parameters and the underlying model masses and couplings~\cite{Gowling:2021gcy,Giese:2021dnw}.
 
    \item {\bf Gravitational waves:} Future GW observations will provide an exciting new window into the early universe. In the run-up to these observations, the theory community will have to work towards a better understanding of the various potential sources (e.g.~sound waves~\cite{Guo:2021qcq,Gowling:2021gcy} and magnetohydrodynamic turbulence~\cite{Brandenburg:2021bvg} during phase transitions~\cite{Croon:2020cgk}, non-minimal cosmic-string models unifying the Nambu--Goto approximation with the Abelian-Higgs picture~\cite{Matsunami:2019fss,Hindmarsh:2021mnl}, GWs generated by nontrivial dynamics during infation or sourced by small-scale scalar fluctuations of large amplitude~\cite{Cook:2011hg,Garcia-Bellido:2017aan,Bartolo:2018evs,Fumagalli:2020nvq}) and continue the development of techniques for foreground removal~\cite{Martinovic:2020hru,Poletti:2021ytu,Boileau:2021sni}, signal reconstruction~\cite{Karnesis:2019mph,Caprini:2019pxz,Flauger:2020qyi} as well as for the analysis of anisotropies~\cite{Suresh:2021rsn}, polarisation and cross-correlations with other cosmological probes~\cite{Domcke:2019zls,Ricciardone:2021kel,Braglia:2021fxn}.
    
\end{itemize}
\section{Dynamical Spacetimes}
{\bf Coordinators:} Rafael A. Porto and Philipp Moesta.\\
 {\bf Contributors:} Tim Dietrich, Badri Krishnan, Paolo Pani, Adam Pound, Fernando Quevedo, Radu Roiban and Mao Zeng.
 \\

The nascent field of GW science will be a truly interdisciplinary subject enriching different branches of physics, from testing gravity deep in the strongly coupled regime to constraining astro-particle physics phenomena in the early and late universe. Yet, the associated theoretical challenges are enormous \cite{Buonanno:2014aza,Porto:2016zng,Porto:2017lrn}. 
Faithful waveform models are a compulsory ingredient for successful data analysis and reliable physical interpretation of the signals. This is critical, for instance, to study the nature of compact objects in binary systems as well as probing physics beyond the standard model both with cosmological and astrophysical sources. However, while present waveforms may be sufficient for detection and crude parameter estimation with LIGO, Virgo and Kagra detectors \cite{Abbott:2020niy,Aasi:2013wya}, they are too coarse \cite{Purrer:2019jcp,Ferguson:2020xnm} for {\it Precision Physics} with GW data from the next generation of broader-band observatories such as LISA and ET. It is in this respect that we find ourselves in an analogous situation to particle physics in the pre-LHC era: for~key processes within planned empirical reach, we are currently dominated by theoretical/numerical uncertainty. To~move forward and profit the most from future GW observations, source modelling covering all the regimes of the relativistic dynamics, and increasing in accuracy in parallel with detector-sensitivity, will be sorely needed. In what follows we review the associated opportunities and major challenges in the era of GW astronomy.

\subsection{Opportunities}

\subsubsection{Compact Objects}

One of the main goals of GW observations is to elucidate the nature of compact objects in the universe. Not only future GWs will probe the dynamics of strongly-interacting matter in the unique environment inside neutron stars \cite{Dietrich:2020eud}; they will also reveal whether black holes behave as dictated by classical GR, e.g. no-hair theorems \cite{Isi:2019aib}, vanishing Love numbers \cite{LeTiec:2020bos,Chia:2020yla}, absorptive properties \cite{Poisson:2004cw}; or perhaps even open the door towards quantum gravitational effects on horizon scales, e.g. fuzzballs~\cite{Bianchi:2020bxa}. Moreover, while black holes and neutron stars may be the most common GW sources, other type of compact objects could still roam around in the cosmos so far undetected, with a vast zoo of exotic alternatives --- such as clouds of ultralight particles around black holes \cite{Arvanitaki:2010sy, Baumann:2018vus,Baumann:2019ztm}, boson stars \cite{Sennett:2017etc}, etc., only weakly constrained by present GW data~\cite{Cardoso:2019rvt}. That is the case mainly because unraveling internal structure entails distinguishing between the different imprints of minute finite size corrections (and/or frequency-dependent effects) during the inspiral regime, or the details of the merger and ringdown phase in various possible scenarios. Not only this requires large signal-to-noise ratios for the different regimes of the dynamics, as expected from LISA and ET, but we must also provide a (very) accurate description of the {\it standard model background} evolution of black holes and neutron stars in GR. This~turns out to be one of the major challenges in the study of dynamical spacetimes, as we discuss shortly. 

\subsubsection{Dark Matter}

The detection of GWs has also initiated a new era for dark matter searches \cite{Bertone:2019irm}. For instance, various types of exotic compact objects may very well constitute a fraction of the dark matter abundance in the universe, including halos of ultralight particles \cite{Hui:2016ltb} as well as black holes from primordial origin \cite{Sasaki:2018dmp,Carr:2020xqk,DeLuca:2021wjr}. 
For the former, in particular, precision GW data from binary systems carrying a boson cloud may be a natural laboratory to flesh out the masses and spins of putative new light states in nature \cite{Baumann:2018vus,Baumann:2019ztm}.  For the latter, constraints on the parameters of the sources, for example the detection of sub-solar mass black holes, could provide evidence for the primordial nature of these objects. In addition, environmental effects due to dark matter can also modify the waveforms for the coalescence of black holes \cite{Kavanagh:2020cfn}. For all these cases, an accurate reconstruction of the GW signal is essential to identify the nature of the sources.

\subsubsection{General Relativity}

Black holes and compact stars, together with the production of GWs, are among the most spectacular predictions of GR. It is therefore natural to use them as probes of the most fundamental principles of Einstein's theory: the equivalence principle, the number of gravitational polarizations or degrees of freedom that mediate the gravitational interaction, the speed of GWs and their propagation, the existence of extra hair around black holes or neutron stars, etc. \cite{Will:2014kxa}. In general, extensions of GR modify both the conservative dynamics of a binary system, e.g. by affecting its binding energy, as well as the dissipative sector, e.g. by introducing new radiative degrees of freedom or modifications to the absorptive properties on horizon scales. These effects can be chiefly used to search for alternative theories of gravity. In some cases, when the new degrees of freedom are relevant at scales either comparable or larger than the typical compactness of the bodies but shorter than the typical merger scale, modifications of GR  may be parametrized using an EFT formalism \cite{Goldberger:2004jt,Porto:2005ac,Goldberger:2005cd,Porto:2007qi,Sennett:2019bpc}. In other cases, alternatives to GR are deeply connected to screening mechanisms that make them phenomenologically viable both near compact objects or in the solar system as well as at cosmological scales, e.g. \cite{deRham:2014zqa}. Hence, testing gravity with GWs and compact objects will also help us elucidate its behavior on large scales, connecting also to the dark matter and dark energy problem, e.g. \cite{Creminelli:2017sry}.

\subsection{Hunt for Accuracy}

The non-linearities of Einstein's field equations and various scales involved make the enterprise of solving the two-body problem in gravity --- both in the strongly-coupled as well as the perturbative regime of small velocities, weak fields or small-mass ratios --- a real {\it tour de force}, which has required over the years a concerted effort involving various analytic and numerical methodologies \cite{Blanchet:2013haa,Lehner:2014asa,Porto:2016pyg,Barack:2018yvs,Schafer:2018kuf}.

\subsubsection{Perturbative Regime}

\hspace{0.4cm}{\it Post-Newtonian/Minkowskian expansion.} The state-of-the-art in Post-Newtonian (PN) theory is approaching the 4PN order for non-spinning \cite{Damour:2016abl,Porto:2017dgs,Marchand:2017pir,Foffa:2019yfl,Marchand:2020fpt} and spin-dependent \cite{Bohe:2013cla,Levi:2016ofk,Pardo:2020hxc,Cho:2021mqw} effects, with partial results known at higher PN orders \cite{Foffa:2019eeb,Blumlein:2020pyo,Blumlein:2021txj}. 
For some key contributions, this entails computations at the next-to-next-to-next-to-next-to leading order (N$^4$LO) beyond Newton's equations. In the jargon of particle physics this is equivalent to a {\it four loop} calculation which, although seemingly unrelated, has more in common with the actual derivation in GR that one might think. Starting with the pioneering work in \cite{Goldberger:2004jt}, the framework of QFT \cite{Rothstein:2003mp} as well as scattering amplitude methodologies \cite{Bern:2019prr}, often restricted to earth-based accelerators such as the LHC, have been successfully adapted to the (classical) inspiral problem in Einstein's gravity, e.g. \cite{Goldberger:2009qd,Galley:2015kus,Blanchet:2019rjs,Cheung:2018wkq,Kosower:2018adc,Bjerrum-Bohr:2018xdl,Bern:2019crd,Kalin:2019rwq,Kalin:2019inp,Kalin:2020mvi,Goldberger:2020fot,Jakobsen:2021smu,Mougiakakos:2021ckm,Herrmann:2021lqe,DiVecchia:2021bdo}. In particular, the application of these ideas --- notably the EFT machinery \cite{Porto:2016pyg} --- has contributed with high-accuracy calculations in the PN regime relevant to GW science; but also recently in the Post-Minkowskian (PM) expansion, due the existence of a map between gauge-invariant observables of the scattering problem (e.g. deflection angle) and those for bound orbits (e.g. periastron advance and binding energy) \cite{Kalin:2019rwq,Kalin:2019inp}. After the breakthrough at N$^2$LO (3PM) \cite{Bern:2019nnu,Kalin:2020fhe}, subsequent progress furthered our knowledge of the conservative dynamics in PM theory, currently closing on the spinless 4PM order \cite{Bern:2021dqo,Dlapa:2021npj}, as well NLO results on spin \cite{Bern:2020buy,Chung:2020rrz,Guevara:2019fsj,Liu:2021zxr,Kosmopoulos:2021zoq}  and tidal effects \cite{Kalin:2020lmz,Cheung:2020sdj}. Radiative corrections have also been computed at leading order in \cite{Dlapa:2021npj,Jakobsen:2021smu,Mougiakakos:2021ckm,Herrmann:2021lqe}. These PM results incorporate an infinite tower of (special-)relativistic contributions to GW observables, which can ultimately improve waveform modeling \cite{Antonelli:2019ytb}, in particular for hyperbolic encounters as well as eccentric~orbits. 

\vskip 4pt

{\it Small-mass-ratio limit.} 
Decades of progress have enabled simulations of inspirals and waveforms for generic orbits around a spinning black hole at the leading order in an adiabatic expansion~~\cite{Burke:2019yek, Chua:2020stf,Fujita:2020zxe,Katz:2021yft}. The first sub-leading order (or first ``post-adiabatic" 1PA) involves conservative and dissipative effects, as well as corrections due to the companion's spin. Calculations of the conservative self-force at first order are now mature, e.g.~\cite{vandeMeent:2018rms}, and there has been significant recent work on the effects of the secondary's spin, e.g. \cite{Skoupy:2021asz}. At the frontier in the field, ongoing work at second order~\cite{Upton:2021oxf,Pound:2021qin,Warburton:2021kwk} has recently culminated in the calculation of complete 1PA waveforms~\cite{Wardell:2021} in the restricted case of quasi-circular, non-spinning binaries. Self-force theory also has surprising application at the opposite end of the mass-ratio spectrum. Although originally intended only to model small-mass-ratio inspirals, self-force calculations have proved to be unexpectedly accurate for all mass ratios, demonstrating remarkable accuracy in comparison with numerical simulations, e.g.~\cite{vandeMeent:2020xgc,Warburton:2021kwk}.\vskip 4pt

{\it Ringdown.} The post-merger phase, during which the result of the collision settles into its final configuration, is also amenable to a perturbative appproach. For the case of Kerr black holes, the ringdown can be described in terms of quasi-normal modes \cite{Kokkotas:1999bd}. Unlike the inspiral phase, where finite-size effects contribute only towards the end, the final stages carry direct information about the nature of the source through GW {\it spectroscopy} \cite{Isi:2019aib}, allowing for several consistency tests of GR \cite{Capano:2021etf} as well as new searches for exotic compact objects \cite{Cardoso:2019rvt}.

\subsubsection{Numerical Relativity} While perturbative methods employ analytical approximations to 
solve Einstein's equations for different parts of the binary coalescence, 
numerical relativity makes usage of a numerical discretization to solve the field equations, and is therefore general enough to study gravitational fields when they are at their strongest. There has been tremendous progress in numerical relativity over the last decades, e.g., the first successful simulations of binary black 
holes~\cite{Pretorius:2005gq,Baker:2005vv,
Campanelli:2005dd} or binary neutron star systems~\cite{Shibata:1999wm}. 
Among others things, it is now possible to simulate binary black hole systems for several hundred orbits~\cite{Szilagyi:2015rwa} and cover large regions of parameter space through dedicated simulation campaigns. These have led to numerous databases and catalogs, e.g.~\cite{Ajith:2012az,Boyle:2019kee,Dietrich:2018phi,Kiuchi:2019kzt}, that are freely available, further strengthening the exchange within the community. The community has also managed to perform simulations beyond pure GR, e.g.~\cite{Okounkova:2019zjf}, or including exotic compact objects such as boson stars~\cite{Liebling:2012fv}. Linking signals other than GWs to the properties of the sources is also key ingredient for multi-messenger studies. To that end, increasing details of the micro-physics have been added to the simulations, e.g. inclusion of magnetic fields, neutrinos, viscosity, finite-temperature, and composition effects. This allows us to study electromagnetic signatures connected to out-flowing high-density material \cite{Hotokezaka:2012ze,Lehner:2016lxy,Radice:2016dwd}.

\subsection{Future Challenges}

 All of the developments described above have extended the knowledge of the dynamics of spacetime and GW emission sourced by binary compact objects to high level of accuracy, both in a perturbative scheme and through simulations. Yet, even though in combination  they have led to waveform models that may be sufficient for the capabilities of present detectors, we are still falling short of what will be required by the increase in sensitivity and expected number of cycles projected for third-generation GW observatories \cite{Purrer:2019jcp,Ferguson:2020xnm}. 
 
 In addition, the minute corrections from tidal deformations, first entering at 5PN order in the inspiral regime, may be swamped by our lack of analytic understanding of the general relativistic dynamics when these contributions are manifest, thus hindering the potential for new discoveries in GW astronomy. 
This calls for invigorating the theoretical efforts towards reaching the accuracy needed to properly interpret GW signals in future detectors. Indeed, this is an area where the synergies between different methodologies has blossomed, bridging between seemingly unrelated lines of research under the umbrella of GW science.  

\subsubsection{Analytic Toolbox}

\hspace{0.4cm} {\it Traditional methods.} The PN expansion has been the standard weapon of choice for over a hundred years to tackle the two-body problem with weakly gravitating slowly moving sources, using traditional methodologies to solve Einstein's equations \cite{Blanchet:2013haa,Schafer:2018kuf}. Presently, the status in PN theory with standard tools is at the fourth order \cite{Damour:2016abl,Marchand:2017pir} in the conservative part, with partial results also in the radiation sector \cite{Marchand:2020fpt}. A technical challenge in the derivation at 4PN --- which led to discrepant results before agreement was reached --- is the presence of radiation-reaction tail effects inducing non-local in time (conservative) interactions, yielding  divergent contributions in a PN scheme. The existence of intermediate divergences, also in the near zone derivation, resulted in various ``ambiguity parameters'', which were initially resolved through comparison with self-force computations \cite{Damour:2014jta,Bernard:2015njp,Bernard:2016wrg}. Inspired by the use of dimensional regularization in EFT-based methodologies (already accounting for the correct finite contribution from tail terms  \cite{Galley:2015kus}), the ambiguities were ultimately tackled via a delicate matching condition between near and far zone computations \cite{Marchand:2017pir}, although still relying on additional regularization schemes. 

Together with the increasing complexity in  PN computations, these difficulties demonstrate the challenges ahead using traditional methods. This has led various practitioners to advocate instead for a mixture of results from different fields, or a {\it Tutti Frutti} \cite{Bini:2019nra,Antonelli:2020aeb}, incorporating also self-force calculations.\vskip 4pt

Self-force theory~\cite{Barack:2018yvs}, which is tailored to the regime of small mass ratios but strong fields and fast motion, provides another perturbative method for modelling compact binaries. In~the small-mass-ratio limit, there is an overarching challenge: extreme-mass-ratio inspirals will persist in the LISA band for thousands of orbits, which are expected to be highly eccentric, inclined, and spinning. Hence, waveform models must accurately track the phase for its entire duration in band (and likely through orbital resonances). It is expected that 1PA calculations will be necessary and sufficient for this purpose~\cite{Pound:2021qin}. Self-force computations face two particular challenges in pursuing this goal. First, while waveforms can ultimately be generated rapidly enough for data analysis purposes~\cite{Katz:2021yft}, doing so requires pre-computing the self-force and waveform amplitudes at a dense sample of points in the binary parameter space. Current first-order calculations in Kerr spacetime are prohibitively expensive and must be radically sped up to span the parameter space. The second, more daunting challenge is that second-order self-force calculations must be extended from a Schwarzschild to Kerr backgrounds, and from quasi-circular orbits to generic ones. The same challenges appear also for the study of intermediate-mass ratios. 

Besides extensions into the regime of comparable masses, self-force theory has also played an important role in creating synergies between methods, often facilitated by extremely high-order analytical PN expansions,~e.g.~\cite{Bini:2019zjj}, by the first-law of binary mechanics~\cite{Fujita:2016igj}, and recently via a simple mass-ratio dependence \cite{Bini:2019nra,Antonelli:2020aeb}, uncovered in studies of scattering processes in PM~dynamics \cite{Vines:2018gqi,Kalin:2019inp,Kalin:2019rwq,Damour:2019lcq}. Yet, various obstacles lie along the way in order to enable higher order PN/PM results through these methods. For example, scattering calculations in self-force theory are in a nascent stage, e.g.~\cite{Long:2021ufh}. Similarly, fully determining the 5PN dynamics in this fashion requires second-order self-force results \cite{Bini:2019nra}, which may remain out of reach for several years.\vskip 4pt

{\it Modern approaches.} Despite historical achievements, the standard PN formalism \cite{Blanchet:2013haa,Schafer:2018kuf} may be heading towards a level of intricacy which may render further progress somewhat difficult. Many of the challenges of traditional PN methods we mentioned before can be ameliorated by utilizing an EFT approach \cite{Porto:2016pyg}, which recasts the binary dynamics into the computation of a series of Feynman diagrams using dimensional regularization. This approach naturally handles the point-particle approximation and separation into far and near zones (or potential and radiation regions) together with its associated divergences, without resorting to the various ``ambiguities'' introduced in other methods \cite{Damour:2016abl,Marchand:2017pir,Porto:2017dgs}. 

The EFT framework has reached the state-of-the-art both for the spin-independent contributions in the conservative sector at 4PN order \cite{Galley:2015kus,Foffa:2019yfl}, as well as complete spin effects to NLO \cite{Pardo:2020hxc,Cho:2021mqw}. More recently, the EFT camp has achieved the milestone goal of the spinless conservative dynamics at 5PN order \cite{Foffa:2019eeb,Blumlein:2020pyo}. While the EFT formalism provides a systematic framework which may be pushed to higher orders \cite{Porto:2016pyg}, the breaking of manifest (special-)relativistic invariance introduced by the PN expansion, and the use of gauge-dependent objects such as the gravitational potential, produces lengthy intermediate results containing spurious pieces. Moreover, additional complexity arises due to the proliferation of Feynman diagrams in a PN scheme. 

Remarkably, both these issues have been recently addressed by shifting attention towards scattering processes within the PM expansion instead, both through  EFT-based \cite{Kalin:2020mvi,Liu:2021zxr,Mougiakakos:2021ckm,Jakobsen:2021smu} as well as amplitude-based \cite{Bjerrum-Bohr:2018xdl,Bern:2019crd,Bern:2020buy,DiVecchia:2021bdo} methodologies. On the one hand, the former repurposes the powerful EFT machinery for the two-body problem in gravity \cite{Goldberger:2004jt,Porto:2005ac} towards computing PM scattering data --- dramatically cutting down the necessary steps and number of Feynman diagrams/integrals involved --- combined with a dictionary which directly relates gauge-invariant observables for hyperbolic- and elliptic-like motion \cite{Kalin:2019inp,Kalin:2019rwq}. On the other hand, the latter combines recent and ongoing breakthroughs in quantum scattering amplitudes and relations between gauge and gravity theories \cite{Bern:2019prr}, with classical physics extracted through a Hamiltonian via a matching computation \cite{Goldberger:2004jt, Cheung:2018wkq} or suitable limits \cite{Kosower:2018adc}.\vskip 4pt

The rapid progress in analytic computations in gravity in recent years gives us hope that the perturbative dynamics of binary systems may be ultimately tackled via a combination of all these approaches. Needless to say, not only in the conservative sector but equally important incorporating also radiative effects, both in the dissipative and conservative sectors, either through a multipole PN expansion \cite{Blanchet:2013haa,Porto:2016pyg}, small-mass-ratio limits \cite{Barack:2018yvs}, or a PM scheme \cite{Dlapa:2021npj,Jakobsen:2021smu,Mougiakakos:2021ckm,Herrmann:2021lqe,DiVecchia:2021bdo}. 
The~challenge for the latter includes also extending the connection between scattering and bound orbits to include radiation-reaction tail effects \cite{Galley:2015kus,Dlapa:2021npj}, in particular to study orbits with small eccentricity. Another challenge, shared by QFT-based approaches, is the computational complexity involved in multi-loop integrals \cite{Foffa:2016rgu,Damour:2017ced}, notably those in the PM regime \cite{Bern:2021dqo,Dlapa:2021npj}. In this regard, the field of precision gravity will benefit from similar efforts in calculating integrals for precision physics with particle colliders, e.g. \cite{Kreer:2021sdt}. Moreover, in combination with self-force data, the PM expansion could help us elucidate the analytic structure of the scattering problem in the complete kinematic regime, from the near-static to the ultra-relativistic limit. This invites us to the challenge of finding patterns which may ultimately help us identify constraining physical principles, allowing us to ``bootstrap'' higher-order results from lower-order ones without explicit calculations, perhaps even {\it solving} the two-body dynamics in the perturbative regime. 

\subsubsection{Simulations}

Numerous challenges have to be overcome in the next years to use simulations for the interpretation of future GW and multi-messenger data. One of the main bottlenecks is that numerical relativity studies come with large computational costs and are significantly slower than perturbative approaches. An example where this problem becomes 
most pronounced is the investigation of large mass ratio 
systems, e.g., mass ratios of 1:100, since both, the orbital 
timescale as well as the Courant factor scale with the 
total mass ratio~\cite{Lousto:2010ut}. 
To finally cover the large-mass ratio regime, new 
approach and methods have to be developed. However, even for comparable mass ratios one needs large computational resources to achieve sufficient accuracy 
for the direct interpretation~\cite{Lovelace:2016uwp}, the construction of waveform 
models~\cite{Ossokine:2020kjp,Pratten:2020fqn}, or 
simply to test perturbative methods~\cite{Szilagyi:2015rwa,Dietrich:2017feu}. 
Therefore, similarly to upgrades of high-performance computing facilities also numerical relativity codes and methods have to be improved continuously. Such improvements include among others, the usage of more sophisticated and parallelizable methods 
to discretize the field equations, e.g., discontinous Galerkin approaches~\cite{Bugner:2015gqa,Kidder:2016hev}, or new gauge choices to ensure more accurate simulations~\cite{Bernuzzi:2009ex,Dumbser:2017okk}. 

The overall situation becomes even more challenging if one does not only solves Einstein's equations, but want to study systems 
involving matter fields, e.g. for neutron stars. The simplest extension would be to incorporate general-relativistic hydrodynamics. However, in the regime after the merger of two neutron stars, a time when 
temperatures and magnetic field strengths increase, 
and rising densities allow us to probe parts of the supra-nuclear 
equation of state not accessible by any other physical 
process or experiment~\cite{Radice:2016rys}, one has to deal with both 
 general-relativistic magneto-hydrodynamics and 
general-relativistic radiation hydrodynamics. Indeed, the correct simulation of neutrino radiation is challenging 
since one would need to solve the relativistic Boltzmann equations, 
which results in a seven-dimensional parameter space, 
where in addition to three spatial and one time dimensions also the 
three-dimensional momentum space has to be covered. 
To date, even the biggest high-performance computing facilities are unable 
to provide enough resources for such simulations so that various approximation 
methods for the description of neutrinos have been 
employed~\cite{Sekiguchi:2011zd,Foucart:2020qjb}. Likewise, the inclusion of magnetic fields is also challenging, since one needs to resolve small-scale turbulent effects, such as 
the Kelvin-Helmholtz or magneto-rotational 
instabilities~\cite{Siegel:2013nrw}. 
Resolving the necessary length scales is currently out of reach, and 
even the most expensive simulations have resolutions that are orders 
of magnitudes too low~\cite{Kiuchi:2017zzg}. 
However, an accurate interpretation and description of the magnetic field 
effects are essential to connect GW observations with coincident detections of gamma-ray bursts.

\subsection{Gravitational Wave Science}

The future of GW astronomy is loud and bright with branches on many areas of theoretical physics,\footnote{\url{https://agenda.infn.it/event/22947/}}
including searches for physics beyond the standard model, e.g. \cite{Baumann:2018vus,Baumann:2019ztm}, constraints on the properties of dark matter \cite{Bertone:2019irm} and production mechanisms in the early cosmological evolution of the universe, either from inflationary dynamics, phase transitions, etc. \cite{Caprini:2018mtu}. For instance, there are many scenarios of physics beyond the standard model resulting in a stochastic GW background in the range of frequencies of future detectors such as LISA and ET, but also in the MHz to GHz, which has attracted a community effort to move forward towards high frequencies \cite{Aggarwal:2020olq}. In~addition, observations of GWs will enable an exquisite determination of the expansion rate of the universe, thus shedding light on the nature of dark energy or whether gravity is modified at large distances. Moreover, there is also strong synergy with various aspects of multi-messenger astronomy, notably neutrino and nuclear physics. See~other sections in this document for more details. 

At the same time, as we have emphasized throughout, much of the discovery potential in GW science hinges upon high-precision theoretical predictions yielding more accurate waveform models, which rely on both numerical and standard perturbative tools in GR in combination with sophisticated computational methods which have been instrumental in particle physics. The synergy between different approaches has created new bridges between various communities and we expect that the study of dynamical spacetimes and compact binaries will continue to provide novel techniques and key results in the years to come, thus enabling foundational investigations in physics through precision data with the next-generation of GW observatories. 
\section{Nuclear Astrophysics}
\label{sec:nuclearastro}
{\bf Coordinators:} Tetyana Galatyuk and Tanja Hinderer. \\[2pt]
 {\bf Contributors:} Tetyana Galatyuk, Tanja Hinderer, discussions at the EuCAPT symposium.
\subsection{Introduction}
Nuclear astrophysics has two interrelated goals: to (i) use astrophysical environments as unique laboratories for studying subatomic physics, and (ii) understand the role of nuclear processes in shaping the cosmos. Major driving questions in this field include the following: 

\smallskip

\textbf{ 1. Fundamental understanding of strong interactions:}\\
\emph{ The phase diagram of QCD, nature of quark deconfinement, and role of chiral symmetry breaking?}\\ 
Quantum chromodynamics (QCD), the gauge theory of strong interactions, has very different features than electroweak interactions and gravity, for instance, the coupling becomes weak at short distances. This gives rise to unusual phenomena such as quark confinement and the importance of the vacuum, for instance, for creating nearly all of the mass of baryonic matter in the universe. The underlying theoretical Lagrangian is relatively simple, however, the complex emerging phenomena are not captured by current first-principles calculations, while perturbative, numerical, and holographic methods encounter various complications and are applicable only in parts of the parameter space.
Consequently, our current knowledge about the nature of the deconfinement and chiral phase transition, the existence of the QCD critical point, and properties of matter at high temperature and density remains insufficient~\cite{HotQCD:2018pds,Cuteri:2021ikv}. Remarkable new insights have come from studying matter in regimes of temperature and density far from well-established knowledge with heavy ion collisions~\cite{Andronic:2017pug,HADES:2019auv}; astrophysical collisions of NSs probe the complementary swath of parameter space of dense matter at low temperature, as illustrated in Fig~\ref{fig:phases}.

\smallskip

\textbf{ 2. Emergence of nuclear structure from fundamental building blocks:}\\
\emph{How do subatomic constituents assemble and interact? dependence on density, isospin, and temperature?}\\
The multitude of many-body phenomena of interacting nucleons is a major scientific frontier~\cite{DOE}. Ab-initio calculations from fundamental QCD theory are challenging, and effective descriptions for nuclear matter remain limited due to the large amount of unknowns such as the nuclear three-body forces and the dependence of the interactions on density, temperature, and isospin asymmetry (excess of neutrons over protons)~\cite{Bortignon:2016wey,Nazarewicz:2016gyu}. Information from nuclear experiments, together with various new theoretical insights has led to progress, however, because of the great complexity of subatomic systems, current theoretical models have yet to accurately describe even simple nuclei~\cite{Coraggio:2021jvf}. In astrophysical environments such as NSs described below, the significantly higher densities  achieved greatly enhances the multi-body phenomena, gives rise to superfluidity and superconductivity, and enables studies of the parameter dependencies. 

\smallskip

\textbf{ 3. Formation and abundances of heavy elements:}\\
\emph{Origins and formation of heavy elements? Impact on cosmic evolution? }\\
Nuclear reactions in the early universe created the simplest elements such as hydrogen and helium, while elements in the periodic table up to iron and nickel are produced through nuclear fusion in stars. About half of the remaining elements only form through neutron capture processes that require a high density of free neutrons~\cite{Rauscher:2014fea}, which occurs in supernovae or NS binary mergers. Binary mergers involving at least one NS are the only known sites with the conditions needed for the synthesis of the very heaviest elements through the rapid neutron capture (r)-process. The path of the r-process through the chart of nuclides remains uncertain, as it passes through a large swath of unexplored parameter space of nuclides having high numbers of neutrons and protons~\cite{Mumpower:2015ova}. In this regime, fundamental properties such as magic numbers corresponding to shell closures, possible islands of stability, termination points by fission, and the resulting elemental abundances all remain poorly known. Theoretical predictions are complicated by the sensitive dependence on a large number of parameters including neutron density, temperature, neutron capture cross sections, magic numbers, beta decay properties, and initial composition, among numerous others. The multimessenger signals from populations of NS mergers are a highly anticipated input for significant advances on this problem~\cite{Barnes:2020uht}.

\subsection{Interdisciplinary information connected by theory}

\begin{figure}
    \centering
    \includegraphics[width=0.4\textwidth]{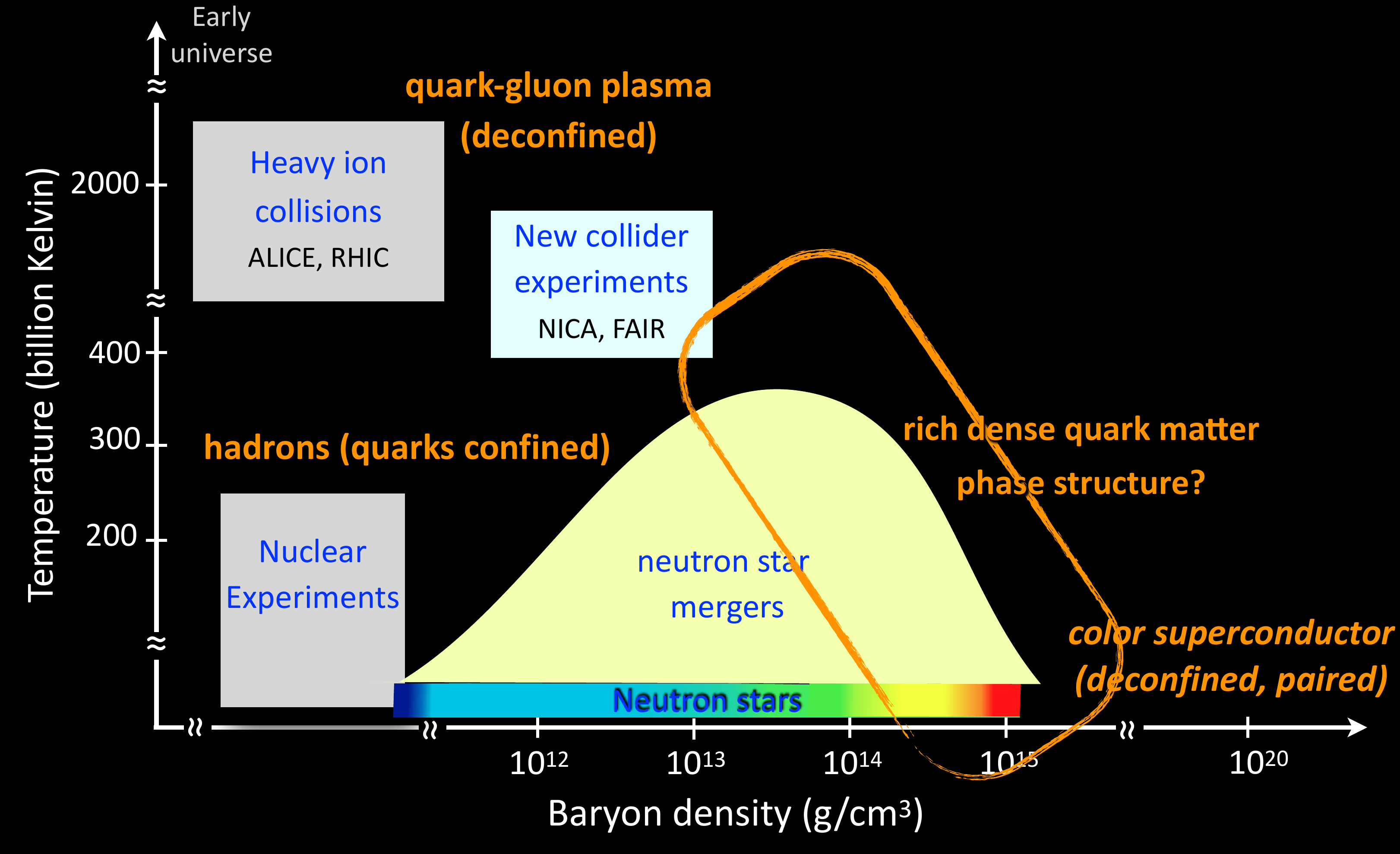}
    \caption{\emph{Schematic QCD phase diagram.} The phase structure, nature of the transitions, and possible critical points remain an open question in most of the parameter space. Neutron stars explore complementary regimes of subatomic physics to heavy-ion collisions and nuclear experiments. }
    \label{fig:phases}
\end{figure}

\hspace{5mm} {\bf Neutron stars}. Neutron stars (NSs) comprise matter compressed by strong gravity to up to several times nuclear density, and ranging over nearly ten orders of magnitude from the outer parts of the interior to their cores~\cite{Lattimer:2015nhk} and encompassing extremely rich physics on different scales. Schematically, a NS consists of a crust formed by a lattice of nuclei until it becomes favorable for free neutrons to drip out from the nuclei and eventually form a uniform liquid of mainly neutrons. However, in the inner cores, at densities above a few times the normal nuclear density, the neutron wavefunctions overlap significantly, and thus new degrees of freedom become important, and novel phases of degenerate quark matter may appear~\cite{Baym:2017whm,Alford:2007xm}. Coincidentally, at this density range heavy hadrons such a hyperons are also expected to appear, and an intermediate Bose-Einstein condensate of such particles may form before a deconfinement transition~\cite{Blaschke:2018mqw}. All of these conditions make NSs unique testbeds for unexplored regimes of QCD, as illustrated in Fig.~\ref{fig:phases}, as well as for nuclear multi-body phenomena and their variations with density. Several global properties of NSs directly depend on the microphysics of their interiors and can be probed as follows.  

\emph{Radio and x-ray measurements}. The maximum mass that a NS can sustain before gravitational collapse to a black hole depends on the properties of its matter~\cite{Lattimer:2012nd}. Masses can be measured to high accuracy with radio timing in binary pulsars, with observations of high-mass pulsars yielding constraints on the equation of state, which encodes the information on subatomic microphysics.  
The NS radius for a given mass also depends on the equation of state~\cite{Lattimer:2012nd}. Masses and radii can be measured for several pulsars through pulse-profile modeling of x-ray data, as already done for a few systems e.g. with the NICER telescope~\cite{2016SPIE.9905E..1HG,Riley:2021pdl,Riley:2019yda}.

\emph{Gravitational waves from binary inspirals}. The gravitational waves (GWs) from NSs in inspiraling binary systems of two NSs or with a black hole companion carry signatures of NS matter. This occurs through phenomena such as spin-induced deformations and tidal effects, which depend on the properties of matter and lead to small but potentially measurable imprints in the signals. For instance, constraints on the equation-of-state dependent tidal deformability parameter~\cite{Flanagan:2007ix} have been obtained with recent events~\cite{LIGOScientific:2018cki,LIGOScientific:2020aai,LIGOScientific:2018mvr}. 

\emph{Gravitational waves from tidal disruption}. In binary systems of a NS and a black hole companion, depending on the parameters such as mass ratio, spins, and equation of state, the NS may plunge into the black hole or get tidally disrupted when the black hole's tidal field overcomes the self-gravity binding the material~\cite{Foucart:2020ats,Shibata:2011jka}, which leads to a sharp termination of the GW signal~\cite{Shibata:2011jka,Foucart:2020ats}. The shutoff frequency associated with the disruption provides a complementary signature of the properties of cold, strongly-interacting matter.

\emph{Continuous gravitational wave signals}. Such signals could be generated by elastically- or magnetically-driven deformations of rotating NSs such as mountains on surface or unstable oscillation modes.  They also have the potential to lead to unique insights into subatomic physics, and probe complementary aspects of NSs than accessible with binary systems~\cite{Sieniawska:2019hmd}.

\smallskip

{\bf Binary neutron star mergers}. Unlike binary black holes, which always merge into a remnant black hole, the mergers of two NSs result in three possible outcomes depending on the parameters such as masses, spins, and equation of state, among other subdominant ones~\cite{Burns:2019byj,Baiotti:2016qnr}: (i) prompt collapse to a black hole, (ii) formation of a transient hypermassive, differentially rotating remnant which eventually collapses to a black hole, (iii) ultimately a stable NS remnant.
The cases (ii) and (iii) are reminiscent of heavy ion collisions, with the temperature rising upon merger and matter accessing new regimes in the QCD phase diagram. The ensuing behavior is highly dynamical, involving differential rotation, oscillations, and dissipation, with various microphysical processes (for instance small-scale magnetohydrodynamic instabilities, neutrinos, among others) becoming simultaneously important.  
The GWs are unique probes of this process. In cases (ii) and (iii), where the merger results in a material object, the main observable is the frequency associated with the rotation of the dynamical remnant. The frequency spectrum also involves a number of additional features related to other effects, whose interpretation is more complicated.

The merger of two NSs occurs at higher frequencies than the range in which current detectors are most sensitive. Moreover, as in heavy ion collisions, to fully interpret the results of the postmerger requires accurate information about the initial state and 'control parameters' (e.g., energetics and asymmetry) of the collision. These will differ for each binary system and must be extracted from accurate parameter measurements of the corresponding inspiral signals (masses, spins, eccentricity, cold equation of state). 
While in favorable special cases, the presence of a prominent postmerger GW signature may be detected in the next years after further upgrades to exising facilities, detailed studies of this interesting regime will likely remain outside the capabilities. However, tracking binary NS signals through the merger and beyond is among the key science targets for the next decade's planned detectors: the Einstein Telescope~\cite{Punturo:2010zz,Maggiore:2019uih} (Europe, already on the ESFRI roadmap) and Cosmic Explorer (US) detectors~\cite{Reitze:2019iox}. These instruments will have an order of magnitude better sensitivity and wider bandwidth, detect $\sim 10^5$ binary events per year, and enable measuring nearby inspiral to high accuracy, probing the new physics in the high-frequency regimes encountered during the merger and beyond, and advance alerts of imminent NS mergers or disruptions that are critical for detailed multimessenger studies.  

\smallskip

{\bf Kilonovae}. Matter outflows from tidal disruptions or in binary NS mergers lead to a wide range of electromagnetic counterparts (and neutrino signals). Of greatest importance for nuclear astrophysics is the so-called kilonova associated with the r-process nucleosynthesis of heavy elements~\cite{Lattimer:2019iye,Metzger:2019zeh,Barnes:2020uht,Wu:2018mvg,Watson:2019xjv,Arcones:2016euo}. Once the tremendous gravitational compression ceases, the subatomic matter from the NS quickly undergoes nuclear processes which produce unstable nuclides having an enormous number of closely spaced energy levels, and undergoing radioactive heating and decays. The emitted radiation ranges covers a broad spectrum peaked in the ultraviolet at very early times and dominated by infrared emission at later times. Due to the complexity of processes and variables involved, large uncertainties remain in our understanding and interpretation of kilonovae~\cite{Barnes:2020nfi}. The kilonova associated with the NS GW event GW170817 is to date the scientifically richest observation of such a process and its time evolution~\cite{Smartt:2017fuw,Kasen:2017sxr,Tanvir:2017pws,Watson:2019xjv,Domoto:2021xfq}. Besides being of key interest for heavy element formation~\cite{Arnould:2020rpz}, the kilonova lightcurves also provide information about the matter outflows from the merger or disruption, which in turn depend on the progenitor properties, e.g.~\cite{Shibata:2017xdx,Foucart:2012nc,Raaijmakers:2021slr,Barbieri:2019kli,Capano:2019eae,Dietrich:2020efo}, among numerous other quantities, e.g.~\cite{Wollaeger:2017ahm}.

\smallskip

{\bf Core-collapse supernovae}. Supernovae offer an interesting window into matter at finite temperature and baryon density~\cite{Burrows:2012ew,Janka:2012sb, Arnould:2020rpz}, are production sites of heavy elements, and  birthplaces of NSs and stellar-mass black holes. Their GW signals encode information on the explosion mechanism, the possible proto-NS, and the equation of state~\cite{Andersen:2021vzo,Bizouard:2020sws,Pajkos:2020nti,Abdikamalov:2020jzn}. The GW detection rates are expected to be low~\cite{Srivastava:2019fcb} but would be accompanied by EM and neutrino counterparts. A main component of the signals will be burst-like, which complicates the detection and interpretation compared to signals from binaries, though nevertheless, the GWs will contain precious information for nuclear astrophysics.

\smallskip

{\bf Terrestrial probes}. The field of nuclear astrophysics crucially relies on interdisciplinary connections of the insights from astrophysical systems described above with those from terrestrial experiment, for instance:  

\emph{Nuclear experiments}. Examples of nuclear experiments include the PREX and CREX studies~\cite{PREX:2021umo,Kumar:2020ejz} which measure the neutron skin thickness of neutron rich nuclei. Nuclei with a symmetric number of protons and neutrons are energetically preferred, thus, added extra neutrons are pushed outwards and form a 'skin' around the symmetric center. The skin thickness depends on the repulsive multi-body forces between neutrons. Similar forces are also relevant for neutrons in NSs, albeit at much higher density and isospin asymmetry~\cite{Piekarewicz:2019ahf,Reed:2021nqk}.

In addition, there are a number of other nuclear experiments which characterize various interactions and properties of subatomic multi-body systems, and also have connections to the physics of NSs. 

Understanding the complex physics of kilonovae is aided by new experimental measurements, for instance the masses of exotic nuclides, and the synthesis of neutron-rich nuclei with rare isotope beam facilities~\cite{Horowitz:2018ndv}. Another key component for kilonovae modeling and interpretation of the lightcurves comes from methods of atomic physics~\cite{Gaigalas:2019ptx} for describing radiation from level transitions in very heavy nuclides. 

\emph{Heavy ion collisions}.
Heavy ion collisions are elucidating QCD phase structure. The experimental approach is to probe at utmost precision different regions of the QCD matter phase diagram. The existing data from the high-energy collision experiments at LHC, RHIC-BES, SPS, AGS, SIS18 allow accurate statements about the bulk properties of the matter created in heavy-ion collisions to be made. However, the challenge remains to locate the onset of partonic matter, to detect the QCD critical point and to study matter properties by isolating their unambiguous signals. Extensive upgrade programs at LHC and RHIC will scrutinize the conditions of matter at small baryo-chemical potential (higher energy collisions produce more equal numbers of baryons and anti-baryons leading to a lower net baryon density). New collider and fixed-target facilities with lower energies are planned in Europe (FAIR,  NICA, SPS) and Asia (J-PARC) will come online in the next years to probe unexplored regimes using rare diagnostic probes not accessible so far in the high baryo-chemical potential region \cite{Galatyuk:2019lcf}.

\subsection{Theoretical challenges and opportunities}

Significant experimental advances in the coming years will increase the sensitivity of current detectors~\cite{KAGRA:2013rdx}, new telescopes will come online enabling multimessenger studies~\cite{Foley:2019evo}, and vigorous efforts on next-generation GW detectors, especially the Einstein Telescope, are underway. Theorists must respond to the larger and more diverse datasets, and establish efficient stategies for interpreting the outputs for advances on fundamental questions. In addition, continued theoretical developments are essential to quantify the science case and inform design choices of future facilities and telescopes.

A major challenge with nuclear astrophysics in general is the enormous complexity of the processes, the multitude of scales and physical phenomena involved,
and the highly nonlinear dependence of the observables on the underlying physics. This requires a concerted interdisciplinary effort from different communities. A continued effort to better foster such links, stimulate interactions, and establish strategies for combining and jointly interpreting information from various datasets and theoretical insights will be key to ensure substantial scientific advances in this field.

For GWs, theory plays an additional, crucial role for the measurements themselves because the data analysis is based on cross-correlating theoretical models with the data to determine the source parameters~\cite{Cutler:1994ys,LIGOScientific:2019hgc}. Robust measurements thus require models that include all relevant physical effects, are accurate over a wide range of possible parameter space, and computationally efficient enough for Bayesian inferences requiring millions of waveforms to be generated for the Markov Chain Monte Carlo explorations of the likelihoods for each event. Developing such models is a challenging task because it requires solving the nonlinear Einstein Field Equations coupled with relativistic magneto-hydrodynamics for the dynamics of the spacetime and matter, and additional schemes to include e.g. neutrino transport. The early inspiral can be approximated by analytical models based on a tapestry of perturbative schemes, while the merger and postmerger is only accessible with numerical relativity simulations. Despite significant recent progress, many theoretical challenges remain. For instance, analytical models are still missing several important physical effects during the inspiral and make simplyfing restrictions. Numerical relativity simulations require enormous time on supercomputers~\cite{Dietrich:2020eud} and remain limited in length, resolution, paramter space coverage, and are missing more realistic microphysics inputs. More details on the tremendous computational challenges in numerical relativity are discussed e.g. in~\cite{computationalchallenges,Duez:2009yz,Lehner:2014asa}.

Computational challenges are also a pressing problem for the data analysis. For instance, in next-generation detectors such as the Einstein Telescope, NS binary signals will linger for several hours to days in the sensitive band~\cite{Samajdar:2021egv}, and several tens of thousands of such events per year are expected~\cite{Maggiore:2019uih}. With current methods and infrastructures, the data analysis will become prohibitively expensive. New tools such as deep learning~\cite{Huerta:2019rtg} could help address this critical issue [c.f. the section Astrostatistics].

\medskip

{\bf 1. The QCD phase diagram and quark deconfinement}\\
\emph{Understanding and modeling observational signatures of phase transitions, facilitating broader connections}

 Degenerate quark matter potentially relevant to NSs and their mergers is far from the realms of established theoretical and experimental knowledge. It may exhibit strong pairing correlations between color, spin, and flavor degrees of freedom ('color superconductivity'), and may be characterized by a rich phase structure~\cite{Alford:2007xm}. Intermediate phases such as quarkyonic matter or Bose-Einstein condensates of hyperons, pions, or kaons may also possibly occur~\cite{Blaschke:2018mqw}. Many open questions remain, for example: is color superconductivity realized in nature? What are the properties of novel phases and the nature of the deconfinement transition?  What insights can emerge for QCD as a whole? How are deconfinement and chiral phase transition related? To make progress on these questions, it is critical to understand possible observational signatures of phase transitions and properties of quark matter. This remains a current challenge due to the large set of unknowns.

If strong first-order phase transitions occur in cold NSs, depending on the size of the energy jump and the onset density, they can manifest as distinctive features in the mass-radius and -tidal deformability curves~\cite{Alford:2015gna}. Mapping out these relationships is among the aims with upcoming GW measurements of populations of NSs, and x-ray observations of pulsars. Strong phase transitions also impact the NS's quasi-normal mode spectrum, and give rise to characteristic modes directly associated with the transition~\cite{Rodriguez:2020fhf}. The mode frequencies depend on characteristics of the transition and could become resonantly excited by tidal interactions in a binary inspiral, leading to distinct GW signatures in a single event that may potentially be measurable with next-generation detectors. 
It is also possible that a phase transition only occurs in the  finite temperature regimes encountered in mergers that avoid prompt collapse to a black hole. The main GW signature is expected to be a sudden change in the rotation frequency due to the rapid change in pressure upon reaching the transition~\cite{Most:2018eaw}; subdominant imprints are also expected though more difficult to measure and interpret~\cite{Bauswein:2020ggy}. 

Theorists must also prepare for the possibility that the transition to dense quark matter is a crossover~\cite{Baym:2017whm}. A crossover transition would make it significantly more difficult to identify the presence of quark matter; for instance, due to degeneracies with the uncertain nuclear interactions and energetics in the equation of state. Observables such as radius, tidal deformability, and main frequency of the postmerger spectrum, may thus be insufficient to robustly reveal smooth changes in the composition. This could bias the nuclear physics implications derived from the data. Other probes will be needed to distinguish the composition, for instance, out-of-equilibrium properties such as transport coefficients, which differ for nucleonic and quark matter~\cite{Alford:2007xm}. The transport coefficients of NSs and their postmergers are, however, likely to lead to only small effects~\cite{Alford:2017rxf}, though a more detailed study of possible signatures and prospects for measuring them in inspirals, postmergers, EM counterparts, and continuous GWs remains among the open problems for theory. Furthermore, neutrino signatures of the postmerger may differ for quark matter, which is another area requiring further work to determine observational consequences. 

Regardless of the nature of the phase transition, theoretical models for the GWs from NS binary systems that include more realistic physics are urgently needed for data analysis. This requires concurrent developments on the microphysics side to provide more realistic inputs on the possible matter properties. Examples of such aspects include finite temperature effects, going  beyond a phenomenological pressure-density relationship, and transport coefficients, among others. Interactions between the GW and nuclear physics communities will be essential to communicate what information is needed and what formats of the results would be most useful, e.g. as inputs for numerical simulations and for analytical studies. Stimulating such exchanges and networking is a major goal for the nuclear astrophysics division within EuCAPT. An important broader theoretical goal, also among the key aims of this division, is to connect information and insights gained from heavy ion collisions, theory, and NSs from the different channels. The aim is to understand the dependence on temperature and density, and to develop a deeper understanding of the deconfinement mechanism and possibly shed light on its relation to chiral symmetry breaking.

\medskip

{\bf 2. Multi-body phenomena in nuclear matter and emergent subatomic structure}

Much remains to be understood about the interactions of nucleons beyond merely two-body forces but also including higher-order interactions in dense environments. The impact of isospin asymmetry on the interactions, their density dependence, as well as the role of temperature, which becomes important in the cases of material postmerger objects, remain open challenges. NSs involve much higher densities than achieved in nuclei on earth, which greatly enhances the importance of interactions. Moreover, the enormous differences in density from their surface to their cores is enables studying density dependencies. The main quantity encoding these properties is the equation of state, which in turn affects observables such as radius and tidal deformability. The interpretation is complicated by the fact that the equation of state depends on a number of variables such as the symmetry energy, multi-body interactions, and composition. Quantities such as characteristic quasi-normal mode frequencies or transport coefficients provide complementary and more direct information on various properties but are more difficult to measure. The multi-body phenomena also include superfluidity and superconductivity, whose impact on observables will require further work. Past studies have considered possible signatures in the context of pulsar glitches, however, the potential with GWs from binary inspirals and mergers to probe such phenomena and their parameter dependence remains to be understood in more depth. Several basic consequences such as the splitting of the quasi-normal modes into two branches in the presence of superfluidity have already been revealed, yet the possible impacts on GW signals remains to be more comprehensively explored.



\medskip

{\bf 3. R-process nucleosynthesis}

As outlined above, many theoretical challenges remain in the modeling.  On the nuclear physics side, a main issue is to reduce the enormous uncertainties about the properties of nuclides in unexplored regimes of the nuclear chart and the path of the process. On the astrophysics side, much work remains to be done on understanding kilonovae and the dependence of the lightcurves on the multitude of physical processes involved and the dependence on progenitor parameters. As mentioned above, this is a vibrant interdisciplinary area involving experiments, atomic and subatomic physics, and long numerical simulations of post-merger or -disruption matter outflows that incorporate the remnant's strong gravity, relativistic magneto-hydrodynamics, radiative transport, neutrino schemes, and realistic initial conditions from mergers~\cite{Rosswog:2019rha}, which altogether contribute to elucidating the process of heavy element nucleosynthesis. 



\medskip

{\bf 4. Modified gravity and beyond standard model fields}

Theorists must also prepare for possible consequences of modified gravity or beyond standard model particle physics on nuclear astrophysics. This would impact the GWs, modify the structure of NSs, or give rise to extra/dark fields coupled to the subatomic matter. All of this would impact the interpretation of measurements~\cite{NSnonGR,Yazadjiev:2018xxk,Danchev:2020zwn,Ciancarella:2020msu,Palenzuela:2013hsa}. Understanding the resulting features and possible biases will require significant further theoretical studies. 

\medskip 

{\bf 5. Science case and design choices for new instruments}

Explorations of the nuclear astrophysics with strong-gravity sources have recently begun with GW detectors, EM follow up networks e.g.~\cite{EMfollow}, NICER/x-ray and radio telescopes, neutrino detectors. Next-generation instrumental capabilities are highly anticipated for deriving a fundamentally transformative understanding from the larger and higher precision datasets. The NICER telescope will measure NS radii for more pulsars in the next years, new radio telescopes ~\cite{SKA} will provide more mass measurements of binary pulsars, and the envisioned next-generation Einstein Telescope~\cite{Punturo:2010zz} and Cosmic Explorer~\cite{Reitze:2019iox} detectors will provide the next step for nuclear astrophysics with GWs. Significant further theoretical studies are essential to fully realize the scientific potential with these detectors, and the need for theoretical advances is already pressing to concreteley quantify the science potential based on design choices and optimizations in preparation for concrete funding investments and to pave the way for the science exploitation. The next-generation GW detectors will also enable earlier alerts and more accurate rapid sky localization of events, which will be important for detailed characterizations of the kilonova spectrum. New EM telescopes will come online in the 2020s~\cite{BlackGEM,Foley:2019evo}, yet, the visions for the landscape for the 2030s, when the third-generation GW detectors will become operational, requires further efforts (e.g.~\cite{Ciolfi:2021gzg}). Ideas for the longer-term future are also already being developed, including, for instance, supplemental high-frequency instruments ~\cite{Ackley:2020atn}, ultrahigh-frequency GW detectors~\cite{Aggarwal:2020olq}, or covering the lower-frequency window of binary inspirals~\cite{ArcaSedda:2021dte,Canuel:2017rrp}.

In summary, theoretical work is timely and essential for developing the science case and design choices, as well as the crucial models, tools and strategies for interpreting the wealth of data obtained with current and planned future instruments, in order to make full use of their science potential. Theoretical efforts are further indispensable combining interdisciplinary information and deriving from it  breakthroughs on longstanding frontiers in physics and astrophysics.



\section{Cosmic Accelerators}
\label{sec:cosmicaccelerators}
{\bf Coordinators:} Sera Markoff and  James Matthews. \\[2pt]
 {\bf Contributors:}
 \v{Z}eljka Bo\v{s}njak, Alexandre Marcowith,   Athina Meli, Viviana Niro, Foteini Oikonomou, Enrico Ramirez-Ruiz, Nikolina \v{S}ar\v{c}evi\'c, Jacco Vink,
Walter Winter.

\begin{figure}[h]
\includegraphics[width=\textwidth]{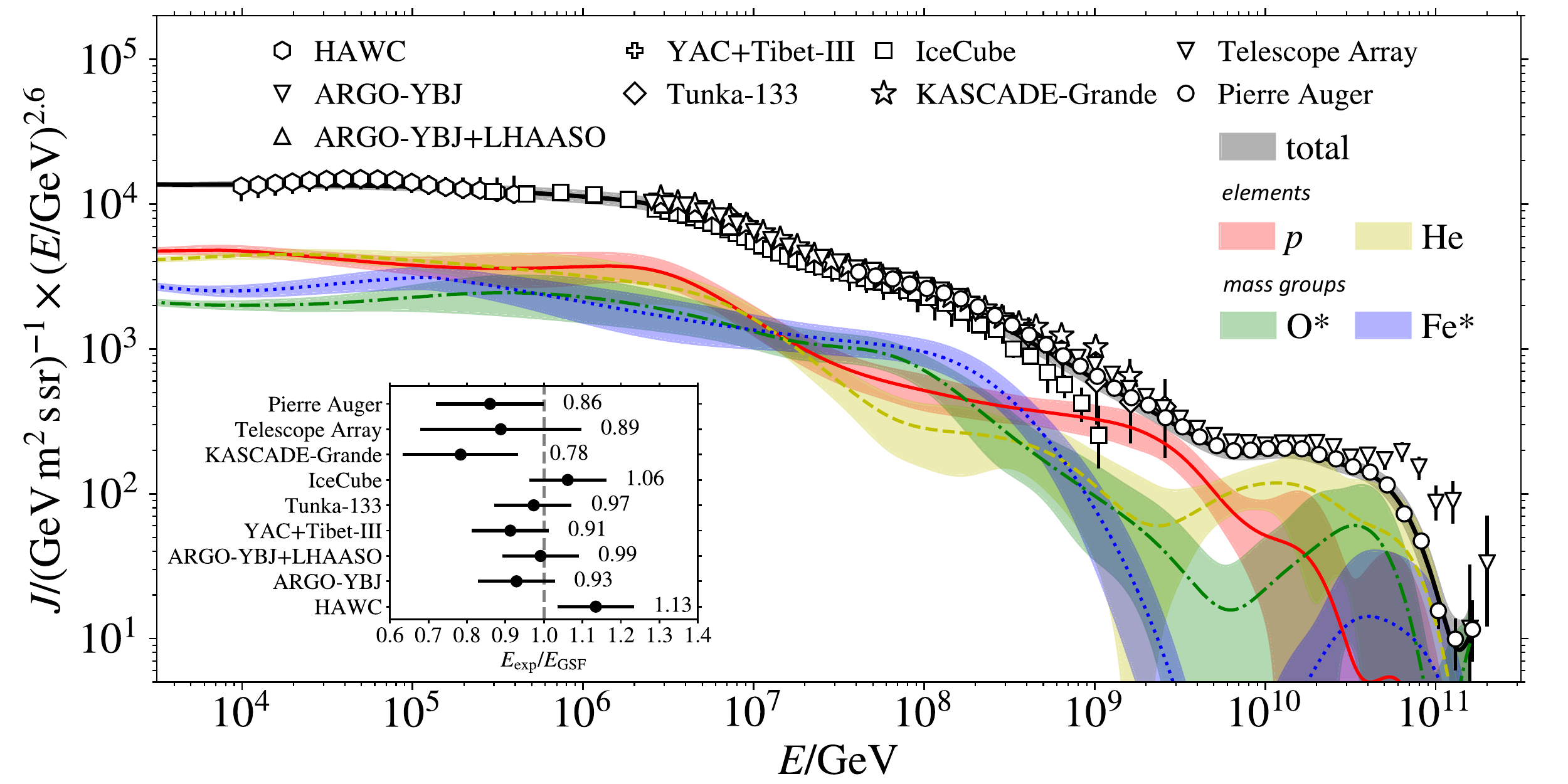}
\caption{Combined fit of the cosmic-ray energy spectra and its mass composition measured by various experiments, presented in the rapporteur summary for the Cosmic Ray Indirect session at the 2019 ICRC (see \cite{Schroeder_ICRC_2019} for details). While taking into account the experimental uncertainties, the fit allows for a constant
shift of the energy scale of each experiment (inset).  The interpretation shown here is model-dependent but is meant to illustrate how spectrum and composition are both important constraints for theory.}
\label{fig:CRsummary}
\end{figure}

\begin{figure}[h]
\includegraphics[width=\textwidth]{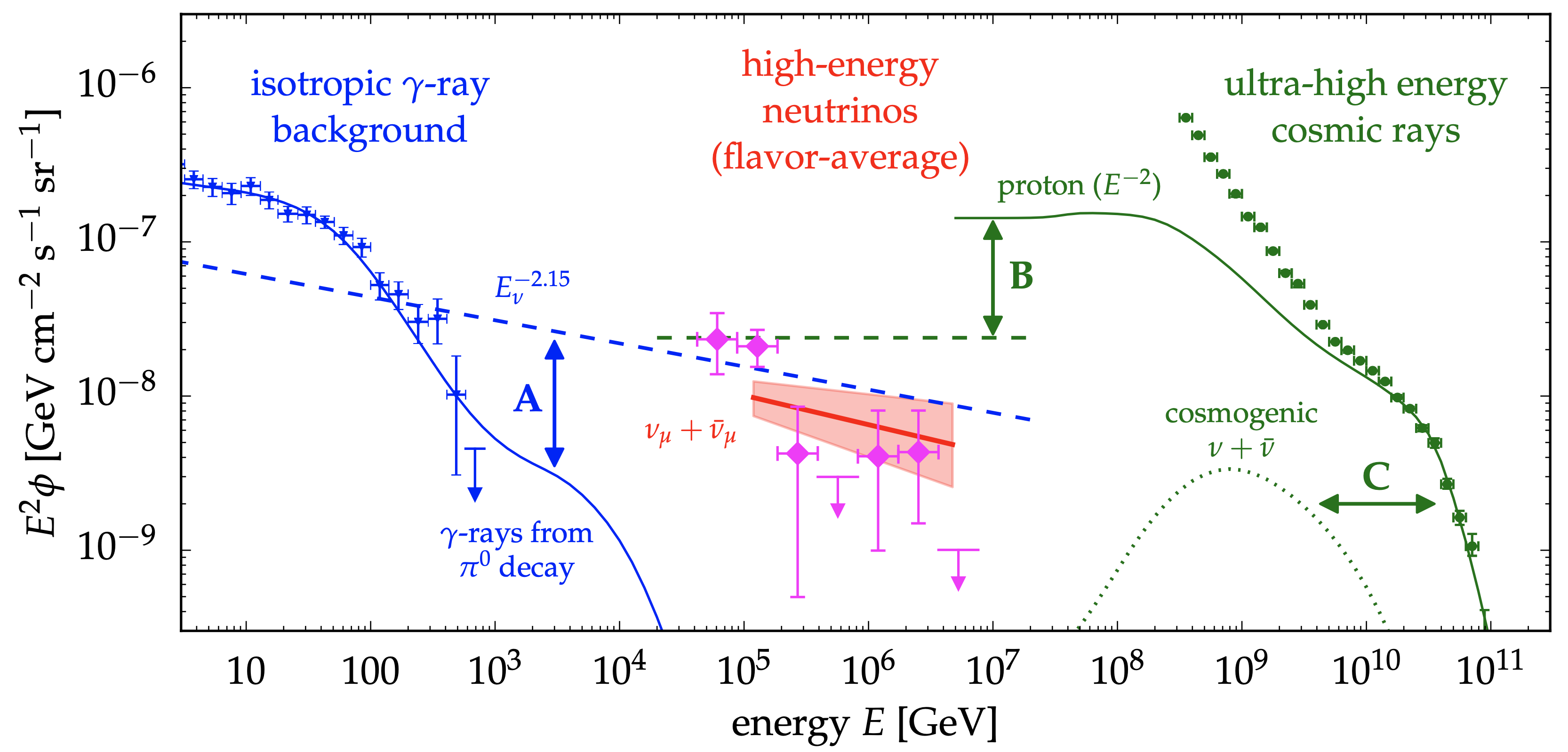}
\caption{Energy squared times the flux ($\phi$) of neutrinos, per flavor (red and magenta data) compared to the flux of unresolved extragalactic \gr\ emission (blue data) and ultra-high energy cosmic rays (green data).  Dashed lines indicate two neutrino limits based on multi-messenger models.  See Astro2020 decadal white paper for more details \cite{Vieregg_decadal_2019}.}
\label{fig:CRdiffuse}
\end{figure}

\vspace{.2in}
Over a century after the discovery of cosmic rays (CRs) by Victor Hess established the existence of high energy particles emanating from space \cite{hess_cosmic_1912}, we still have not definitively identified the primary sources, either within our own Galaxy or beyond. The latest evidence for astrophysical hadronic acceleration has come via several different facilities:  i) 
IceCube's\footnote{\url{https://icecube.wisc.edu}} discovery of high-energy astrophysical neutrinos \cite{IceCube_2013science}, ii) the Pierre Auger Observatory's\footnote{\url{https://www.auger.org}}  detection of a dipole anisotropy for CRs at ultrahigh energies \cite{pierre_auger_collaboration_observation_2017} and iii) The identification by Fermi-LAT\footnote{\url{https://fermi.gsfc.nasa.gov}} of likely ``pion-bumps" around 200~MeV in the spectra of a few supernova remnants \cite{ackermann13}.  The most recent total cosmic ray spectrum and (model-dependent) composition presented at the ICRC2019 conference is shown in Fig.~\ref{fig:CRsummary}, while Fig.~\ref{fig:CRdiffuse} shows the diffuse fluxes from \gr s, neutrinos and cosmic rays. Together these two figures represent key benchmarks for any model purporting to explain the source of cosmic rays, however there are numerous other constraints from the electromagnetic spectrum that also need to be self-consistently factored in, and so far no one scenario provides a satisfactory explanation for all data.  Unambiguously identifying cosmic accelerators thus remains the perennial challenge, however the ultimate goal is to clearly understand the driving physical processes and how each contributes to the global hadronic energy content, and to be able to account for the total diffuse fluxes in all species.  In this report we attempt to highlight the cutting edge of the field, which strives to bring multi-messenger (MM) signals into a consistent theoretical framework while overcoming challenges of small number statistics,  scale-separation and a limited ability to directly localise sources and/or or regions within sources where particles are accelerated.  After a general discussion of the key open theoretical  questions, we will showcase a few select topics which 
exemplify the challenges and promise of the coming decade.  

\subsection{Current theoretical landscape}
\label{sec:theory}

Cosmic accelerators provide significant opportunities   for theoretical studies, as we need to uncover  the processes by which charged particles (CRs, from electrons/pairs to ions) are accelerated within a given set of physical conditions.  A range of different (but related) acceleration mechanisms have been proposed, including shock acceleration, magnetic reconnection, stochastic acceleration in shear/turbulent sites, magnetospheres, or other alternatives (see recent reviews in e.g., \cite{matthews_particle_2020}, and references therein), and more than one may be operating within a given source. A clear understanding of these mechanisms, and distinguishing between them, is critical to the study of a wide range  of exotic phenomena, including active galactic nuclei (AGN), gamma-ray bursts (GRBs), supernova remnants  (SNR) and pulsar wind nebula (PWN) and potentially other stellar-remnants.  In addition, a complete understanding of acceleration represents a fundamental physical problem, with a cross-disciplinary nature that requires key developments in many areas of the field.

As with much of modern astrophysics, the current theoretical landscape is aided by the use of numerical simulations, which are mainly divided into either kinetic/particle or fluid approaches, as well as hybrid methods. Fluid approaches can simulate the larger-scale evolution and the bulk plasma properties of a wide variety of potential CR accelerators. 
Grid-based hydrodynamic (HD) and magneto-hydrodynamical (MHD) simulations often include also special relativistic (SR) effects and/or general relativity (GRMHD) or alternatives (see reviews by, e.g \cite{davis_simulations_2020,komissarov_numerical_2021}). With advances in high-performance computing (HPC) over the last years, it is increasingly possible to resolve and track the formation and evolution of particle acceleration sites within such large-scale fluid simulations. However making predictions of MM and multi-wavelength (MWL) spectra requires a strong connection with plasma physics, in order to understand from first principles how acceleration sites form {\sl in situ} and transfer kinetic and/or magnetic energy to particles, as well as how the energised particles can in turn influence their acceleration sites (CR feedback).  It is not yet computationally possible to self-consistently cover the full range of scales from ion skin depths up to the `astrophysical' scales of, e.g., kpc-scale AGN jets. 

At much smaller scales, recent years have also seen spectacular progress in using particle-in-cell (PIC) methods to simulate the acceleration from first principles (see reviews by, e.g. \cite{pohl_pic_2020,nishikawa_pic_2020}), with application to both shock acceleration, e.g. \cite{spitkovsky_particle_2008} and magnetic reconnection, e.g. \cite{sironi_relativistic_2014}. When applied to shock acceleration, PIC simulations are able to account for both the nonlinearity inherent to the particle acceleration process and the generation of plasma instabilities at the shock \cite{marcowith_microphysics_2016}. Particle acceleration in magnetic reconnection sites is a younger and less mature topic, but with significant recent progress; in this case, PIC simulations are used to study the formation of current sheets and the energisation of particles from both direct acceleration in the reconnection site as well as a variety of Fermi-type processes \cite{blandford_magnetoluminescence_2017}. As with shock acceleration, plasma physics is critical. For fast reconnection and efficient particle acceleration to occur, tearing and plasmoid instabilities must break up the current sheet \cite{sironi_plasmoids_2016}, and turbulent reconnection is therefore likely to be important. 

PIC simulations can explore how the acceleration efficiency and particle distribution functions depend on local plasma properties, but they still experience several limitations, mainly due 
to computational expense.  These limitations include rather small spatial/temporal resolutions, artificially decreased proton-electron mass ratio, often 2D dimensionality while turbulence is a 3D phenomenon, the lack of self-consistent cooling terms, and general difficulties including heavier nuclei and concerning composition. Hybrid kinetic/fluid approaches exist, with techniques including Vlasov-Fokker-Plank (VFP) code coupled to MHD \cite{bell_vfp-review_2006,reville_universal_2013} and MHD-PIC, e.g. \cite{bai_magnetohydrodynamic-particle--cell_2019}. Although these methods are likely to become more sophisticated in the coming decade, connecting large-scale fluid simulations to observations will probably continue to rely on an additional layer of `subgrid' models or parameterisations for particle distribution functions, radiation and radiation transfer. These models are in some sense generic whether employed within simulations or semi-analytical descriptions of a plasma flow. However, while significant progress is being made on incorporating particularly leptonic-based radiative signatures within fluid simulations (e.g. \cite{vaidya_particle_2018, EHT_PaperV}), the computational cost of including an additional hadronic fluid and keeping track of the interactions and secondary particle production is a major hurdle. 

An unsolved, long-standing question in the field concerns the origin of ultrahigh energy cosmic rays (UHECRs), protons and ions with energies $\sim1-100\,{\rm EeV}$ (see Fig.~\ref{fig:CRsummary} and \ref{fig:CRdiffuse}). UHECR phenomenology is a rich topic with obvious theoretical interest (and many challenges), which, by its very nature, must stretch particle acceleration theories to their limits. The classic work by Hillas \cite{hillas_origin_1984} established that sources must be extremely powerful to accelerate UHECRs, naturally favouring relativistic shocks in, e.g., GRBs and AGN as candidate sites. However, there are a number of problems with reaching the UHECR regime at relativistic shocks. Simulations have shown Fermi acceleration can develop at these shocks if micro-turbulence is triggered \cite{spitkovsky_particle_2008,2009MNRAS.393..587P}. However, UHECR acceleration is inhibited by the inability of turbulence at the shock to grow quickly enough to large enough scales, and maximum energies are thus severely limited, perhaps to orders of magnitude below the EeV regime \cite{lemoine_electromagnetic_2010,2013ApJ...771...54S,reville_maximum_2014,bell_cosmic-ray_2018}. This limit could potentially be relaxed in mildly relativistic shocks, powerful non-relativistic shocks and/or relativistic magnetic reconnection events. However, each of these alternatives need detailed investigation using a range of techniques, since it is not possible to probe the UHECR regime using the PIC simulations that have been so successful at studying particle acceleration on smaller scales / at lower energies.  

A central challenge for the theory community in the coming decades is thus to confront the various `separation of scales' problems in order to derive more physically sound models for CR accelerators.  Multiple techniques will continue to be necessary to build a holistic picture -- PIC, (GR)MHD and semi-analytic approaches need to be used as complementary tools. Furthermore, the fundamental questions regarding MM observations remain:  what are the origins of CR nuclei, and which acceleration mechanisms are responsible?  Are the \gr-emitting regions the same as those of astrophysical neutrinos?  How do CRs escape from their sources?  Can we predict the composition of CRs needed as input for transport in the whole chain from progenitor composition, jet formation, particle acceleration, and escape? In the following, we will connect the generic issues raised here with current/upcoming observations, and we will qualify some of the issues of specific object classes.

\subsection{Key observations requiring new theoretical frameworks}

We assume that direct CR observations will not yet yield unambiguous identifications within the coming decade, but will continue to provide important constraints on total spectra, anisotropies and composition. At the same time, a new generation of more sensitive MWL facilities have recently, or will soon, come online; see recent reviews/examples regarding facilities in radio \cite{norris_extragalactic_2017}, 
X-ray \cite{nandra_hot_2013}, and \gr\ \cite{funk2015,rieger_gamma-ray_2019}. Particularly relevant here are the Cherenkov Telescope Array (CTA)\footnote{\url{https://www.cta-observatory.org}} \cite{cherenkov_telescope_array_consortium_science_2019}, the Square Kilometer Array (SKA)\footnote{\url{https://www.skatelescope.org/}} and its pathfinders \cite{dewdney_square_2009,van_haarlem_lofar_2013,norris_radio_2013} and global (and eventually space-based) very long baseline interferometry (VLBI) such as the next-generation Event Horizon Telescope\footnote{\url{https://eventhorizontelescope.org/}} concept \cite{ngEHTwhitepaper_2019}. Each of these next-generation instruments represents a major leap in terms of sensitivity and imaging resolution.  With this in mind, we see the greatest potential and synergy with theoretical developments in the areas described in the following sections.

\subsubsection{Pevatron Identification in the Galaxy: Supernovae and Beyond}

The standard CR paradigm states that supernova remnants (SNRs) are the primary sources of CRs up to at least a few PeV, the energy of the so-called `knee' in the CR spectrum \cite{bell_particle_2014}. This paradigm is mainly based on the fact that the total CR power in the Galaxy, $\sim 10^{41}$~erg\ s$^{-1}$, is comparable to the total power that can be provided by SNRs, $\sim10^{42}$~erg\ s$^{-1}$. This equivalence is, however, mostly based on the local CR flux around 1~GeV, and does not explicitly address the ability of SNR to actually accelerate protons up to the knee.  The latter requires stringent conditions: a strong magnetic field amplification (MFA) at the blast wave shock and a CR mean free path comparable to its Larmor radius, the so-called Bohm regime. The leading candidate mechanism to achieve this is the CR-driven non-resonant hybrid (NRH) instability  \cite{bell_turbulent_2004}. More generally, the nonlinear plasma physics inherent to shock acceleration is a topic of great theoretical interest in itself, with strong progress made in the last decade thanks to a combination of numerical simulations and purely theoretical work, e.g.,  \cite{marcowith_microphysics_2016,bell_cosmic_2019,bai_magnetohydrodynamic-particle--cell_2019}.

MFA is indeed on-going in all historical SNRs, associated with X-ray filaments due to synchrotron radiation of TeV electrons \cite{vink12}, with magnetic field strengths over an order of magnitude above the mean Galactic strength $\sim 5 \upmu \mathrm{G}$. However, the SNR paradigm still faces major difficulties. Many historical SNRs have shock speeds and circumstellar densities that are too low to facilitate the rapid MFA needed \cite{bell_cosmic-ray_2013} to obtain PeV energy, and indeed so far no historical SNRs show evidence for \gr\ emission above 100 TeV. These facts have led the community to consider alternative PeVatron sources: 
\begin{itemize}
\item Supernov\ae~(SNe)/young SNRs: PeV CR acceleration can start right after the supernova explosion \cite{Marcowith_2018} and last a few months/years after in young, rather than the historical SNRs covered above.  
The fast shock velocity and high magnetic field strength result in rapid CR acceleration to very high energies due to the onset of the NRH instability. MFA in SNe is indeed observed at radio wavebands as it is the case for the well-monitored SN 1993J \cite{1998ApJ...509..861F}.
A drawback of this model is that only a small ($\sim 10$\%) fraction of the supernovae have sufficiently high circumstellar densities.
\item Massive star clusters (MSCs): 
CR can be accelerated in MSCs because of the size of the accelerator combined with the potential for repeated episodes of Fermi acceleration by magnetic turbulence and shocks  \cite{Bykov_2018}. 
\gr\ emission from massive star clusters tends to support this possibility \cite{Aharonian_2019}, and more recently the Large High Altitude Air Shower Observatory (LHAASO)\footnote{\url{http://english.ihep.cas.cn/lhaaso/}} collaboration detected PeV photons coming from several star-forming regions in the Galaxy \cite{lhaaso21}. A question remains, however, whether these regions are PeVatrons due to the collective effects of many powerful sources together (supernovae and winds driving, e.g., superwinds or superbubbles), or whether these regions stand out simply due to the sum of  individual source contributions.
\item Compact objects: unrelated to the power provided by supernova, there are models connecting PeV acceleration to compact objects:  pulsars and PWNe, X-ray binaries (XRBs) \cite{kantzas_new_2021} and Sgr A* itself \cite{hess_sgra2016}. In these cases, the acceleration conditions can be quite different and we must consider relativistic physics and the possibility of high magnetisation.  For pulsars and PWNe the current paradigm is that they primarily accelerate leptons, but a hadronic component is sometimes discussed \cite{fang_pulsars_2012}. The LHAASO results  indeed establish the Crab pulsar/PWNe as a PeVatron, most likely of leptonic origin. Moreover, the Crab nebula seems to be rather exceptional in its properties and may not be a good model for the general population of pulsars/PWNe.  However Tibet AS$\gamma$\footnote{\url{http://www.icrr.u-tokyo.ac.jp/em/index.html}} has reported the observation of diffuse 100 TeV- 1 PeV \gr s in the Galactic plane~\cite{Amenomori:2021gmk} and in addition to the Crab pulsar, LHAASO has just reported the detection of 11 additional PeV \gr\ point sources in the Galactic plane~\cite{lhaaso21}.  Other than the Cygnus Cocoon superbubble \cite{Abeysekara_HAWC_2021} and the XRB SS~433 \cite{Abeysekara_HAWC_2018} identified by the High Altitude Water Cherenkov Detector (HAWC\footnote{\url{https://www.hawc-observatory.org/}}; \cite{tepe_hawc_2012}), most new Galactic PeVatrons remain unidentified, though known pulsars fall within most error circles.  
\end{itemize}

\noindent
When considered jointly with the IceCube limits on the diffuse neutrino flux from the Galactic plane (Fig.~\ref{fig:CRdiffuse}) as well as the apparent complexity in the Galactic CR spectrum (e.g. \cite{pamela11,lipari20}), it is starting to seem more likely than not that the Galaxy contains many classes of Pevatrons.  The challenge to theory is how to uniquely characterise their acceleration properties, including composition particularly around the high-energy cutoff, and power, to allow discerning between possibilities.  Resolving this will require a more self-consistent theoretical treatment than has so far typically been applied, starting with tackling the likely necessary non-linear physics to achieve Pevatron energy, at the same time making simultaneous predictions for CRs, neutrinos and EM radiation up to the \gr\ band for individual sources.  The improved sensitivity of the next generation instruments coming online will also enable some of the first MM variability studies to test and hone the model physics.  These individual models then need to be convolved with population synthesis models to account for diffuse fluxes.  And finally, the same source models need to be employed within propagation models as described in Chapter\ref{sec:travelingmessengers}, to obtain a complete picture of the final expected composition measured on Earth.   

\subsubsection{Relativistic Jets from Supermassive Black Holes}
AGN jets are still considered primary candidates for UHECRs, and the last decades have brought significant progress but also some stark observational challenges, resulting in a rapid evolution of our ideas about their potential to accelerate hadrons to the highest observed energies (for a recent review also covering XRB jets see \cite{romero_particle_2018}).  Just in the last few years, neutrinos have been statistically associated with the 
AGN blazar TXS 0506+056 on the basis of a simultaneous \gr\ flare~\cite{IceCube:2018cha,IceCube:2018dnn}, and there are several indications (up to $3\sigma$) for neutrinos from different classes of AGN in the point source search~\cite{Aartsen:2019fau}.  
In addition, the recent claimed association of a neutrino from the Tidal Disruption Event (TDE) AT2019dsg \cite{Stein_2021} suggests that TDEs are potentially capable of high-energy particle acceleration, where similar to AGN the exact neutrino production site is under debate~\cite{Winter:2020ptf,Liu:2020isi,Murase:2020lnu}.  The non-observation of neutrino multiplets~\cite{Ahlers:2014ioa,Murase:2016gly,Capel:2020txc} suggests that the bulk of neutrinos detected do not originate in the ``usual suspects'', i.e., high-luminosity sources, but are instead relatively weak and abundant, see e.g. \cite{Palladino:2018lov,Petropoulou:2019zqp} for AGN blazars.  Stacking limits also suggest that high-luminosity sources produce at most~$17\%$ (high luminosity blazars), $1\%$ (jetted TDEs), $26\%$ (non-jetted TDEs) of the diffuse neutrino flux~\cite{2016ApJ...824..115A,2019ICRC...36..916H,2019ICRC...36.1016S}, whereas recent source associations and other arguments~\cite{Palladino:2018evm} may suggest multiple contributions. For example, taking these associations at face value, \cite{Bartos:2021tok} conclude that roughly one third of the neutrinos come from TDEs, one third from AGN, and one third from unknown sources.  The observed UHECR composition seems to indicate a charge-dependent maximal energy of the accelerator, which points towards modest radiation densities in the sources so that sufficient intermediate mass nuclei can escape (e.g., \cite{Kimura:2017ubz,Rodrigues:2020pli}). This might lead one to consider low-luminosity classes such LINERS, BL Lacs and FRI-like systems in general, or later stages of TDE evolution. However, the power constraint derived from the Hillas energy still requires large kinetic or magnetic powers, so alternatively regions of jets in which large amounts of energy are dissipated without significant radiation losses might be favoured, such as shocks in the outer lobes or hotspots of radio galaxies \cite{eichmann_ultra-high-energy_2018,matthews_ultrahigh_2019}. On the other hand, if neutrinos and UHECR are to be produced in the same region, sufficient radiation densities are required for efficient neutrino production, leading to a nuclear cascade in the source; see e.g. \cite{Biehl:2017hnb} for TDEs. The high baryon loading required to match neutrino observations (and possibly also UHECR models) is energetically challenging, potentially requiring kinetic energies well beyond what are typically thought to be sub-Eddington accretors; the best-studied case for neutrino production in this context is TXS 0506+056, see e.g.~\cite{Gao:2018mnu,Keivani:2018rnh}. The observed neutrino-source associations have led to extensive leptohadronic modelling of the spectra of the sources aimed at determining the physical conditions in these environments, and to an initiative to compare and benchmark a large number of the currently available codes that perform leptohadronic modelling~\cite{Cerruti:2021hah}.

A parallel debate is ongoing relating to the sites of particle acceleration and mechanisms, based on direct imaging. VLBI monitoring of blazars has established the presence of what appears to be a shock region that is stable over decades, e.g., \cite{Homan_MOJAVEXII_2015}, that in nearby sources like M~87 coincides with a transition from parabolic to conical profile \cite{nakamura_parabolic_2018}.  This profile change is generally interpreted as corresponding to the transition from magnetic acceleration in a collimation zone to kinetic/free expansion \cite{Marscher_2008}.  This transition occurs typically $\sim10^4-10^6 r_g$ downstream from the black hole, which taken at face value would be in tension with the sub-hour flares seen in the \gr s in many blazars.  Such fast variability, could support classes of models that occur  close to the black hole as in a stagnation or magnetospheric gap, e.g., \cite{rieger_gamma-ray_2019}, but could also be explained via smaller `jets-in-jets' or plasmoids within the larger-scale flow e.g., \cite{giannios_reconnection-driven_2013, petropoulou_blazar_2016}.   

\begin{figure}[h]
\includegraphics[width=\textwidth]{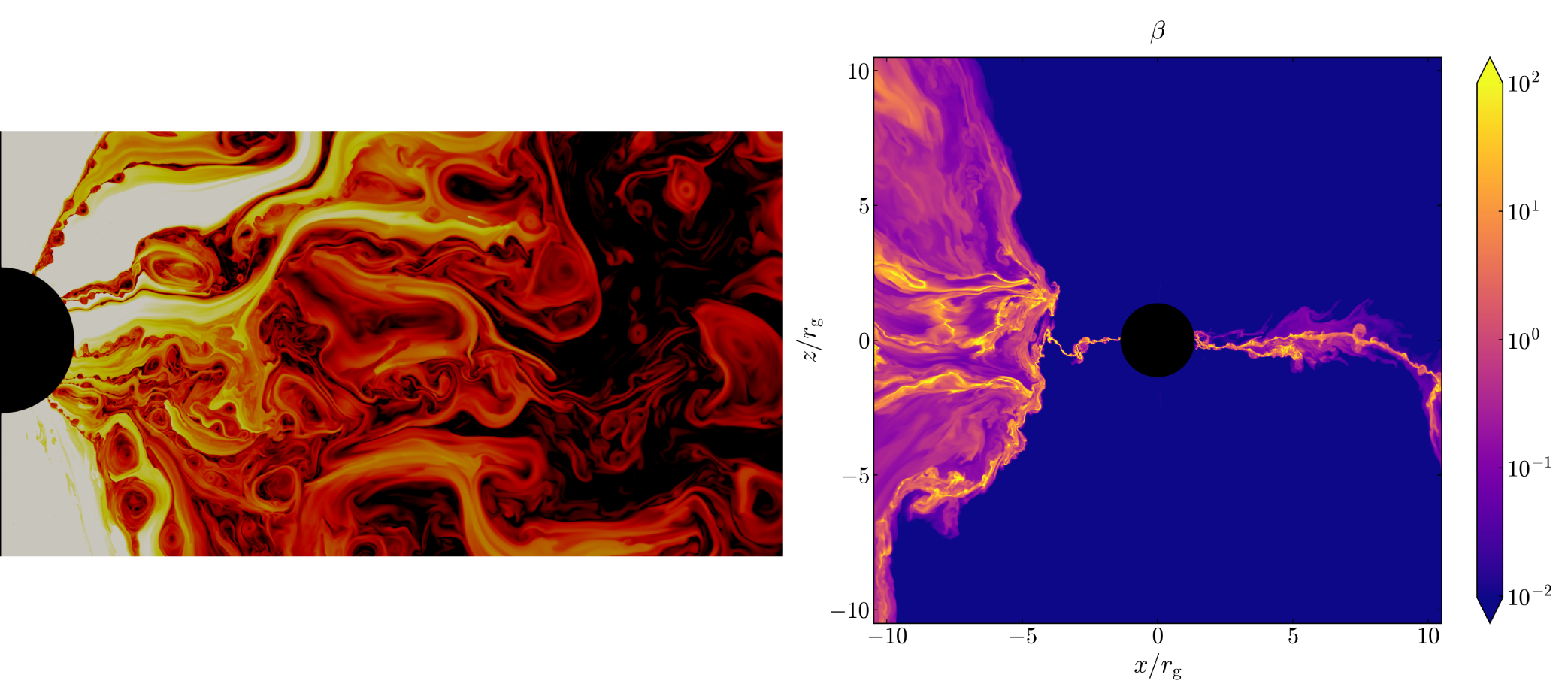}
\caption{Extremely high resolution global GRMHD simulations where promising sites of particle acceleration, plasmoids, have been resolved. Both snapshots show color scale in  $\log{\beta}$ (where plasma $\beta=U_g/U_B$):  (l) from a 2D resistive GRMHD in \cite{Ripperda_Bacchini_Philippov_2020} showing plasmoids forming in current sheets at the jet/disk interface nar the black hole as well as in the disk, (r) 3D ideal GRMHD from Ripperda, Liska et al., subm., showing plasmoids forming during a flare where the accretion flow near the black hole reduces to a thin current sheet.}
\label{fig:plasmoids}
\end{figure}

The combination of global VLBI imaging of the inner few to $\sim 10^3 r_g$ now possible with the Event Horizon Telescope (EHT; \cite{EHTPaperI,EHT_3c279,EHTPaperVII_2021}) simultaneously with radio through VHE \gr\ observations opens a new door towards resolving these questions, particularly in light of advances in the simulations. New GRMHD simulations at very high resolution (with and without explicit resistivity; \cite{Ripperda_Bacchini_Philippov_2020}; Ripperda, Liska, Chatterjee et al., ApJL, subm., and see Fig.~\ref{fig:plasmoids}) 
are able to resolve current sheets from reconnecting anti-parallel magnetic fields becoming unstable and forming plasmoids containing energised particles, that merge and grow into astrophysically-scaled hotspots seen at larger scales, e.g., \cite{Nathanail_Fromm_Porth_Olivares_Younsi_Mizuno_Rezzolla_2020}.  So far this has only been achieved for a few simulations but as higher resolution becomes more feasible in the next decades, between GPU acceleration \cite{Liska_HAMR_2019} or adaptive mesh refinement \cite{Ripperda_Bacchini_Philippov_2020}, or likely both, we can expect to see simulations that not only self-consistently launch and follow jet evolution to observable scales in 2d, as in \cite{Chatterjee_Liska_Tchekhovskoy_Markoff_2019}, but also extend to 3D while being able to resolve smaller scale instabilities and plasmoid formation.  These simulations will quickly move beyond the current state of the art, where prescriptions for lepton distributions from kinetic simulations are already incorporated as subgrid models to enable predictions of broadband spectra and VLBI images, e.g. \cite{Moscibrodzka_Falcke_Shiokawa_2016,ball_electron_2018, EHT_PaperV}, and MWL lightcurves \cite{Chatterjee_MWL}.  We can expect more groups to include PIC-techniques within the MHD simulations, e.g. \cite{bai_magnetohydrodynamic-particle--cell_2019, Sironi_Rowan_Narayan_2021} particularly now that plasmoids are being resolved. Two fluid approaches can extend these works to considering ions and electrons separately, but self consistently including hadronic acceleration globally would 'break' the fluid assumption.  However if localised regions such as plasmoids dominate the hadronic acceleration one may be able to make approximations to local acceleration without attempting to include the feedback on the dynamics globally.  In that case there will be simulations that can make predictions for EHT and next-generation global/space VLBI images at the same time as predicting MWL images, spectra and variability, plus neutrino and pion-induced \gr\ fluxes, at least for low-luminosity sources.  These will bring us one step closer as a field to definitively testing the various competing ideas about acceleration processes and sites within AGN and by extension TDE jets, and help resolve the longstanding questions and degeneracy regarding hadronic acceleration, power and feedback. 

\subsubsection{Gamma-ray bursts}

Gamma-ray bursts (GRBs) offer a unique laboratory for the coupling between dynamics and particle acceleration because, between the explosion and afterglow, one can probe the
physics of a blast wave over a very wide range of Lorentz factors.  This range spans from over a few $10^2$ during the prompt phase to $\beta\Gamma<1$ in the late afterglow, and does so nearly in realtime, with the EM data tracking physical conditions at the shock.  Modelling the evolution self-consistently using samples of sources and statistical inference techniques is already yielding surprising results (e.g., \cite{Aksulu_GRBs_2021}).

Furthermore, the recent burst of discoveries of TeV emitting GRBs has raised challenges for standard emission models, such as \cite{hess_190829A} where the spectrum seems to favour synchrotron emission extending from X-ray to TeV, and raising questions about how those photons manage to escape the source unattenuated. Clearly magnetic fields are dynamically important, and thus the cutting edge involves the development of sophisticated (magneto-)hydrodynamical codes for numerical simulation of GRB sources. Well-resolved three dimensional GRMHD simulations are becoming increasingly common and the symmetry-breaking involved in transitioning from two to three dimensions has been crucial to understanding jet formation and propagation. 

There are still debates about the acceleration properties and processes within GRBs, with many of the outstanding questions overlapping with relativistic jets in AGN and XRBs. Shock acceleration is probably the main mechanism that accelerates ions as well as electrons to high energies, with particle acceleration taking place both at the reverse shock at early times \cite{laskar2013} and the forward shock thought to produce the GRB afterglow. Both the acceleration and radiation mechanisms that cause the prompt \gr\ emission are still unclear, with internal shocks and magnetic reconnection important to consider. Understanding how the jet energy is dissipated is intrinsically linked to the particle acceleration process and has clear parallels with black hole jets, along with the question of whether the jets are comprised of ultra-relativistic protons, or if they are predominantly leptonic pairs. TeV sources should be far more plentiful in the former case. 

The same shocks which are thought to accelerate the electrons responsible for the radiative GRB signatures  should also accelerate protons. Based on their large powers, GRBs have long been discussed as UHECR sources, with maximum particle energies in principle reaching the UHE ($\gtrsim$EeV) regime. However, both the internal and the external reverse shocks are mildly to highly relativistic, and so the problems with UHECR acceleration at relativistic shocks outlined in section~\ref{sec:theory} apply. Further theoretical and modelling work is needed to understand whether these problems can be alleviated. Beyond UHECRs, other multimessenger signals come from the resulting cascades from hadronic collisions, and for short duration GRBs, the  gravitational waves produced by the NS-NS merger. In the latter case, the insight gleaned from simulations of the merger can be used to inform the propagation and dissipation physics of the jet and its particle populations. In summary, GRBs provide a fertile ground for theory that has many physical parallels with AGN jets in terms of particle acceleration mechanisms and energy dissipation, but with the advantage that the evolution of the particle populations can be investigated in detail in the time domain.

\subsubsection{Diffuse/unified contributions from populations and other sources}

Convolving individual source models with the entire population will be a key focus in the coming decade. For the Galaxy it is important to note that as these populations begin to reveal themselves in the TeV/PeV, there will be strong synergy with the population synthesis models developed by the stellar evolution community.  If we can identify the number and types of PeV sources, and place limits on their duty cycles, we can use these results to better constrain the overall Galactic population of stellar remnants, explosions and binaries.  These in turn can directly yield predictions for, or be tested against, the diffuse emission (Fig.~\ref{fig:CRdiffuse}), total CR distribution/composition (Fig.~\ref{fig:CRsummary}), which will also require more sophisticated treatments of local abundances, as well as help anticipate stellar feedback, and the balance of pressures in starburst galaxies.  For AGN, a better understanding of which sources classes are CR accelerators and under what conditions will help us better constrain the extragalactic diffuse fluxes, as well as constrain the powers released by black hole jets into their surroundings, with links to cosmological feedback. In all these cases, the propagation physics of CRs in magnetic fields and the MM connection between CR ions, neutrinos and MWL are critical pieces of the puzzle (see Chapter~\ref{sec:travelingmessengers} for further details).

There are also new potential CR sites being discovered via recently initiated X-ray and radio facilities, for instance a 15 Mpc intergalactic medium emission filament, a warm gas bridge, infalling matter clumps and (re-) accelerated plasma was discovered in the Abell 3391/95 galaxy cluster system, by combined SRG/eROSITA data with ASKAP/EMU and DECam data. Additionally, observations of the Coma cluster with the SRG/eROSITA, revealed a rich substructure in the X-ray surface brightness distribution with most salient features
which can be naturally explained by a recent (on-going) merger with the NGC 4839 group. 
In particular, the data identified a faint X-ray bridge connecting the NGC 4839 group with the cluster is  a convincing proof that NGC 4839 has already crossed the main cluster. Furthermore, the presence of two shocks with the subsequent electron acceleration supports the radio halo formation observed.

\subsection{Critical paths forward for the coming decade}

In this {\sl Cosmic Accelerators} section we have attempted to describe the cutting edge and challenges in theoretical and numerical studies of astrophysical particle accelerators, as well as identifying some key outstanding questions and synergies with observations. Our effort is far from exhaustive and the topic summarised here is clearly related to a number of other sections of this {\sl White Paper}, particularly Chapter~\ref{sec:travelingmessengers}. It is clear that particle acceleration is a fertile and extremely active area of theoretical astrophysics, straddling multiple different sub-fields and including all possible messengers (neutrinos, CRs, photons and gravitational waves). The physics at work is often nonlinear, self-regulated and complicated, and covers a vast range of length and energy scales, thus posing unique challenges to theoretical and numerical efforts.

Given the complexity and scale-separation problem(s) described above, it is clear a variety of numerical and other theoretical methods must be used in a complementary fashion to improve our understanding of {\sl Cosmic Accelerators}. It follows that interdisciplinary, collaborative work will be needed to efficiently leverage European expertise on these various subtopics. We therefore suggest that significant effort is devoted to community building, facilitating collaboration between scientists and institutions with different specialisations and cross-pollination of ideas and methods; the importance of this is only strengthened by the impact of the Coronavirus pandemic on scientific collaborations. The EuCAPT initiative itself can play an important role here, but we also emphasise the need for dedicated conferences that span different sub-fields. Specifically, we suggest workshops styled on those hosted by the US-based {\sl Aspen Center for Physics}\footnote{\url{https://www.aspenphys.org/}} and {\sl Kavli Institute for Theoretical Physics}, \footnote{\url{https://www.kitp.ucsb.edu/programs/current}} as well as interdisciplinary conference centres such as the \textit{Lorentz Centre}\footnote{\url{https://www.lorentzcenter.nl/}} in the Netherlands, plus conferences such as the Particle Acceleration conference held in Calabria, Italy in 2018 \cite{perri_particle_2020}. Financial and logistical support for similar programs will help foster the creativity and collaboration needed to tackle physics problems that cannot, at this stage, be solved with computing brute force alone.   

In parallel there is an urgent need for more investment in high performance computing (HPC) facilities since much of "European" work is being carried out on, e.g., USA facilities.  If we want to maintain our competitive edge we need investment particularly also in GPUs and developer/support positions to help the community transition to accelerated coding. In the context of upcoming European infrastructure like the European Open Science Cloud (EOSC) the resources to model large MM data sets as mentioned in Chapter~\ref{sec:astrostatistics} will require linking currently disparate methods into one self-consistent approach to simultaneously predict images and MM fluxes.  In order to test theoretical models statistically against the data, tools need to be developed that allow forward-folding these models into different instrument responses simultaneously for all messengers.

\section{Traveling and Interacting Messengers}
\label{sec:travelingmessengers}
{\bf Coordinators:} Daniele Gaggero and Kumiko Kotera. \\[2pt]
 {\bf Contributors:}
Rafael Alves Batista, Pasquale Blasi, Mauricio Bustamante, Silvio Sergio Cerri, Armando di Matteo, Carmelo Evoli, Stefano Gabici, Dario Grasso, Claire Gu\'epin, Alexandre Marcowith, James Matthews, Foteini Oikonomou, Christoph Pfrommer, Federico Urban, Arjen van Vliet.\\

Cosmic particles travel from their acceleration regions to the Earth, interacting with the traversed media at various scales via micro-physics processes. These interactions leave imprints on astrophysical environments, and multi-messenger hints on the long-sought origins of the highest energy particles. In this Chapter, we discuss the theoretical challenges related to cosmic-ray transport, interactions and the multi-messenger perspectives for astrophysics and fundamental physics.

\subsection{Micro-physics of cosmic-ray transport}\label{section:microphysics}

How do energetic charged particles interact with electromagnetic fields and, simultaneously, feed back onto the same fields they are interacting with? What are the plasma processes supplying cosmic rays and regulating their journey through turbulent media? How does the plasma environment react to the vibrant presence of such traveling cosmic rays? 
These are fundamental questions that lay deep at the roots of cosmic-ray research, usually referred to as {\it``the micro-physics of cosmic rays''}. 
Answering these questions is essential for the description of CR transport in a wide range of astrophysical phenomena: from the acceleration of cosmic rays (CRs) in diverse objects, supernova remnants (SNRs) and pulsar-wind nebulae (PWNe), the generation of ultra-high-energy cosmic rays (UHECRs) in active galactic nuclei (AGN) and gamma-ray bursts (GRBs), to their transport and self-confinement at different astrophysical scales, the inter-stellar medium (ISM) of the Galaxy as well as within the circum-galactic medium (CGM), the inter-galactic medium (IGM) and the intra-cluster medium (ICM) of galaxy clusters. 

As such, they feed not only on a purely theoretical thirst for knowledge, but also on our need to provide an explanation for a plethora of astrophysical observations that challenge our current understanding and, at the same time, help us in planning the  next-generation experiments.  
The plasma processes that regulate CR transport and describe the instabilities they excite are also at the basis of the description of laser-plasma wake-field acceleration processes -- which might play a fundamental role in future Earth-based particle accelerators, and is also the same basic physics underlying particle motion in controlled-fusion devices, just to mention a few examples outside of the astrophysics domain.
In order to make significant progress in understanding the micro-physics of cosmic rays, it is crucial to complement theory with the state-of-the-art numerical simulations. This also means to combine in a synergistic way a set of different approaches and techniques that have been historically employed separately to attack the problem ({\it e.g.}, local kinetic approach vs global phenomenological models) or that have been developed within different research fields (e.g., advanced numerical methods and massive parallelization techniques commonly employed by the plasma physics community). To this end, we envision the following steps to be taken by the cosmic-ray community.
 
\subsubsection{Kinetic approaches and ``effective'' models}  An actual understanding of the basic processes underlying CR propagation and ``CR-turbulence feedback'' ({\it i.e.}, including CR-driven instabilities) requires a self-consistent kinetic approach. One necessary step to take is represented by {\it ab-initio} simulations in which CRs are treated kinetically, either via Lagrangian (``particle-in-cell'', PIC) or Eulerian (``Vlasov'') methods, consisting of both full-kinetic and hybrid fluid-kinetic approaches. These simulations will cover a broad range of plasma regimes (plasma $\beta$ parameter, ionisation fraction, etc.) and turbulence properties, (type of injected fluctuations and turbulence level) in order to characterise the micro-physics of CR transport in different environments.
The results of such computations will provide a unique opportunity to understand the nature of the mutual interaction between CRs and MHD turbulence in the Galaxy. 
In the GeV-TeV domain, the problem is likely dominated by the non-linear interaction between CRs and self-generated turbulence. In this energy range, the free energy stored in the CR population (due, e.g., to CRs with drift velocities exceeding the local Alfv\'en speed and/or to CR anisotropies) is large enough so that it efficiently drives Alfv\'enic turbulence via resonant (streaming~\cite{1969Kulsrud}) and non-resonant (Bell~\cite{Bell2004}) instabilities, as well as generate background ion-cyclotron modes~\cite{Shalaby:2021}. All these instabilities are expected to contribute to a local enhancement of magnetic-field fluctuations, and off which CRs efficiently scatter, thus determining a decrease in their own mean free path (``self-confinement'').
At higher energies, {\it i.e.}, above TeV, CR scattering is instead more likely dominated by their interaction with pre-existing magneto-hydrodynamic (MHD) turbulence. However, we are still unable to pinpoint which modes dominate the scattering rate in this regime. In fact, the anisotropy of the Alfv\'enic cascade {\it \`a la} Goldreich-Sridhar~\cite{GoldreichSridharAPJ1995} -- the amount of power that goes into the cascading turbulent fluctuations at any given scale being biased by a {\it mean} magnetic-field direction -- is expected to enter the picture by significantly lowering the efficiency of gyro-resonant scattering of CRs on Alfv\'en modes~\cite{ChandranPRL2000}. At the same time, depending on the plasma regimes under consideration, other MHD modes that do not exhibit such anisotropic behavior in their cascade~\cite{ChoLazarianPRL2002} can significantly contribute to the overall turbulence budget, thus possibly re-establishing a more effective scattering rate of CRs on pre-existing turbulent fluctuations in the $\gtrsim$~TeV energy domain~\cite{FornieriMNRAS2021}.
This first-principle approach paves the way to the development of {\it effective models} for the propagation of CRs based on their energy and on the relevant environmental parameters that will be employed in global CR-transport simulations~\cite{2019Thomas}. Such semi-analytical effective models are needed not only to create a useful visual picture of what we expect to happen, but also to extrapolate the results obtained via PIC/fluid-PIC simulations, which are usually limited to scales that are not directly those of astrophysical interest, to astrophysical situations.
 
\subsubsection{Global approaches and comparison with the data landscape} 
In order to produce predictions to be tested against local and non-local CR data, {\it``global CR-transport simulations''} are ultimately required. 
These simulations solve the full kinetic CR-transport equation (i.e., evolving the density and energy spectrum of each CR species taking into account diffusion, advection, reacceleration, losses, and also accounting for the complex network of interactions among these CR species, from spallation cross sections to decay times of unstable nuclei) on the scale of the whole Galaxy (i.e., including the inhomogeneous 3D structure of its magnetic field, density, gas phases, and CR sources). It is necessary that such global approach go beyond a phenomenological treatment of CR diffusion. One necessary step to take is to consider first-principle diffusion coefficients, gaining insights both from an analytical approach, possibly going beyond the standard quasi-linear theory (QLT) or its weakly non-linear extensions by developing a fully non-linear framework, and by implementing the semi-analytical (effective) models discussed above. 

\paragraph{} In order to pursue the above steps, one needs to employ state-of-the-art numerical methods and massive parallelization strategies in both full-kinetic and hybrid fluid-kinetic plasma codes, as well as in  cosmic-ray transport numerical and semi-analytical solvers (such as \texttt{GALPROP} \cite{Galprop1,Galprop2,Galprop3}, \texttt{DRAGON} \cite{Evoli:2008dv,Gaggero:2013rya,Evoli2017I,Evoli2017II}, \texttt{PICARD} \cite{Picard1,Picard2} and \texttt{USINE} \cite{Usine}). These strategies allow to optimise the performance of such codes when running on a large number of processors,
and are essential to afford the always larger and more detailed simulations required by the points above. From the ``kinetic approach'' point of view, one indeed needs to perform a large set of simulations in order to explore the relevant parameter space and to simultaneously allow for the necessary scale separation. 
As far as the ``global approach'' is concerned, on the other hand, it is required to efficiently manage the complex 3D structure of the whole Galaxy and, at the same time, to push the resolution below to the mean separation between CR sources (e.g., below $\sim100$pc; and, eventually, it will be required to (nonlinearly) couple the complex transport equation for all CR species with some evolution equations for the electromagnetic-field fluctuations). 

UHECR acceleration and propagation encompass its own set of specific challenges in order to couple ``kinetic" and ``global" approaches, with large separations of scales between the thermal and non-thermal populations of leptons and hadrons, {\it e.g.}, energies, gyroradius, or mean free paths for various energy-loss and interaction processes. Particle acceleration to these extreme energies involves, {\it e.g.}, shock acceleration~\cite{Marcowith:2016vzl} or magnetic reconnection~\cite{Zweibel09}; local 3D PIC simulations become computationally affordable to study these processes~\cite{Zhang:2021akj}, and allow to infer characteristic timescales to be used in phenomenological approaches. At the source level, the development of global PIC/fluid-PIC simulations is critical to achieve a joint description of spatial properties together with some micro-physics effects, as explored for instance in the context of global PIC simulations of pulsar magnetospheres~\cite{Cerutti:2014ysa, Guepin:2019fjb}.

While most of these strategies have been already developed and implemented in several numerical frameworks by the plasma physics community (e.g., in controlled fusion and space plasma research, as well as in the context of astrophysical plasmas), they are still overlooked by global CR-transport codes. Our recommendation is to further move in this direction, and significantly upgrade and optimize the suite of numerical tools available to the cosmic-ray community. 

\subsection{Impact of cosmic rays on their environment}

A crucial aspect we want to emphasize is the impact of CRs on the formation and evolution of galaxies and galaxy clusters. In fact, CRs provide the most important source of ionization in molecular clouds, supply dynamical feedback by driving galactic winds, and excite vigorous instabilities by escaping from their source region. A realistic modeling of these processes, both in the Galactic and extra-Galactic context, is a major challenge for the astrophysical community.

\subsubsection{Interactions of low-energy cosmic rays with interstellar matter}\label{sec:CRionization}

Low-energy CRs are the only ionizing agents capable of going through large gas column densities and penetrate interstellar clouds \cite{2020SSRv..216...29P}. Hence they influence star formation \cite{2007ARA&A..45..565M} and constitute the first link of a chain of reactions leading to a rich interstellar chemistry \cite{2012A&ARv..20...56C}. They are also responsible for the spallogenic nucleosynthesis of light elements in the Universe \cite{2018ARNPS..68..377T}. The Voyager data \cite{2013Sci...341..150S, 2019NatAs...3.1013S} marked a breakthrough in this context, because the spectrum of such particles inside the heliosphere is heavily affected by solar modulation \cite{2013LRSP...10....3P}. 
Thanks to these data, the local interstellar spectrum of CR nuclei and electrons is now known down to energies of few MeV (or few MeV/nucleon) \cite{2016ApJ...831...18C}.
Surprisingly, the ionization rate inferred from diffuse clouds observations is $1-3$ orders of magnitude higher than that expected for a spatially uniform distribution of low energy CRs in the Galaxy, as measured by Voyager 1 and 2 \cite{2018MNRAS.480.5167P, 2016A&A...585A.105L}.
The presence of the naturally expected spatial fluctuations in the intensity of low-energy CRs in the Galaxy \cite{2021arXiv210500311P} and the specific location of the Solar System inside the local bubble \cite{2019ApJ...879...14S} might justify the discrepancy. 
Such low energy CRs propagate {\it a priori} as test particles and  have a mostly ballistic trajectory along large scale magnetic field lines. However, this simple picture can be modified either because some turbulent power can be imparted into small resonant scales and/or by the interaction of these CRs with large scale compressional perturbations. Here again, dedicated studies of the turbulence-particle interaction microphysics at such scales in the different environments of interest (molecular clouds, accretion disks...) seem to be important to conduct. 

\subsubsection{Feedback driven by cosmic-ray pressure on star formation}\label{sec:CRpressure}

CRs, magnetic fields, and turbulence are in pressure equilibrium in the midplane of the Milky Way \cite{1990ApJ...365..544B}, implying that CRs play an important dynamical role in maintaining the ISM energy balance. 
In fact, if the CR and magnetic midplane pressures exceed that of the thermal plasma, their buoyancy overcomes magnetic tension 
and opens up the magnetic field into the halo: this enables CRs to be transported ahead of the thermal plasma along these field lines into the halo. 
If the CR flux exceeds the gravitational attraction of the disk, the multiphase ISM is accelerated and forms a strong galactic outflow as shown in vertically stratified boxes of the ISM \cite{Simpson2016,2016ApJ...816L..19G}, in isolated galaxy simulations \cite{2012MNRAS.423.2374U,Pakmor2016b,2021ApJ...910..126S}, and in cosmological simulations of galaxy formation \cite{2014ApJ...797L..18S,2020MNRAS.497.1712B,2020MNRAS.492.3465H}. Those demonstrate that CR feedback can regulate the star formation rate, and modify the structure of galactic disks. 
However, its strength depends once again on the microscopic CR transport properties discussed above. If the CR pressure is less than that of the thermal plasma, the latter can radiate away the excess energy to approach equipartition as a dynamical attractor solution. Most importantly, the far-infrared (FIR)-radio correlation and the FIR-gamma-ray correlation of star-forming galaxies enables testing CR calorimetry and thus to calibrate the CR feedback efficiency across the star formation sequence \cite{Werhahn2021b,Pfrommer2021}, provided the local CR data is reproduced in 3D MHD simulations of galaxy formation \cite{Werhahn2021a}.

Moving to the galaxy cluster environment, the observed star formation rates at the centre of cool cores are reduced to levels significantly smaller than those expected from unimpeded cooling flows. According to the standard paradigm, radio lobes that are inflated by relativistic AGN jets drive turbulence in the intra-cluster medium (ICM) and heat the surrounding thermal plasma at a rate that balances radiative cooling. Data suggests that the cooling plasma and nuclear activity are coupled to a self-regulated feedback loops. A promising heating mechanism can be provided by the streaming instability triggered by CRs that escape from the lobes as they are conducted along the magnetic fields into the ICM \cite{2008Guo,2017JacobPfrommerII}. 
We conclude that CRs may play a critical role in galaxy formation and the evolution of galaxy clusters by sharing energy and momentum with the thermal plasma, provided that we can cast the CR microphysics into efficient macroscopic fluid theories which enable predictive simulations.
 
\subsubsection{Cosmic-ray propagation near their sources}

The processes controlling CR escape from their accelerators depend on the source evolution stage and the surrounding ISM properties \cite{Nava:2016, Dangelo:2016, Nava:2019, Brahimi:2020}. Once they have escaped, CRs 
can back-react over the surrounding medium of the source \cite{Telezhinsky:2012, Celli2019} via processes described above. 
On the other hand, high CR anisotropies near sources can non-resonantly excite Alfv\'en modes \cite{Bell2004} or drive background ion-cyclotron modes in the comoving CR frame \cite{Shalaby:2021}. Besides, the effect of neutrals on higher-order CR anisotropies is starting to be explored \cite{Reville:2021}. These instabilities are a local source of magnetic turbulence which increases the CR scattering rate and decreases their mean free path, so that they are potentially responsible for gamma-ray halos around some pulsars \cite{Evoli:2018, Sudoh:2019lav, Giacinti:2019nbu, Dimauro:2020}. The generation of magnetic turbulence and this reduced diffusivity around source have multiple consequences: a dynamical feedback over the ISM \cite{Schroer:2020}, explain a part of the CR grammage \cite{Dangelo:2016}, explain detailed features in the CR spectrum \cite{Ivlev2018}. An essential way to select the correct theory will be to perform dedicated observations of the non-thermal emission around cosmic-ray sources from radio to the multi-TeV domain. Aside this, dedicated simulations of CR escape and feed back on the local medium are necessary to understand the sub-structures of the ISM. 

\subsection{Tracing back messengers to probe high-energy events}

At the highest energies, traveling messengers, {\it i.e.}, CRs and their daughter neutrinos and gamma rays, carry hints on their unknown origins. Deciphering their observational signatures requires to understand and to accurately model their propagation media and their interaction processes with these media. This exercise needs to be conducted globally, by combining different messengers at each stage of the modeling.

\subsubsection{The multi-messenger challenges}

The reported transient photon-neutrino associations have not met simple theoretical explanations ~\cite{IceCube:2018dnn,Stein:2020xhk}. In the future, with increased expected rate of coincident detection and increased false positive rate of associations, the challenge for theory will be to direct the follow-up efforts of steerable telescopes with limited observation time \cite{Guepin17}. Models also still fail to converge on the possibility of a common origin of observed high-energy neutrinos, gamma rays, and UHECRs, as seems to be indicated by the coincident levels of their diffuse fluxes~\cite{Waxman:1998yy,Murase:2018utn}. The different messengers may be connected through their arrival directions. For UHECR and neutrino cross-correlations, the theory is well established; for example neutrino-UHECR correlations can only be expected for sources with negative source evolution~\cite{Palladino:2019hsk}. For joint gravitational-wave (GW) and HE neutrino emitters, current limitations stem mainly from the angular resolution of GW instruments, but also on the modeling of astrophysical GW backgrounds \cite{2019ApJ...870..134A}. 
To increase the detectability of multi-messenger transient sources and to interpret the observed diffuse multi-messenger data and their cross-correlations, theoretical input is needed in terms of cosmic-ray acceleration capabilities, spectral indices, composition, and neutrino production efficiency, for different sources (see also Sections~\ref{sec:cosmicaccelerators} and \ref{sec:particlesfromstars}). 

Multi-messenger analyses play a crucial role in the Galactic environment as well. In the GeV - TeV domain, the presence of a progressive hardening of the hadronic spectrum towards the inner Galaxy has been widely debated in the community \cite{Gaggero:2014xla,Acero:2016qlg,Yang:2016jda}. The search for a Galactic neutrino component --- currently investigated by the ANTARES and IceCube collaborations --- together with the study of the multi-TeV gamma-ray diffuse emission, are expected to improve significantly our understanding of this feature. This anomaly certainly poses a challenge on the theoretical side, and may be deeply connected to the issues discussed in the Section \ref{section:microphysics}. In fact, several interpretations based on either a strongly anisotropic transport regime \cite{Cerri:2017joy}, or non-linear transport effects \cite{Recchia:2016bnd} were recently put forward. More refined theoretical investigation, and further multi-messenger studies at high energy will provide the opportunity to grasp the nature of this relevant spectral feature.

\subsubsection{Arrival directions and magnetic fields}

Several anisotropy signals have been discovered recently in the UHECR arrival directions~\cite{Aab:2017tyv,Abbasi:2018qlh}. In order to unambiguously interpret these signals 
and to identify the sources, a good understanding of source physics is again needed. With improved charge separation (AugerPrime) and increased statistics (TAx4, GRAND, and POEMMA), major advancements are within reach. In parallel, the understanding of the propagation of UHECRs will be paramount.

The propagation of high-energy particles can be substantially affected by Galactic and extragalactic magnetic fields (GMF and EGMF). The GMF is poorly known despite recent advances~\cite{Pshirkov:2011um,Jansson:2012pc,Adam:2016bgn,Unger:2017kfh}. 
Projects like IMAGINE~\cite{Boulanger:2018zrk} 
will make significant improvements by building comprehensive models with observables other than the Faraday rotation measures, polarisation, and synchrotron emission maps. 
EGMFs are even more uncertain, especially in cosmic voids, which fill most of the volume of the Universe. New perspectives should be offered by SKA and other radio instruments, {\it e.g.}, \cite{Vazza_2018}. Alternatively, high-energy gamma-ray observations may also play an important role, since the electromagnetic cascades they induce in the intergalactic medium can be used to constrain magnetic fields~\cite{Durrer:2013pga, Batista:2021rgm}. It has been claimed that electromagnetic cascades are quenched by plasma instabilities arising due to the interactions of the electron-positron pairs with the intergalactic medium~\cite{Broderick:2011av, Vafin:2018kox, Vafin:2019bdb}. This issue cannot be tackled at the moment due to the difficulty in performing the PIC simulations in environments with densities as low as in cosmic voids (see Section~\ref{section:microphysics}). Nevertheless, it seems possible to fully bracket this source of uncertainties to obtain estimates on magnetic-field properties using gamma-ray observations~\cite{Yan:2018pca, AlvesBatista:2019ipr}.

The effect of EGMFs on UHECR propagation is commonly investigated employing simulated cosmological volumes~\cite{Dolag:2004kp, Hackstein:2017pex, AlvesBatista:2017vob}, but this meets difficulties regarding three-dimensional simulations of particle propagation (discussed below). This problem becomes especially important considering the vast parameter space of EGMFs allowed, which renders impractical the systematic exploration of the uncertainties. Conversely, UHECRs can also be used as probes of magnetic fields, under certain assumptions about the sources of UHECRs~\cite{Boulanger:2018zrk,Bray:2018ipq,AlvesBatista:2018owq,Eichmann:2019ong,VanVliet:2021sbc}, but large uncertainties on the UHECR chemical composition and the GMF models are limiting factors. 

\subsubsection{Radiative interaction backgrounds and hadronic interactions}
        
The interactions of high-energy particles and the generation of secondaries in the sources, their vicinity, and in the Universe is fairly well understood from a theoretical perspective. Nevertheless, 
the distribution of background photons and baryons provide a high level of uncertainty. 
While the cosmic microwave background (CMB) is known to a high precision, the extragalactic background light (EBL) and the cosmic radio background (CRB) are less well known~\cite{Batista:2015mea, AlvesBatista:2019rhs}, although they pervade the Universe and serve as interaction target for cosmic rays, gamma rays, and charged leptons. In the sources and their environment, background radiation and baryon levels are inferred from observations and source structure and evolution models. They condition the escape of cosmic rays and gamma rays, and the production of secondary multi-messengers. Theoretical progress on source physics (dynamics, evolution and radiation) will provide crucial input on this subject.

Another uncertainty for the propagation of UHECRs comes from the limited knowledge of cross sections for nuclear interactions, most notably the cross sections for disintegration of nuclei when interacting with background photon fields. 
For many nuclear isotopes, no measurements of the cross sections are available, or only the total photoabsorption cross section is given without data on photodisintegration into specific channels. While estimates of these cross sections can be provided by phenomenological models (see e.g.~Refs.~\cite{Puget:1976nz, Koning:2012zqy, Fasso:2003xz}), the correctness of these estimations is unclear without experimental verification. A different choice of photodisintegration model can have a considerable impact on the expectation for the UHECR spectrum and composition at Earth~\cite{Batista:2015mea, AlvesBatista:2019rhs}. 
     
\subsubsection{Numerical modeling challenges}\label{section:numerical_propagation}

A number of numerical tools exist 
that employ various techniques ranging from solving transport equations to full Monte Carlo (MC) simulations~\cite{Batista:2016yrx, Aloisio:2017iyh, Heinze:2019jou}. While the former approach tends to be faster for a fixed setup, the latter provides more flexibility. However, the problem becomes complex in the presence of magnetic fields. In this case, a three-dimensional treatment of particle propagation is often required. This is possible in a full MC approach, but it can be computationally expensive. Particles emitted by arbitrarily distributed sources have to be tracked over large distances, while continuously being deflected by magnetic fields and interacting with the gas and radiation present along the way. 
While there are some solutions to this problem implemented in codes like \texttt{CRPropa}~\cite{Batista:2016yrx, Jasche:2019sog},
this difficulty makes it prohibitive to fully probe the multidimensional parameter space.

Another challenge occurs due to the large energy range of astrophysical messengers. High-energy particle interactions can lead to the production of lower-energy secondary particles, which, in turn, can produce more particles, causing a cascade of increasingly larger numbers. Because this process redistributes the energy of the primary particle over several decades in energy among the secondaries, the performance of MC methods is often sub-optimal. However, a bulk-transport approach may not capture all the properties of the medium through which the particles travelled. This is, for instance, the case with electromagnetic cascades in the intergalactic medium, where an ultra-high energy primary can cause a cascade down to GeV energies. The simulations of such cascades in full detail (including effects of magnetic fields, for example) can, therefore, be problematic. Codes like \texttt{CRPropa}~\cite{Batista:2016yrx} and \texttt{EleCa}~\cite{Settimo:2013tua} nevertheless do provide a framework for such simulations and improvements on the efficiency can be expected in the near future.
 
\subsection{Connections with fundamental physics}

The study of charged cosmic rays and the associated non-thermal emission has offered the opportunity to investigate the presence of signals potentially connected to the annihilation or decay of particle dark matter. Many channels have been studied in detail over the latest decade. In the domain of charged cosmic particles, we highlight the positron, antiproton and anti-deuteron/anti-helium channels: in fact, the paucity of anti-particles, mostly produced by secondary interactions, are expected to improve the signal-to-noise ratio in favor of a hypothetical dark matter signal. In the gamma-ray domain, we mention in particular the study of dwarf galaxies and the quest for a signal towards the inner part of the Milky Way. Several claims were debated over the years in some of those channels, and in many cases strong constraints on the properties of the hypothetical dark matter particle (mainly mass and annihilation cross section) were placed (see \cite{Cirelli:2012tf} for a comprehensive review).
We remark that the challenge in these indirect searches for new physics is the correct modelling of the astrophysical fluxes, still hampered by the unclear aspects mentioned above regarding the nature of cosmic-ray transport (see \cite{Gaggero:2018zbd} for a specific review on this inter-disciplinary interplay). Therefore, a significant advance in the understanding of the fundamental problem of CR transport will also improve the capability to identify a potential exotic signal. We refer to Section~\ref{sec:DM} for a more comprehensive discussion about New Physics searches.

In a broader context, we point out that the long-acknowledged power of high-energy cosmic rays~\cite{Anchordoqui:2018qom, AlvesBatista:2019tlv}, gamma rays~\cite{AmelinoCamelia:1997gz, Stecker:2001vb, Stecker:2003pw, Abdalla:2020gea}, and neutrinos~\cite{Ahlers:2018mkf, Ackermann:2019cxh, Arguelles:2019rbn} to probe fundamental physics stems from their high energies---which allow them to probe physics at new energy scales---and their long traveled distances---which allows minute new-physics effects to accumulate and compound. For cosmic rays and gamma rays, data-driven tests have been performed for decades; for neutrinos, the IceCube discovery of TeV--PeV astrophysical neutrinos finally turned proposals into data-driven tests.  The theory landscape of fundamental-physics models that can be tested is vast; it includes, {\it e.g.}, the violation of fundamental symmetries~\cite{Colladay:1998fq, Kostelecky:2008ts} and the existence of new interactions with baryonic and dark matter.  These tests are weighed down by particle-physics and astrophysical uncertainties that can be large, but that must be factored in to yield realistic sensitivities.
For the future, we advocate for a two-pronged theory strategy, made possible by the growing body of experimental observations, that maximizes the chances of making a discovery or of setting new, robust limits.  First, tests of fundamental physics should always factor in all relevant particle-physics and astrophysical uncertainties; {\it e.g.}, in UHECR hadronic interaction models, in the neutrino mixing parameters, and in the shape of particle spectra.  Additionally, fits of proposed models to experimental data should, inasmuch as possible, freely explore the full space of model parameters associated to them, rather than fix some parameters to preconceived values; otherwise, valid descriptions of the data could be overlooked.

\section{Neutrino properties}
\label{sec:neutrinophysics}

\textbf{Coordinators:} Olga~Mena and Thomas~Schwetz.\\ \textbf{Contributors:}
Gabriela~Barenboim, Mauricio~Bustamante, Pilar~Coloma, Marco~Drewes, Ivan~Esteban, 
Julia~Harz, Christiane S.~Lorenz, Silvia~Pascoli.

\bigskip

\begin{figure*}
	\centering
	\includegraphics[width=1.0\textwidth]{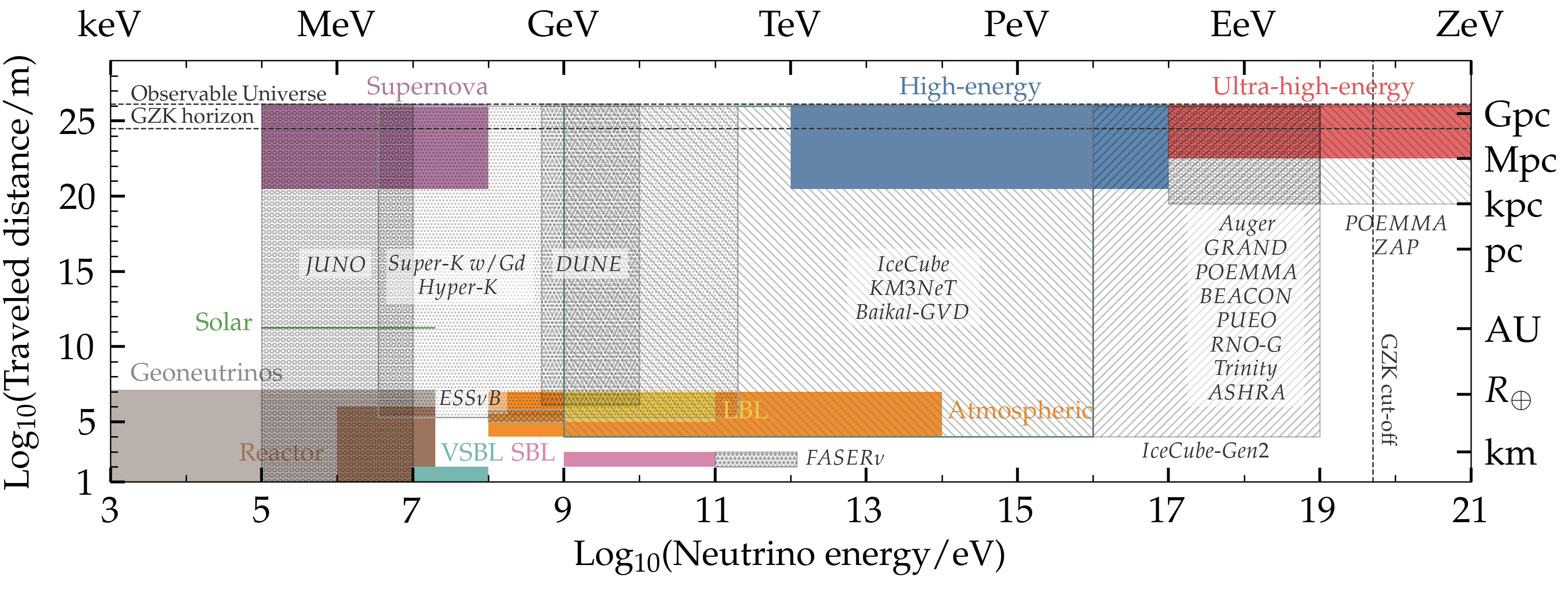}
	\caption{\textit{Neutrino sources distributed in energy and distance, and present and future experiments that target them.  Figure provided by M.~Bustamante, reproduced from Ref.~\cite{snowmass_mb}.}}
	\label{fig:scales}
\end{figure*}

Neutrino physics is living a Golden Age. In 2015, the Nobel Prize in Physics was awarded to Takaaki Kajita and Arthur B. McDonald \emph{“for the discovery of neutrino oscillations, which shows that neutrinos have mass. [...] New discoveries about the deepest neutrino secrets are expected to change our current understanding of the history, structure and future fate of the Universe'' } (see \cite{Fukuda:1998mi,Ahmad:2002jz,Ahmad:2001an} for essential publications). From neutrino oscillation experiments, we know that neutrinos have masses, implying a departure from the Standard Model (SM) of Particle Physics. Neutrinos have a crucial impact in Particle Physics, Cosmology and Astrophysics. Figure~\ref{fig:scales} shows the wide breadth of neutrino energies and baselines available to us, and the current and future detectors that target them.  In this section we focus on this timely and highly interdisciplinary topic, with the goal of identifying main open questions and key challenges for theory and phenomenology. 

\subsection{Origin of mass} 
\label{sec:neutrinoMass}

One fundamental question in physics is the origin of elementary particle masses; while those of the charged fermions stem from the Higgs mechanism, that behind neutrino masses remains a puzzle. 
Thanks to the overwhelming evidence for neutrino oscillations we know that in Nature at least two massive neutrinos exist, as required by the two distinct squared mass differences measured, the atmospheric $|\Delta m^2_{31}| \approx 2.55\cdot 10^{-3}$~eV$^2$ and the solar $\Delta m^2_{21} \approx 7.5\cdot 10^{-5}$~eV$^2$ splittings~\cite{deSalas:2020pgw,Esteban:2020cvm}. Since the sign of the largest mass splitting remains unknown, two possible mass orderings are possible, the 
\emph{normal} and the \emph{inverted} orderings \cite{deSalas:2018bym}. 
In the normal ordering, $\sum m_\nu \gtrsim 0.06$~eV, 
while in the inverted ordering, $\sum m_\nu \gtrsim 0.10 $~eV, 
with $\sum m_\nu$
representing the sum of the three light neutrino masses (see the right panel of Fig.~\ref{fig:mass}).  Furthermore, we know, from cosmology and nuclear decay searches, that the scale of neutrino mass is below $1$~eV, many orders of magnitudes smaller than for any other fermion in the SM.  From a theoretical perspective, there must be an underlying theory in Nature, maybe involving sterile neutrino states, responsible for the mass of neutrinos, probably different from the 
standard Higgs mechanism that gives mass to the charged fermions. Major future challenges to unravel these fundamental question are summarized in what follows.

\begin{itemize}
\item 
\textbf{Are neutrinos Dirac or Majorana particles?}
This a fundamental question in particle physics. Neutrinos are Dirac fermions if total lepton number ($L$) is conserved. They are instead of Majorana nature if there is lepton number violation (LNV): a neutrino will be indistinguishable from its antiparticle. It is essential to determine if lepton number is a fundamental symmetry of Nature or not and the answer could guide us in understanding the origin of neutrino masses and of their smallness.

From the model-building perspective, unless $L$ conservation is explicitly imposed, LNV is a generic feature of physics beyond the SM and a Majorana neutrino mass term for right-handed neutrinos should be present in the theory, as it is not forbidden by any SM gauge symmetry. Nevertheless, in spite of strong arguments in favor of the Majorana nature of neutrinos~\cite{Bilenky:2020vjk}, the possibility of $L$ conservation and Dirac neutrinos is not excluded experimentally. An experimental signal of the putative Majorana character of the neutrinos is neutrinoless $\beta\beta$ decay: the observation of this process would be a major discovery proving LNV and would point to the Majorana neutrino character. 
The non-observation of neutrinoless $\beta\beta$ processes provides at present bounds on the so-called “effective Majorana mass” of the electron neutrino, $m_{\beta \beta}$; see the middle panel of Fig.~\ref{fig:mass}. The predictions depend on the neutrino mass spectrum and unknown Majorana phases. Current experimental bounds constrain $m_{\beta \beta}<0.061-0.165$~eV at $90\%$~CL~\cite{KamLAND-Zen:2016pfg}, depending on the nuclear matrix elements. There is a strong experimental effort to reach sensitivities in the range  $m_{\beta \beta}<0.010-0.020$~eV~\cite{Agostini:2017jim,Giuliani:2019uno} to cover the values predicted in case of inverted ordering for dominant light neutrino mass exchange.

\begin{figure*}
	\centering
	\includegraphics[width=1.0\textwidth]{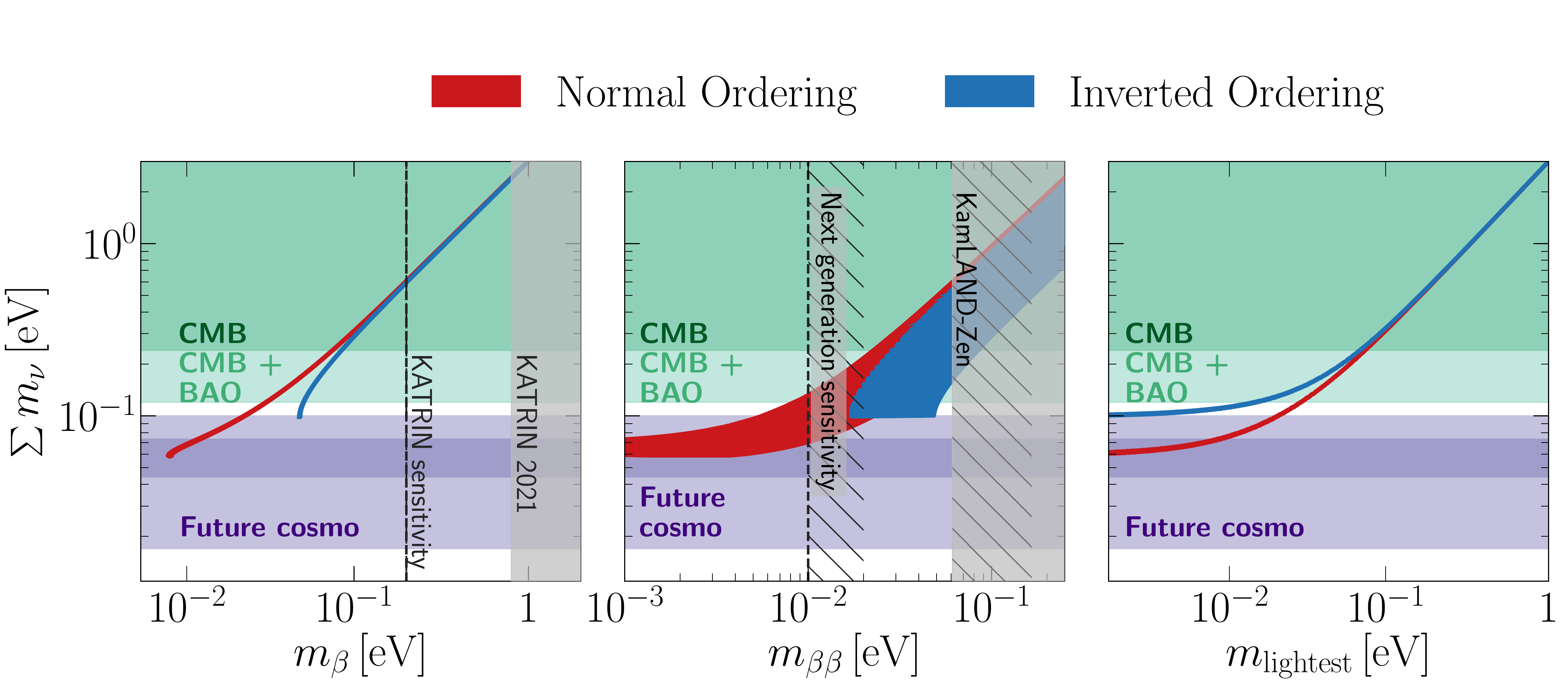}
	\caption{\textit{Theoretical predictions, current and future bounds for the three quantities characterizing the observables for neutrino masses: beta-decay ($m_\beta$), neutrinoless double beta decay $m_{\beta\beta}$ and the cosmological measured quantity $\sum m_\nu$ are depicted in the left, middle and right panels respectively. The hatched regions in the middle panel reflect the effect of uncertainties in the nuclear matrix elements in the bounds on neutrinoless double beta decay searches. Figure provided by I.~Esteban, adapted from \cite{Esteban:2020cvm}; see text for further references.}}
	\label{fig:mass}
\end{figure*}

\item \textbf{Can we develop new economic ideas and strategies to pin down the underlying interaction in case of an observation of neutrinoless double beta decay?}
While the observation of neutrinoless $\beta\beta$ decay would prove LNV, the fundamental origin of the process (i.e., the responsible operator) will be difficult to identify. Besides the generic light Majorana neutrino exchange it could also be induced by more exotic LNV physics. In a natural theory, such an observation necessarily implies a Majorana mass term for neutrinos \cite{Schechter:1981bd,Takasugi:1984xr}; however, numerically this Majorana mass can be very small~\cite{Duerr:2011zd}. Identifying the mechanism behind LNV in case of a positive observation will be crucial~\cite{Arnold:2010tu,Deppisch:2006hb, Gehman:2007qg,Meroni:2012qf}.

\item \textbf{Are there sterile neutrinos in Nature? 
If so, at which mass scale?} 
In the SM there are three \emph{active} neutrino fields. A generic extension of the SM could be the presence of fermionic SM gauge singlets, often called \emph{sterile} neutrinos or heavy neutral leptons. In this minimal setting, there is no fundamental symmetry establishing a definite number of sterile neutrino species nor their mass scale. Indeed, sterile neutrino masses can span the huge range from sub-eV up to energy scales close to the Planck mass \cite{Drewes:2013gca}. In many neutrino mass models (most notably in seesaw models) the presence of sterile neutrinos is directly linked to generating neutrino masses.

Sterile neutrino species, if present in Nature with masses $\lesssim$~MeV, should be added to the \emph{dark radiation background} in the Universe, represented by the number of relativistic degrees of freedom $N_{\rm eff}$ which, in the canonical three neutrino scenario, equals $3.044$~\cite{Bennett:2020zkv}. Cosmology can not only \emph{weigh}  neutrinos, but also \emph{count} them: recent cosmological analyses~\cite{Aghanim:2018eyx} lead to $N_{\rm eff}=2.99^{+0.34}_{-0.33}$ at $95\%$~CL which strongly constrains the presence of light sterile neutrinos. 
Sterile neutrinos at the eV scale may show up in short-baseline oscillation experiments; experimental hints from such experiments are reviewed in Ref.~\cite{Boser:2019rta} but no consistent picture in terms of sterile neutrino oscillations is emerging yet and such sterile neutrinos would be in tension with cosmology, e.g.,~\cite{Dentler:2018sju,Hagstotz:2020ukm}.
In the MeV range cosmological constraints dominate \cite{Vincent:2014rja}, while
sterile neutrinos roughly in the 10~MeV to TeV range can
be searched for directly with accelerator based experiments \cite{Deppisch:2015qwa,Ballett:2016opr,Cai:2017mow,Agrawal:2021dbo} and indirectly affect a large number of precision tests of the SM \cite{Chrzaszcz:2019inj}.

\item \textbf{Can gravitational wave observations tell us something about the origin of neutrino mass?}
A Majorana mass term could be generated by $B-L$ breaking at a high scale, possibly in the context of GUT theories. The related phase transition in the early universe can produce a possibly observable stochastic gravitational wave background~\cite{Hindmarsh:2011qj,Buchmuller:2013lra,Blasi:2020wpy,Dror:2019syi,King:2020hyd,Gelmini:2020bqg}. Exploring these connections may offer exciting synergies between the emerging field of gravitational wave astronomy and the search for the origin of a Majorana neutrino mass term as well as flavour symmetries in the lepton sector.

\item\textbf{Are flavor models that predict the pattern of neutrino masses and mixings testable?} 
The mass and mixing pattern for leptons is dramatically different than in the quark sector. The origin of those patterns is not understood at present and is sometimes called the SM flavor puzzle. Much effort has been devoted to model-building in order to provide dynamical or symmetry explanations. Can we define an optimal combination of observables to find out whether indeed a specific flavor model is behind all the fermion mixing and masses we observe in Nature?

\end{itemize}

From a phenomenological approach, the next generation of cosmological surveys, neutrinoless $\beta\beta$ decay experiments, and beta decay endpoint measurements offer promising improvements to pin down the absolute neutrino mass scale; see Fig.~\ref{fig:mass}. We are thus at the unique and exciting point where three very different techniques, based on very different physical processes, aim to precisely measure a given physics scale. We must be ready to interpret these future results and be fully convinced of the outcome of each of the three aforementioned neutrino mass searches. Key upcoming challenges in this regard are:
\begin{itemize}
    \item 
\textbf{Particle physicists are skeptical about cosmological neutrino mass determinations.} 
In order to convince the particle physics community that a potential neutrino mass determination from cosmology is robust, one needs to be able to disentangle the cosmological effects of neutrinos from other scenarios that can mimic these effects. For example, we will need to distinguish between neutrinos, dark energy, and curvature. In particular, the combination of upcoming Cosmic Microwave Background, galaxy clustering and lensing measurements may help to break the degeneracies between neutrinos and dark energy~\cite{Mishra-Sharma:2018ykh}.
Perspectives from near future galaxy surveys also benefit from additional probes, such as the halo and void abundances~\cite{Bayer:2021iyb}.
Furthermore, the reliability of neutrino mass results in view of persisting tensions in the $\Lambda$CDM model needs to be addressed.

\item \textbf{Can non-standard neutrino properties explain the $H_0$/$\sigma_8$ tensions?}
Neutrino self-interactions have been discussed as a potential solution of cosmological tensions~\cite{Kreisch:2019yzn}. However, these models are strongly constrained by laboratory data~\cite{Blinov:2019gcj} and by CMB polarisation data~\cite{Choudhury:2020tka,Brinckmann:2020bcn}. In addition, it has been shown that the inclusion of neutrino masses in the $\Lambda$CDM model does not alleviate the $\sigma_8$ tension~\cite{Joudaki:2016kym}; see also \cite{Archidiacono:2020yey}.

\item \textbf{What would happen if different neutrino mass measurements contradicted each other?}  
In the near future, the Karlsruhe Tritium Neutrino (KATRIN) experiment, a tritium beta decay experiment, might directly discover the electron neutrino mass. The current limit is $m_{\beta}<0.8$~eV (90\% CL)~\cite{Aker:2021gma}, and the expected sensitivity is 0.2 eV (90\% CL)~\cite{Drexlin:2013lha}; see the left panel in Fig.~\ref{fig:mass}. Therefore, a neutrino mass detection by KATRIN would stand in contrast to the current cosmological neutrino mass bound of $\sum m_\nu<0.12$~eV (95\% CL)~\cite{Aghanim:2018eyx} (see the green regions in Fig.~\ref{fig:mass}). Inconsistencies among laboratory and cosmological searches would definitely point to a much richer neutrino sector, to a departure from the standard model of cosmology, or to a combination of both. Such a situation would offer exciting possibilities to discover new physics; possibly providing insight in dark sectors of the theory; see e.g.,~\cite{Lorenz:2018fzb,Dvali:2016uhn,Esteban:2021ozz,Chacko:2019nej,Chacko:2020hmh,Escudero:2020ped,Renk:2020hbs}.  

Future cosmological searches will be sensitive to the minimum neutrino mass expected from oscillation experiments, as illustrated by the purple regions  in Fig.~\ref{fig:mass}, and will also be able to measure the expected theoretical deviation in the number of relativistic degrees of freedom $\Delta N_{\rm eff}=0.044$~\cite{Bennett:2020zkv}; see e.g.,~\cite{Basse:2013zua,Hamann:2012fe,Carbone:2011by}. The former cross-check will provide a unique test of neutrino interactions in the very early Universe, even before the formation of light nuclei. It will also be interesting to compare neutrino mass constraints from upcoming CMB surveys \textit{alone} and LSS surveys \textit{alone}, as a difference in these constraints may hint towards new physics in the neutrino sector. For example, current low-redshift constraints of the total neutrino mass sum often favor higher neutrino masses than \textit{Planck}~\cite{Poulin:2018zxs,Ivanov:2019pdj,Palanque-Delabrouille:2019iyz,Lorenz:2021alz,Salvati:2017rsn}.

\item \textbf{If, on the other hand, all the measurements are consistent, what do we learn?} 
How robustly do we know that all experiments are sensitive to the very same quantity (neutrino mass). Can we eventually learn about parameters that are theoretically poorly constrained (for instance nuclear matrix elements in neutrinoless double beta decay)? Will it be possible to determine Majorana phases?

\item 
\textbf{What would be the implications for cosmology/phenomenology/theory of a detection of the relic neutrino background?} 
Such a measurement is exceedingly difficult and would correspond to a revolutionary experimental achievement.
The most promising approach to detect relic neutrinos is to use neutrino capture in a $\beta$-decaying nucleus, as planned at the future 
PTOLEMY project~\cite{Baracchini:2018wwj}. 
The required amount of target material, energy resolution, and background levels are challenging~\cite{deSalas:2017wtt}. 
Such a project would provide excellent sensitivity to the neutrino mass scale and potentially to the mass ordering. 
Are there fundamental limitations to such a measurement \cite{Cheipesh:2021fmg} and can they be overcome?
What would be the implications of the relic neutrino detection for the cosmological model? Can it be used to test other astroparticle physics \cite{Bondarenko:2020vta}?
Synergies with astrophysical simulations and observations are highly relevant for this purpose. 
\end{itemize}

\subsection{Origin of matter}
\label{sec:neutrinoLG}

Our visible Universe is made out of matter (and no antimatter) and can be described from galaxy clusters to atoms by the SM of particle physics. Assuming that after inflation the Universe was created symmetrically, a mechanism in the evolution of the early Universe is needed to explain this matter and antimatter asymmetry, and hence our own existence. 
Explaining the dynamical generation of a matter-antimatter asymmetry (baryogenesis) requires i) baryon number ($B$) violation, ii) $C$ and $CP$ violation and iii) a deviation from thermal equilibrium \cite{Sakharov:1967dj}. Due to the $B+L$ violating SM electroweak sphaleron processes~\cite{Kuzmin:1985mm}, a lepton number asymmetry can be translated into a baryon asymmetry, which is known as 
leptogenesis \cite{Fukugita:1986hr}. Interestingly, the leptogenesis mechanism can have links to neutrino mass generation, implying that both open questions could be tightly connected. Hence, studies of neutrino properties, in particular $CP$ violation and lepton number violation (LNV), may provide a crucial key to understand the origin of matter in the Universe \cite{Chun:2017spz}.
Key questions and future challenges are (see also section~\ref{sec:earlyuniverse} for related discussions):

\begin{itemize}
\item \textbf{Can we establish a link between leptogenesis and light neutrino properties?}
%
Neither LNV~\cite{Dick:1999je} nor CPV at low energy are necessary requirements for successful leptogenesis, while each of them would be considered as ``circumstantial evidence'' in favour of it. 
The search for CPV in neutrino oscillations is a central goal of the upcoming long-baseline neutrino program, including the DUNE and T2HK projects. Under which conditions can we establish robust connections between low-energy observables and leptogenesis? See \cite{Pascoli:2006ci,Moffat:2018smo,Xing:2020erm,Hernandez:2016kel,Drewes:2016jae} for examples along these lines.

\item\textbf{Interplay between Leptogenesis and LNV.}
If a LNV operator higher than dim-5 induces neutrinoless $\beta\beta$ decay, this would imply a strong washout of a pre-existing asymmetry, putting high-scale baryogenesis models under tension~\cite{Deppisch:2015yqa, Deppisch:2017ecm}. 
An observation of neutrinoless double beta decay in tension with a limit on the sum of neutrino masses from cosmology could hint towards such a higher-dimensional operator contribution~\cite{DellOro:2015kys}. How can we make a robust exclusion of the standard mass mechanism in that case? 

\item \textbf{Can we benefit from synergies with LHC and lepton flavor physics?}
%
Low-scale neutrino mass models could lead to LNV signatures at collider experiments, see e.g.,~\cite{Cai:2017mow}. It has been demonstrated that $\Delta L = 2$ washout processes detectable at the LHC could falsify high-scale leptogenesis scenarios~\cite{Deppisch:2013jxa}. Can we develop a robust, falsifiable theoretical framework for LNV and leptogenesis in the context of general BSM theories? It has been shown that also lepton-flavor violation can play a crucial role in the falsification of leptogenesis~\cite{Deppisch:2015yqa,deGouvea:2019xzm}, in flavored leptogenesis scenarios~\cite{Abada:2006ea,Dev:2017trv}, and as spectator effects~\cite{Garbrecht:2014kda,Garbrecht:2019zaa}. How can we optimally make use of the interplay between leptogenesis and lepton flavor physics?

\item \textbf{Low-scale leptogenesis.}
Leptogenesis can proceed through different mechanisms over a broad range of scales as low as a few MeV \cite{Dev:2017wwc,Drewes:2017zyw,Chun:2017spz} and can be realised in different neutrino mass models~\cite{Hambye:2012fh}. This includes, for example, leptogenesis via sterile neutrino oscillations and/or resonant enhancement \cite{Akhmedov:1998qx,Covi:1996wh,Pilaftsis:2003gt,Klaric:2021cpi}. 
Low-scale leptogenesis allows to search for the new particles directly at accelerators \cite{Deppisch:2015qwa,Cai:2017mow,Agrawal:2021dbo}.
With two sterile neutrinos \cite{Asaka:2005pn} the scenario is highly predictive and combining data from high-energy experiments, neutrinoless $\beta \beta$ decay, and neutrino oscillation experiments can in principle constrain all model parameters and test the hypothesis that these particles are the common origin of neutrino masses and visible matter in the Universe~\cite{Chun:2017spz}. 
With three right-handed neutrinos the parameter space is less constrained~\cite{Abada:2018oly,Chrzaszcz:2019inj}. Will it be possible to test this and other non-minimal scenarios? 
Fostering common particle physics, astrophysical and cosmological sterile neutrino searches can effectively remove redundant regions in the a priori huge parameter space.
\end{itemize}


\subsection{Physics Beyond the Standard Model} 
\label{sec:neutrinoBSM}

Neutrino mass implies that the SM has to be extended in some way. This suggests that neutrinos may offer a window to beyond Standard Model (BSM) physics, which can be explored by searching for non-standard neutrino properties and/or over-constraining the standard neutrino paradigm. Key questions and challenges are:

\begin{figure*}
	\centering
	\includegraphics[width=0.7\textwidth]{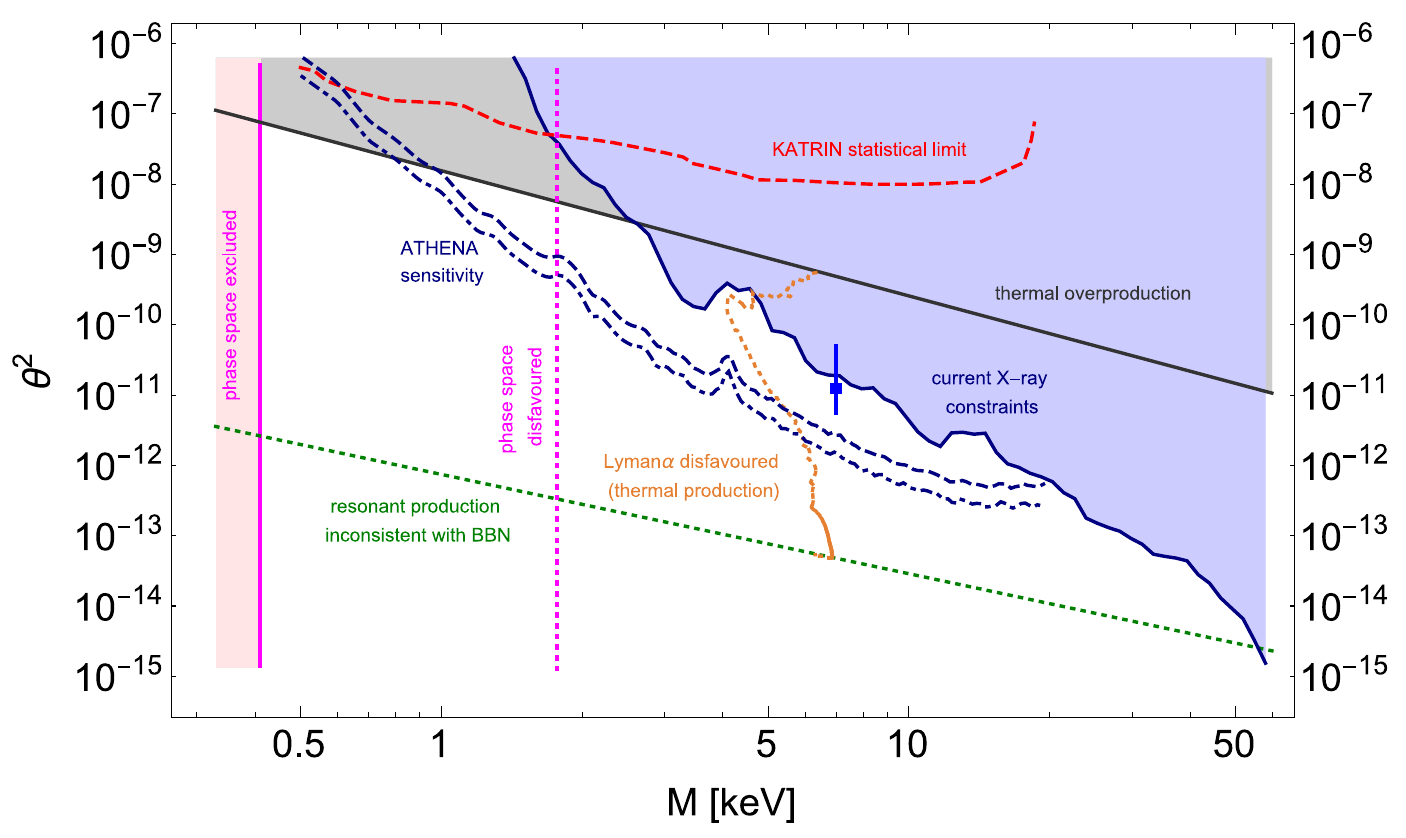}
	\caption{\textit{Constraints on sterile neutrino dark matter in the plane of neutrino mass and mixing. Shaded regions delimited by solid curves are excluded model-independently, whereas dashed or dotted curves require some model assumptions. Figure reproduced from Ref.~\cite{Boyarsky:2018tvu}, where further details and references can be found.}}
	\label{fig:keVnuDM}
\end{figure*}

\begin{itemize}
    \item 
\textbf{Is there a connection between neutrinos and dark matter (DM)?} This could be in the context of minimal models of keV scale sterile neutrino DM \cite{Asaka:2005pn,Asaka:2005an} or in WIMP-like DM models, e.g.,~\cite{Ma:2006km,Lindner:2011it,Escudero:2018fwn,Blennow:2019fhy}. In the latter case, generically such models share many aspects of standard WIMP phenomenology. Identifying a connection between neutrino mass and WIMP DM would be an outstanding achievement. How can such a link be established? 

Much attention has been devoted already to keV sterile neutrino DM \cite{Boyarsky:2018tvu}. This scenario is highly constrained, most notably by X-ray observations, e.g.,~\cite{Roach:2019ctw}, putting traditional production mechanisms 
(Dodelson-Widrow or Shi-Fuller mechanisms) under pressure; see Fig.~\ref{fig:keVnuDM}.
An active field of research is to identify new viable mechanisms to produce keV sterile neutrino DM without conflicting various cosmological and astrophysical bounds; see e.g.,~\cite{Konig:2016dzg,Bezrukov:2019mak,DeRomeri:2020wng,Fernandez-Martinez:2021ypo}. 

\item \textbf{Is there a connection between neutrinos and dark energy?}
The energy scale associated to dark energy at present is roughly $\rho_\Lambda^{1/4}\sim 0.01$~eV, of the same order as neutrino masses, as well as their cosmic density $\rho_\nu^{1/4}$.
This raises the question of where this coincidence comes from, and whether there is a connection between the dark energy and the neutrino sector~\cite{Fardon:2003eh}. While the most simple model for such an interaction is ruled out~\cite{Afshordi:2005ym}, more complicated models remain a viable possibility~\cite{Dvali:2016uhn}. 

\item 
  \textbf{Are current anomalies in the neutrino sector signs of BSM
    physics?}  The long-standing LSND~\cite{Aguilar:2001ty} and
  MiniBooNE \cite{Aguilar-Arevalo:2020nvw} anomalies are difficult to
  explain in terms of sterile neutrino oscillations~\cite{Dentler:2018sju}. It is an important open question whether
  these observations (if experimentally confirmed) are indicating the
  presence of new physics beyond sterile neutrinos; for an incomplete
  list of recent papers see
  \cite{Ballett:2018ynz,Bertuzzo:2018itn,Arguelles:2018mtc,Fischer:2019fbw,deGouvea:2019qre,Dentler:2019dhz,Brdar:2020tle,Chang:2021myh}.  Generically, astroparticle physics provides severe constraints on these type of attempted explanations.

\item \textbf{Over-constraining the 3-flavour paradigm.} Irrespective of anomalies, it will be important to test and over-constrain the standard 3-flavour picture. Besides sterile neutrinos discussed above, another approach in this respect are non-standard neutrino interactions~\cite{Esteban:2018ppq,Farzan:2017xzy,Proceedings:2019qno}. What are theoretically motivated scenarios to search for deviations from the 3-flavour picture?

\item
\textbf{Can we use neutrino experiments for other BSM searches?} Many neutrino experiments operate at the so-called Intensity Frontier and therefore may be used to explore scenarios of new physics in other sectors as well. Powerful constraints on light DM models have been set using neutrino beam experiments~\cite{deNiverville:2011it,Aguilar-Arevalo:2018wea} and neutrino detectors exploring synergies with cosmic ray physics~\cite{Bringmann:2018cvk,Ema:2018bih}. New bounds on millicharged particles have been set using ArgoNeut and Super-Kamiokande~\cite{Harnik:2019zee,Plestid:2020kdm}. In the future it may be possible to extend these searches to probe additional scenarios, including axion-like particles~\cite{Dent:2019ueq, Kelly:2020dda, Brdar:2020dpr}, long-lived particles~\cite{Arguelles:2019ziu}, inelastic and/or boosted dark matter detection~\cite{Agashe:2014yua,Jordan:2018gcd}, dark scalars and new gauge bosons weakly coupled to the visible sector~\cite{Essig:2010gu, Batell:2019nwo}, among others. 

Are there significant advantages for using off-axis near detectors~\cite{Coloma:2015pih,Breitbach:2021gvv}, or for choosing a given detector technology~\cite{Berger:2019ttc}? 
In what regions of the parameter space, or for which new physics models, are neutrino experiments competitive with other fixed target experiments or collider searches?

\item \textbf{Can we use other BSM searches to learn about neutrinos?}
  We give here three examples to answer this question affirmatively:
  DM experiments, $B$ physics, the $g_\mu-2$ measurement. Dark matter experiments may be used to constrain, e.g., new neutrino interactions which could affect the so-called neutrino floor~\cite{Cerdeno:2016sfi,Gonzalez-Garcia:2018dep}. Are there any viable UV-complete models that can lead to a sizable enhancement of the neutrino floor in dark matter experiments?
  Unexpected signals such as the
  recent Xenon excess \cite{Aprile:2020tmw} could find explanations
  involving neutrinos,
  e.g.,~\cite{Babu:2020ivd,Shoemaker:2020kji,Brdar:2020quo}. In
  general, when anomalies in other particle physics experiments appear, can
  connections to neutrino physics be established? Two recent examples
  are anomalies in B-physics \cite{Albrecht:2021tul}, which may find some explanations
  involving sterile neutrinos,
  e.g.,~\cite{Robinson:2018gza,DelleRose:2019ukt,Balaji:2019kwe,Mandal:2020htr,Heeck:2019nmh},
  or similarly the $g_\mu-2$ puzzle \cite{Abi:2021gix}, where possible
  connections to neutrinos have been discussed in
  \cite{Abdullahi:2020nyr,Abdallah:2020biq,Babu:2021jnu}. In many of these cases neutrino astroparticle physics may provide crucial complementary information to test or constrain such explanations.

\item\textbf{Can we use neutrinos to test fundamental symmetries of Nature, e.g., CPT or Lorentz symmetry?}  The CPT theorem is one of the most fundamental predictions of local relativistic quantum field theories. If found, CPT violation will threaten the very foundations of our understanding of Nature.
Current bounds on CPT violation either involve non elementary particles or charged ones. The neutrino sector is different. For example, in seesaw models the light masses are naturally related with high energy scales making neutrinos distinctively sensitive to new physics/new scales. This exclusive mass generation mechanism along with the fact that there is no charge contamination comprised in the test makes neutrinos specially appealing to study CPT violation \cite{Diaz:2009qk,Barenboim:2017ewj,Barenboim:2017vlc}. CPT violation would also have important implications for the generation of the baryon asymmetry in the Universe \cite{Bertolami:1996cq}.  High-energy astrophysical neutrinos provide a unique tool to test CPT/Lorentz invariance; see below.

\item
\textbf{Using high-energy astrophysical neutrinos to test neutrino properties.}
The cosmic neutrinos discovered by IceCube~\cite{Abbasi:2020jmh} have the highest detected energies---up to a few PeV---and travel the longest distances---up to a few Gpc, as illustrated in Fig.~\ref{fig:scales}. Thus, they can probe neutrino properties possibly tiny in size and at energy scales otherwise unreachable~\cite{Ahlers:2018mkf, Ackermann:2019cxh}.  
Tests  performed today include measuring the TeV--PeV neutrino-nucleon cross  section~\cite{Aartsen:2017kpd, Bustamante:2017xuy}, constraining neutrino self-interactions~\cite{Kelly:2018tyg, Bustamante:2020mep}, non-unitarity~\cite{Brdar:2016thq}, decay~\cite{Bustamante:2016ciw}, and CPT/Lorentz invariance~\cite{Arguelles:2015dca, Bustamante:2015waa}.  In the next 10--20 years, new detectors---IceCube-Gen2, KM3NeT, P-ONE, Baikal-GVD---will turn these into high-statistics tests, while synergy with oscillation experiments will shrink key uncertainties~\cite{Song:2020nfh}.  Discovering ultra-high-energy neutrinos~\cite{Beresinsky:1969qj}, with $\gtrsim$~100~PeV energies, would take us further, thanks to proposed large-scale detectors, e.g., IceCube-Gen2, RNO-G, GRAND, POEMMA, PUEO.

\end{itemize}

\section{Particles from Stars}
\label{sec:particlesfromstars}
\noindent {\bf Coordinators:} Aldo Serenelli and Irene Tamborra.\\
{\bf Contributors:} Andreas Bauswein, Juan Calderon Bustillo, Jordi Isern, Michela Mapelli, Marcelo M.~Miller Bertolami, Evan O'Connor, Giulia Pagliaroli, Georg G.~Raffelt, Armen Sedrakian, Shashank Shalgar, Laura Tolos, Francesco L. Villante, Aaron Vincent.

\subsection{Introduction}

The development of observational astronomy has been phenomenal over the last decades. Large scale photometric and spectroscopic surveys have become common place, allowing for  statistical studies of stellar properties at a very large scale. The generalization of observational techniques, in particular asteroseismology, previously restricted to the Sun and a few other objects is paving the way for an unprecedented understanding of the properties of stellar interiors. Even more impressively, the era of multi-messenger astronomy has entered its maturity with the detection of neutrinos of astrophysical origin across energy scales and several tens of gravitational waves events. In particular, the gravitational wave  events represent the coalescence of compact stellar mass objects, such as  stellar mass black holes (BH) or neutron stars (NS), opening a completely new window into the investigation of these objects and fundamental physical theories, such as  gravity. While so far only one of such events has had its optical counterpart detected, this number will increase dramatically in the next decade due to large scale time domain surveys dedicated to  transient events, such as supernovae (SN) explosions, and rapid response networks set up in place for follow-up observations. 

\begin{figure}[h!]
\center
\includegraphics[width=0.82\textwidth]{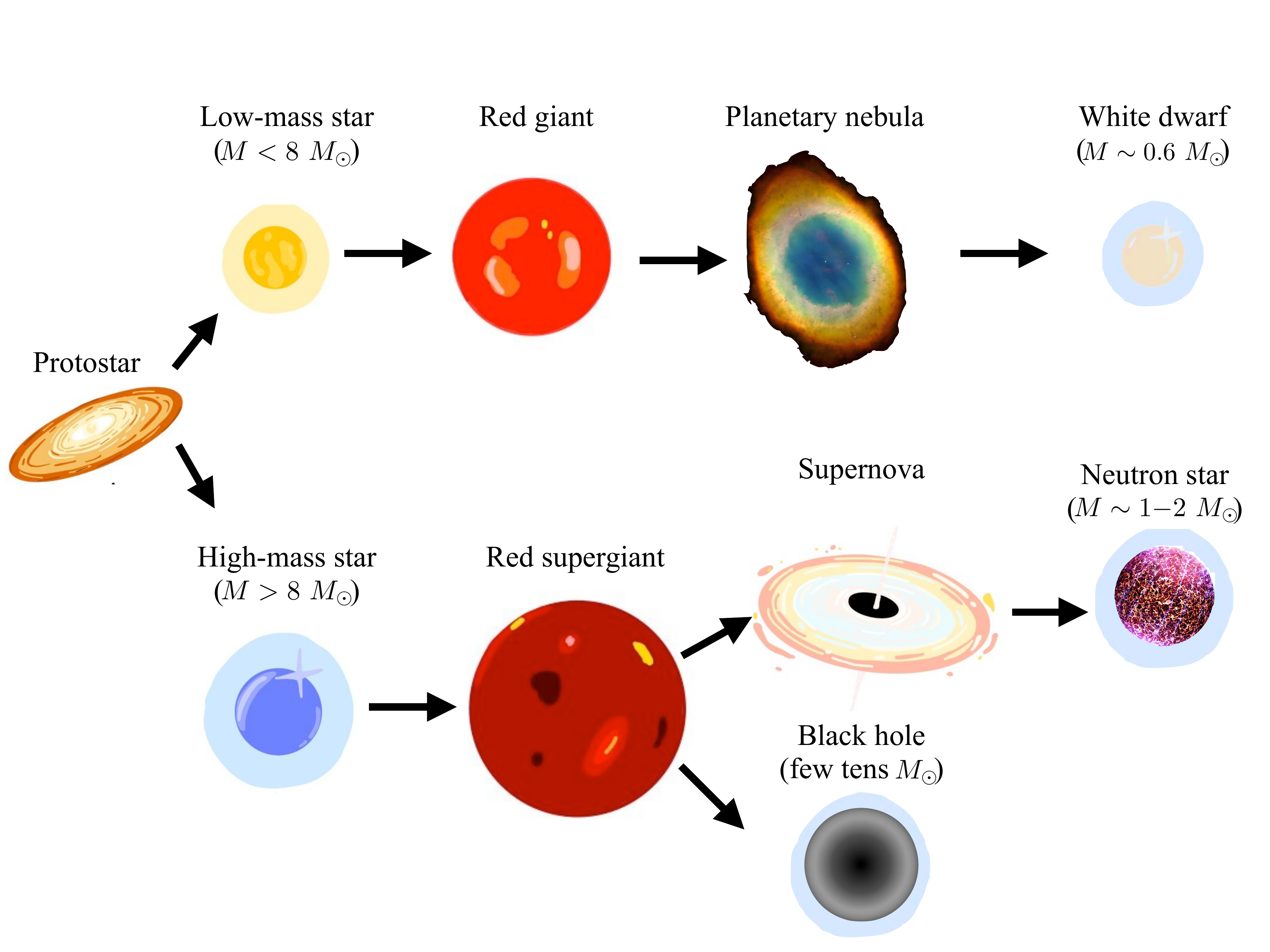}
\caption{Schematic representation of the stellar evolution path for low- and high-mass stars. Typical values for the stellar masses are indicated in parenthesis. Photons, neutrinos, gravitational waves, and non-standard particles are emitted for each branch of the stellar evolution, according to the stellar evolution stage.}
\label{fig:stellar_evolution}
\end{figure}

The sheer size of stars and the  extreme conditions in stellar interiors make them excellent laboratories for particle physics, complementary in many cases to dedicated Earth-based experiments.  The physics scenarios that can be probed depend on the stellar evolution stage (see Fig.~\ref{fig:stellar_evolution}); the sensitivity of the cooling of degenerate cores of red giant stars (RGs) or white dwarfs (WDs) to physics beyond the Standard Model of Particle Physics  is a prime example.

To maximize the scientific exploitation of the wealth of data, present and future, new challenges arise for stellar evolution theory, fundamental physics and also in the statistical methodology, in particular on how to combine heterogeneous sources of data and messengers. At the same time, major progress is required in the theoretical modeling of the source physics. In what follows, we identify  the most pressing  challenges across different classes of stellar mass objects.

\subsection{Sun}
The closest low-mass star to us is the Sun. The Sun is a copious source of neutrinos, despite neutrinos constituting only $2\%$ of the Sun's radiation.
Solar neutrino physics is entering the precision era. The $^8$B and $^7$Be, pp, pep and CNO neutrino fluxes have been directly measured, the first two with better accuracy than theoretical uncertainties.
Planned detectors (e.g. SNO+, JUNO, Jinping, Hyper-Kamiokande, THEIA, DUNE and also dark matter experiments, see \cite{orebi:2021} for a recent review) will further improve the experimental landscape. Precise measurements of the pp flux to 1\% and CNO fluxes to 10\% represent a major experimental challenge, but they would allow to determine stringent constraints on standard and non-standard energy generation \cite{Vinyoles:2015aba} and transfer mechanisms in the Solar interior,  the Solar core composition \cite{Serenelli:2012zw,Agostini:2020lci}, as well as to shed light on the long-standing Solar abundance problem \cite{Vinyoles:2016djt}. It will be possible to test the theory of neutrino  conversion  in the standard 3$\nu$-paradigm and beyond (e.g., new neutrino states, non-standard neutrino interactions). These searches will provide stringent, and in some cases unique, bounds on new physics \cite{Maltoni:2015kca}.

The Sun is also a target for astronomy with GeV-TeV messengers \cite{Nisa:2019mpb}. Hadronic interactions of cosmic rays close to the Solar photosphere produce a bright $\gamma$-ray disk emission whose observed properties are not fully explained by existing models \cite{Ng:2015gya,Linden:2020lvz}. Inverse Compton scattering of cosmic ray electrons (and positrons) with Solar photons produce a $\gamma$-ray halo around the Sun \cite{Orlando:2020ezh}. Determination of these components is important to improve our understanding of the Sun and also for the search of astrophysical neutrinos and dark matter signals.

Direct dark matter searches rely on an elastic dark matter-nucleus interaction to produce a measurable nuclear recoil. In the Sun, once captured  by elastic scattering, dark matter can self-annihilate, producing neutrinos detectable at telescopes such as IceCube  and Super-Kamiokande, or if long-lived particles from dark matter annihilation can escape the Sun before decay, lead to morphologically and energetically distinct gamma ray and neutrino fluxes \cite{Bell:2011sn}. If dark matter is asymmetric, it may accumulate in the star and transport heat in the Solar interior, leading to a Solar structure different than predicted by standard Solar models. Solar models including dark matter heat transport have been used to constrain various asymmetric dark matter  models using current helioseismic and Solar neutrino measurements \cite{Taoso:2010tg,Vincent:2014jia}. However, significant work remains to be done, as the approaches used to describe heat transport are not strictly self-consistent and do not agree with detailed Monte Carlo simulations. This will be especially important in the face of future  Solar neutrino experiments.

Axions or axion-like particles can be produced via Primakov conversion of thermal photons in the plasma. If they are sufficiently long-lived, they may escape the Sun to be reconverted in Helioscopes such as CAST and (Baby)IAXO, leading to the strongest bounds between $m_a = 10^{-4}$ and $1$ eV. Shorter lived particles can provide another means of heat conduction. New vector bosons, sometimes called dark photons, or $Z^\prime$ particles, can be produced in a similar way, as well as via plasmon decay ($\gamma\to\bar\nu\nu$). If these dark photons or their decay products are produced non-relativistically, they also conduct heat, or escape and accumulate in the Sun's gravitational well leading to an enhanced signal at Earth-based detectors \cite{Lasenby:2020goo}. 
Neutrino interactions with ultralight bosonic dark matter can yield signals in the Solar neutrino flux, as could sterile neutrino species \cite{Berlin:2016woy,Lopes:2021wzu}.
Other possible new physics signals from the Sun include mass-dependent Chameleon particle production \cite{Cuendis:2019mgz}. Detailed predictions of the expected fluxes and signals in detectors of such particles require development of non-standard Solar models, including careful studies of theoretical uncertainties (see e.g. \cite{Hoof:2021mld} for a test case).
Finally, obtaining a detailed understanding of Solar physics is crucial in related searches for dark matter and new physics: Solar neutrinos will soon be an irreducible background in dark matter direct detection searches, and Solar atmospheric neutrinos will complicate the search for dark matter annihilation in the Solar core \cite{Arguelles:2017eao,Ng:2017aur}.

\subsection{Evolved stars}

After hydrogen exhaustion, low-mass stars ($M\lesssim 2\,M_\odot$) develop a degenerate helium core that grows more massive as the star ascends the RG branch. Eventually helium ignites, leading to a sudden termination of the RG  branch. The associated discontinuity in the luminosity function of stellar populations, known as the tip of the RG branch (TRGB) is one traditional method of determining relative distances of galaxies (see \cite{Freedman:2020dne, Soltis:2020gpl} for recent studies and references to earlier work).

 The mass of the core at helium ignition depends on the rate at which
 its mass is increased by hydrogen burning, the temperature of the
 surrounding shell, and the rate at which the core is cooled by
 neutrinos. In turn, the luminosity of the RGB star is very sensitive
 to the mass of the helium core. Hence, the luminosity of the
 TRGB  depends  on the cooling efficiency by
 standard neutrino emission through plasmon decay
 $\gamma\to\bar\nu\nu$, a non-standard contribution caused by a
 possible anomalous neutrino dipole moment $\mu_\nu$, or by completely
 new particles, e.g.\ DFSZ-type axions that would be emitted by
 bremsstrahlung through their coupling $g_{ae}$ to electrons. Compared
 to WDs, the density is similar but $T\sim 10^8~{\rm K}\sim 10~{\rm
   keV}$ is larger. The sensitivity on $g_{ae}$ (from bremsstrahlung)
 is similar, whereas the bounds on $\mu_\nu$ are more restrictive \cite{Capozzi:2020cbu, Straniero:2020iyi}. Another unique case
 concerns dark photons where resonance effects enhance the sensitivity
 in the few keV-mass range \cite{An:2013yfc}. The TRGB can be used to
 constrain any new low-mass particle that interacts with photons
 and/or electrons.
The upcoming launching and commissioning of the James Webb Space
Telescope (JWST) will open the possibility to measure the TRGB
luminosity in a much larger sample of stellar populations and holds
the potential to reduce the observational uncertainties in cluster and populations for which observations are  available.

Theoretical predictions of the TRGB luminosity depend on the accuracy of the adopted microphysics in stellar
evolution models. Current uncertainties were expected to be around 0.12~dex in the absolute infrared magnitude of the TRGB \cite{Serenelli:2017}. However, recent determinations of conductive opacities in the slightly more degenerate regime of WDs
envelopes find differences of up to a factor~3, caused by an improved treatment of electron-electron scattering
\cite{2020ApJ...899...46B}. A similar difference in conductive opacities in RG cores would be enough to shift the TRGB magnitude beyond the  previous uncertainties \cite{Serenelli:2017}.  Improved computations of conductive opacities may be needed to make full use of the upcoming JWST observations.
Another place for theoretical improvements of particle bounds are the standard and non-standard cooling rates due to weakly interacting particles. In fact, existing bounds  are based on either old \cite{Nakagawa:1988rhp} or simple \cite{Haft:1993jt,Raffelt:1994ry} estimations of the emission rates.

The evolution of non-degenerate evolved stars can also be affected in observable ways by non-standard energy losses, e.g. by axions coupling to photons. Such possibilities are given, for example, by the sensitivity of the so-called blue loops \cite{Friedland:2012hj,Mori:2020niw} in the evolution of intermediate mass stars or by the lifetime of the horizontal branch phase of low-mass stars \cite{Ayala:2014pea}. Stellar models simulating these evolutionary phases, during which helium is burning under non-degenerate conditions in the stellar cores, are sensitive to uncertain physical phenomena such as the dynamics of convection in the helium cores, the cross section of the critical $^{12}$C$(\alpha,\gamma)^{16}$O nuclear reaction and electron screening. Sustained progress on all these fronts is required over the next decade in the challenge of developing more physically accurate stellar models which could then be used to place robust limits on the properties of astroparticles and, viceversa, to study the impact of astroparticles on stars. This can be achieved by development of hydrodynamic simulations, observational understanding of the interior structure of such stars with asteroseismic techniques, and experimental work. 

\subsection{White dwarfs}

White dwarfs  are excellent laboratories for testing new physics. At
first order their evolution is just a simple gravothermal cooling
process, and the basic physical ingredients necessary to predict their
evolution are well identified
\cite{Althaus:2010pi}. Complementary, there is an
impressively large and solid observational background to check for the
impact of different physical theories.
Searches for hints of new physics in WDs are usually based on
the energy-loss argument \cite{Raffelt:1996wa}. Any additional
energy source or sink provided by new physics affects the cooling rate
of  WDs, leading to observable consequences.  

The cooling rate of WDs can be measured for large WD populations using the luminosity function, i.e.~the number of WDs per unit bolometric magnitude and unit volume as a function of bolometric magnitude. For individual stars, it can be measured using the period drift of non-radial oscillations observed in WDs in specific evolutionary stages \cite{Corsico:2019nmr}. These techniques have provided bounds on the mass of axions,  the neutrino magnetic moment,  the secular drift of the Newton gravitational constant, the density of magnetic monopoles and Weakly Interacting Massive Particles (WIMPs), as well as constraints on properties of extra dimensions, dark forces, modified gravity, and formation of BHs by high energy collisions (see 
e.g.~\cite{Isern:2008nt,Corsico:2013ida,Dreiner:2013tja,Saltas:2018mxc}). White dwarfs can also capture dark matter particles that can heat up the star, affecting its cooling properties \cite{McCullough:2010ai} and possibly  ignite Type Ia SNe \cite{Graham:2018efk,Acevedo:2019gre}.

Historically, both methods have faced observational limitations. As for the luminosity function, challenges lie in building volume (as opposed to brightness) limited samples. Gaia \cite{gaia:edr3} has recently increased the number of observed WDs by
nearly two orders of magnitude and it is also providing accurate
distances and photometry \cite{gentile:2019}. It is now possible, for the first time, to construct luminosity functions of volume limited samples and also  reasonably accurate luminosity
functions of the different galactic populations (thin and thick disks,
halo, bulge). A qualitative improvement will soon be brought about by the Vera C.~Rubin Observatory with multi-band photometric observations. The goal will be to build luminosity functions for WDs of the same mass. Some of these drawbacks can be circumvented by studying the
WD luminosity function in places where the parent population
is well constrained as in Globular Clusters or in populations that have evolved in a
nearly independent way \citep{Isern:2018uce}.  To fully exploit current and future data-sets, a proper handling of statistical and theoretical uncertainties is required. The application of Monte Carlo techniques to build the theoretical WD luminosity functions has proven   to be an extremely powerful tool in this regard \citep{2010Natur.465..194G}.

The main difficulty for using the period drift ($\dot{P}$) of pulsating WDs is its smallness, as $\dot{P} \sim 10^{-13}$ to $10^{-15}$ s/s depending on the class of pulsating WDs. Its determination thus requires years-long monitoring campaigns. However, the number of known pulsating WDs and of precise measurements is experiencing a substantial increase with the space missions Kepler and TESS. This will continue in the future with PLATO. Based on the properties of individual stars, this method is much less dependent on the properties of the parent population. On the other hand, the value of the period drift of a given pulsation mode not only  depends on the cooling rate, but also on the detailed internal structure of the WD \cite{Corsico:2016okh}. As the structure of a WD is the fossil record of the evolution of its progenitor star, uncertainties in the modeling of all the  previous evolutionary phases propagate into the structure of WD models. As in the case of evolved stars, development in stellar physics and models is required to exploit the wealth of data currently available and that continues to grow at an ever faster pace. 

Another avenue with lots of potential is the gravitational wave emission from WD mergers, a prime target of LISA during the second half of the next decade, as binary WDs would produce gravitational waves in the range between 0.1 and 0.01Hz \cite{Lamberts:2019nyk}.  White dwarfs are intrinsically faint and only the local population is known. LISA would allow to map the Galactic population of WDs in very close binary systems, a fundamental piece in our understanding of Type Ia SN progenitors. Moreover, the merging process could give rise to a rich nucleosynthesis site \cite{Thielemann:2018bwk}.

In common to the limitations expressed in the previous section, an additional problem in the use of WDs as laboratories for fundamental physics comes from the lack of accurate calculations of the interactions of new particles with the stellar material (or new interactions of standard particles).  As such most of these studies are based either order-of-magnitude estimations  or on decades-old computations \cite{Nakagawa:1988rhp} (see however \cite{Carenza:2021osu}).

\subsection{Supernovae}
Core-collapse SNe  originate from stars with mass larger than $8\ M_\odot$ and are explosive events that harness the gravitational energy released during the formation of a compact object. 
The total energy released is $\sim 1/10 M_\odot c^2 \sim 2\times 10^{53}$\,erg. 
The bulk of this energy   is ultimately partitioned into  neutrinos ($>99\%$),  gravitational waves ($\sim 10^{46}$ erg), photons ($\sim 10^{48}$--$10^{49}$ erg), and kinetic energy ($\sim 10^{51}$ erg) of the expelled material.
The observation of neutrinos and photons from SN\,1987A confirms the basic picture of the SN mechanism, but the ultimate role of the SN microphysics is the frontier of the field. While recent, sophisticated, three-dimensional simulations are successfully achieving explosions, see e.g.~\cite{Burrows:2019zce, Bollig:2020phc}, theoretical and computational advances in the treatment of progenitors, magnetic fields, and neutrinos are needed to advance our understanding~\cite{Janka:2016fox,Burrows:2020qrp}.

The physics linked to neutrinos is one of the most complicated aspects of SNe \cite{2020LRCA....6....4M,Mirizzi:2015eza,Tamborra:2020cul}. Neutrinos could be unique probes of the explosion mechanism, however uncertainties in the flavor-dependent neutrino interaction rates 
and in the modeling of flavor conversion in the SN core are the main obstacles. Our lack of understanding of neutrino flavor evolution  is  due to its non-linear nature \cite{Mirizzi:2015eza,Tamborra:2020cul}. This is an area of vigorous research at the moment and will continue to be in the future.

Existing neutrino telescopes, such as Super-Kamiokande and IceCube, as well upcoming ones like Hyper-Kamiokande, JUNO and DUNE, will guarantee a high statistics observation of neutrinos from a galactic SN  (or SNe in close-by galaxies) in the near future. A complementary option with respect to traditional technologies is the employment of coherent neutrino-nucleus scattering for the detection of SN neutrinos~\cite{Drukier:1983gj}; the full potential of this  flavor-insensitive technology in this context  remains to be unleashed. These opportunities will be crucial to  gain  insight into the behavior of SN neutrinos as well as on the SN mechanism~\cite{Horiuchi:2017sku,2019ARNPS..69..253M}.

Although challenging, the detection of gravitational waves from  SNe would be game-changing.  Gravitational waves can directly probe the SN dynamics, e.g.~\cite{Kotake:2011yv}.
The detection horizon  is presently limited to within our Galaxy \cite{Abbott:2019prv}, but  statistical confidence can be increased through multi-messenger approaches and more work is needed.
Neutrinos and gravitational waves herald  the impending electromagnetic signal \cite{Kharusi:2020ovw}; for nearby SNe ($<$1\,kpc) neutrinos from the last stages of stellar evolution can also be adopted. These early warnings can  provide astronomers with sky localization in order to capture the shock breakout, which is critical to constrain properties of the progenitor star. Fully self-consistent, whole-star simulations predicting the neutrino and gravitational waves signals, and also following the shock throughout the progenitor star to predict the optical light curve and spectra will enable a stronger relationship between observations and theory~\cite{Nakamura:2016kkl}.  To facilitate this, the community would benefit from  cooperation and data sharing between theorists and observers.

The upcoming detection of the diffuse SN neutrino background, together with the increasing number of SN detections through the advent of time-domain astronomy, will provide unprecedented means to investigate the SN population and open a new frontier for low-energy neutrino astronomy \cite{Mirizzi:2015eza,Nakamura:2016kkl}. From the theoretical perspective, a better understanding of the impact of binary interactions, failed core collapses, and metallicity evolution of galaxies is necessary to interpret the  diffuse SN neutrino background signal, see e.g.~\cite{Nakazato:2015rya,Kresse:2020nto}.

Physics beyond the Standard Model  could be responsible for energy loss as well as affect the interaction rate of neutrinos in the SN core and at Earth-bound detectors~\cite{Raffelt:1996wa}. However, most of the existing bounds rely on a simplified modeling of the SN physics and often neglect the impact of new physics on the neutrino flavor evolution history and on the SN dynamics; these simplifications can have a huge impact on existing bounds, see e.g.~\cite{Suliga:2020vpz}. Future theoretical effort should  be aimed to a self-consistent modeling of non-standard scenarios. Signatures in the cosmic radiation from axion or axion-like particles (ALPs)  produced in SNe are discussed in Sec.~\ref{sec:DM}.

\subsection{Neutron stars}
A massive star undergoing a SN explosion may leave its collapsed core under the form of a NS, an extremely compact object in which over a Solar mass of matter is confined within a radius of $\sim 10$ km. Neutron stars are  unique laboratories of cold dense matter  \cite{Watts:2016uzu}. The determination of the Equation of State (EoS) of the dense interior is a challenge, given that it spans a wide range of densities and isospin asymmetries \cite{Lattimer:2006xb}.  Moreover, a simultaneous description of the EoS and the transport properties of the interior is crucial for a consistent description of the static and dynamic properties of NSs. The simultaneous mass-radius determinations from NICER \cite{Miller:2021qha}, together with the future observations from the eXTP mission \cite{Watts:2018iom} will help to constrain the EoS. Moreover, gravitational waves from binary NS  mergers \cite{TheLIGOScientific:2017qsa} are a new venue for extracting information, not only on the EoS but also on the transport properties. 

The first observation of a binary NS  merger with gravitational waves and electromagnetic radiation has provided invaluable information about stellar properties of NSs  and thus of high-density matter, as well as element formation~\cite{Burns:2019byj}. The electromagnetic counterpart confirms that binary NS  mergers  are important sources of the elements heavier than iron. An early bluish component in the kilonova  light curve implies that weak interactions and neutrinos play a decisive role in shaping the composition of the outflow and thus the nucleosynthesis. This clear imprint of neutrinos immediately raises the question about flavor conversion in binary NS  mergers  (see e.g.~\cite{Wu:2017drk} for preliminary explorative work). Efforts should be devoted to the modeling of binary NS  mergers  with a high level of sophistication; our understanding of the production and interaction of these particles in hot NS matter should also be improved.

Neutron stars are also  interesting sites for testing the nature and properties of dark matter. Whereas the  collapse of a NS due to accreted dark matter can set bounds on the dark particle mass \cite{Goldman:1989nd}, additional constraints can be obtained from stars that accrete dark matter and then collapse into a compact object \cite{Kouvaris:2010jy}. Also, the cooling process and the EoS can be affected by the capture of dark matter \cite{Bertone:2007ae,Cermeno:2017xwb}, and  self-annihilating dark matter accreted onto NSs may change their kinematical properties  \cite{PerezGarcia:2011hh}. Moreover, NSs that contain non-self annihilating dark matter have emerged as an interesting astrophysical scenario to test dark matter \cite{Tolos:2015qra}. Neutron star capture rate computations are becoming more and more refined, but the observational prospects remain largely unexplored, see e.g.~\cite{Bell:2019pyc}. 
 Neutron stars  cool initially by emission of neutrinos, but also particles beyond the Standard Model, e.g.~axions or ALPs~\cite{Sedrakian:2015krq}. A comparison of the theoretical surface temperatures of NSs  with observations puts constraints on the ALP masses and couplings. Axions escaping a NS can generate X-rays excess of hard keV emission observed  from the nearby NSs \cite{Buschmann:2019pfp}. In simple axion models, this possibility is however in tension with the non-observation of X-rays from magnetic WD \cite{Dessert:2021bkv}.  White dwarfs also provide stellar constraints on the ALPs, which in combination with NS limits exclude ALPs with masses above eV~\cite{Sedrakian:2018ydt}. The main goals and challenges ahead are the inclusion of ALPs, sterile neutrinos and alike particles in the studies of binary NS mergers (see e.g.~\cite{Harris:2020qim}), their remnants, post-merger element production,  and proto-NS cooling.

\subsection{Black holes}
The first direct detection of gravitational waves \cite{Abbott:2016blz} opened new perspectives for the study of stellar mass BHs, their mass spectrum and spin distribution. If these BHs have an astrophysical origin, the main challenges concerning their formation come from uncertainties on massive star evolution, SNe and the process of (pulsational) pair instability \cite{Mapelli:2021taw}. The latter is expected to open a gap in the mass function of BHs \cite{Woosley:2016hmi},  
currently challenged by the detection of GW 190521 \cite{Abbott:2020mjq}. On a more general ground, the observation of binary BH mergers has revived the discussion on their formation channels \cite{Dominik:2012kk,PortegiesZwart:1999nm,Miller:2001ez}.
We know $\sim{50}$ candidate gravitational wave events associated with BH mergers, but  a possible electromagnetic counterpart was only claimed  for one of them \cite{Graham:2020gwr}. On top of this, the redshift evolution of  BHs and intermediate-mass BHs remains mostly unexplored: only next generation-ground based gravitational wave detectors  will probe mergers of binary BH mergers at $z\gg{}1$ \cite{2010CQGra..27s4002P,Reitze:2019iox}.

The detection of binary BH mergers  also allows  to study gravity in its strongest regime, enabling new tests of General Relativity \cite{LIGOScientific:2019fpa}. This regime is also optimal to  reveal the existence of  exotic objects mimicking  the BH phenomenology  \cite{Cardoso:2019rvt}, such as boson stars \cite{Liebling:2012fv} or gravastars \cite{Mazur:2001fv}. These objects also open the gate to tests  of new physics, such as exotic fields coupled to gravity, ultralight bosons and dark matter. 
Black holes can  provide hints of  new physics even in the absence of a merger.  Ultralight bosons can form clouds around BHs and undergo a superradiant instability \cite{Brito:2014wla}. For  scalar bosons, the end point is a continuous emission of gravitational waves and extraction of angular momentum from the system. 
Scalar fields can form bosonic stars whose mergers can mimic  BH events \cite{CalderonBustillo:2020srq}. 
The gravitational wave ringdown can also be used to test a vast range of alternative theories of gravity \cite{Carullo:2021dui} and even measure a putative BH electric charge. While current studies are  limited by the sensitivity of our detectors,  LISA will allow for precision studies.

The observation (or absence of) neutrino counterparts of binary BHs can  place bounds on the abundance of dark matter \cite{deVries:2016ljw}. Conversely, the observation of neutrino emission in NS-BH  mergers would allow to constrain their masses. The existence of  stellar mass primordial BHs would have dramatic implications on the existence of WIMPs.
There is, however, no  conclusive evidence for the existence of primordial BHs \cite{DeLuca:2020sae}. Potential future observations of compact objects in either of the lower and higher BH mass gaps shall make it necessary to probe the existence and properties of such objects.

\subsection{Critical paths forward for the next decade}

We identify three main areas where major theoretical progress is needed for the next decade. 

\begin{itemize}

\item Stellar evolution models throughout evolutionary phases and stellar masses. Improvement across all scales, from macroscopic to microscopic processes, are needed. 

\begin{itemize}
\item Dynamical effects  including convection need to be improved through the development of (radiation/magneto)hydrodynamic simulations. The detailed structure of WDs, pre-SN models depend critically on this. 

\item Further development of nuclear physics entering the equation of state  and transport processes in their interiors (particularly for NSs and SNe). 

\item More refined modeling of energy transport due to non-standard mechanisms (e.g., asymmetric dark matter particles) in solar-mass objects, SNe and compact binary merger remnants. 

\item Energy loss rates for general conditions, away from the currently commonly used limits of ideal gas for solar-like stars or fully degenerate gas (WDs and RGs).

\end{itemize}

\item Methods to capitalize on the upcoming wealth of multi-messenger data should be developed. 
\begin{itemize}
\item Strategies to constrain  particle physics scenarios and our understanding of stellar physics  in terms of large statistical samples. 
\item Techniques to  combine the information coming from heterogeneous data-sets  and different messengers are needed (especially for what concerns neutrino and gravitational wave data).
\end{itemize}
\item Computational challenges concerning the  modeling of stellar mass objects, including (neutrino/radiation/magneto) hydrodynamic  as well as modeling of exotic physics scenarios in astrophysical sources need to be addressed in order to correctly interpret the data.  

\end{itemize}

\section{Dark Matter}\label{sec:DM}

\begin{figure}[b!]
\centering
\includegraphics[width=1.\textwidth]{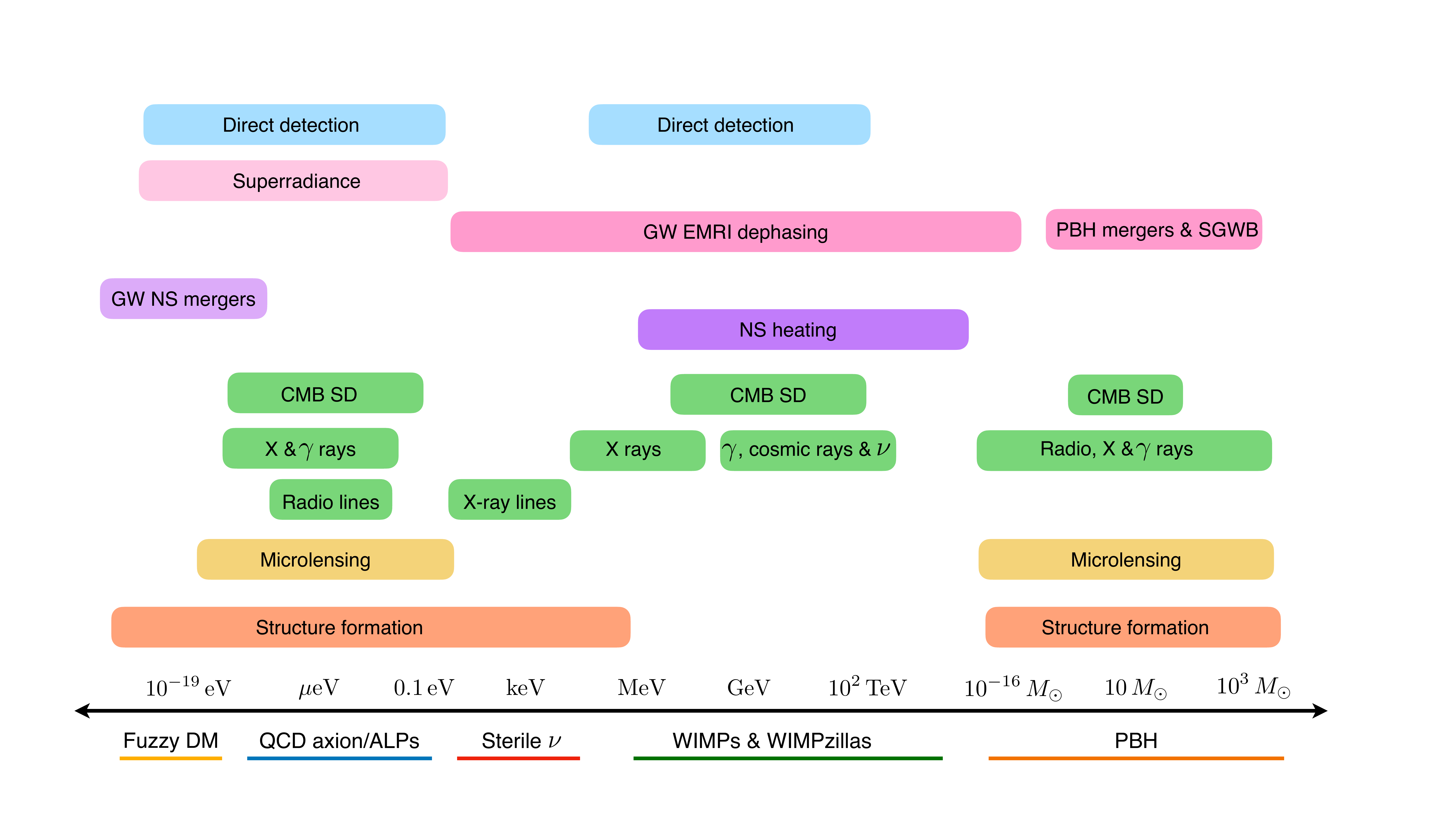}
\caption{Summary of possible constraints on DM. 
We show the available DM mass range with some DM candidates highlighted, and astroparticle observables of different nature that can constrain them. Acronyms: Extreme mass ratio inspirals (EMRI), stochastic GW background (SGWB), CMB spectral distorsions (SD).
}
\label{fig:dm_summary}
\end{figure}

\noindent{\bf Coordinators: } Christian Byrnes, Francesca Calore and David J. E. Marsh \\
{\bf Contributors: } Torsten Bringmann, Riccardo Catena, David Cerdeno, Marco Chianese, Katy Clough, 
Anastasia Fialkov, Nicolao Fornengo, Deanna C. Hooper, Fabio Iocco, Bradley J.~Kavanagh, Alvise Raccanelli, Nashwan Sabti, Laura Sagunski, Pasquale D. Serpico, Sebastian Trojanowski, Edoardo Vitagliano, Arne Wickenbrock.

\subsection{Fundamental questions}
For nearly a century, evidence has been mounting which strongly suggests that the majority of pressureless matter in the Universe is not baryonic in nature. 
Cosmological surveys determine that this ``dark'' matter (DM) component  makes up about one quarter of the Universe's total energy budget today. Whilst the gravitational impact of DM is well measured, its ``microscopic'' nature remains a mystery, despite extensive searches. This Section focuses on theoretical goals for the next decade, amongst researchers working on astroparticle probes of DM.

Of course the main question is clear: ``What is DM?'' and once we find out, we would also want to know its phenomenology, key-parameters values, and the associated structure
formation history (see Sec~\ref{\EU}). The next decade also requires continued development of theoretical tools (including numerical and statistical) to keep up with the precision of data, see Sec.~\ref{\AS} for more details. Before we classify the discussion below by the observable impact of DM, we lead with some broad questions of interest for theorists to ponder:

\begin{enumerate}
\item DM theory parameter space is vast, see Fig.~\ref{fig:dm_summary}. How do we break it up?
\item Is there astrophysical evidence to go beyond the cold and collisionless hypothesis? 
\item If DM is multi-component, how would we know?
\item How is DM produced in the Early Universe, and how does this connect to late Universe observables?
\end{enumerate}

\subsection{Astroparticle observables for dark matter}
In order to make a community roadmap for the next ten years and beyond, we find it useful to think of DM models separated not on the traditional lines of ``weakly interacting massive particle (WIMP)-like'', ``axion-like'', ``primordial black hole (PBH)-like'', etc, but instead according to how DM interacts with its environment (other than gravitationally). 
This approach provides a more direct connection with the phenomenological implications of different classes of DM.
Theorists should have both short and long-term goals with relation to data: while it is important to exploit what will be collected in the next ten years, there are also key projects in the further future that require theoretical legwork now. 

\subsubsection{Cosmic surveys}
The DM density, as measured by surveys, is the most important input constraining theoretical models for DM production, and survey science will continue to be an important probe of  fundamental parameters related to DM. Survey science is particularly synergistic in probing DM and dark energy, and the possible connections of both to the Early Universe.  Precision measurements of the DM density on different scales and at different epochs enables tests of the cold and collisionless hypothesis, and the possible multi-component nature of DM. The cosmic microwave background (CMB) and intensity mapping, by probing the evolution of the baryon temperature, also constrain the interactions of DM with the visible sector. Particular opportunities in the coming years include the following.

\emph{Wave vs.~particle}: A fundamental question about DM, which we may answer in the coming years, is whether it is better described by a collection of particles or a field, showing wave phenomena at astrophysical scales. Within the local environment, the occupation number density of DM is large enough for it to be modelled as wave for particle masses $m\lesssim 0.1 \text{ eV}$, if DM is a boson. %
Astrophysical wavelike behaviour is imprinted as a Jeans scale~\cite{khlopov_scalar}, and leads to formation of interference fringes and solitons~\cite{Schive:2014dra}, which can be probed by structure formation~\cite{Hlozek:2014lca,Rogers:2020ltq} and the high redshift Universe (see below) for masses in the range $10^{-33}\rightarrow 10^{-19}\text{ eV}$. In this mass range the self-interactions of DM can also be constrained (e.g.~\cite{Leong:2018opi,Li:2013nal}). The range where cosmic wavelike effects manifest in the dominant DM component (fuzzy DM, length scales larger than a kpc or so, $m\lesssim 10^{-19}\text{ eV}$) is severely constrained with only a narrow window still allowed. 
The coming years could close the cosmic fuzzy DM window using survey probes of structure formation. Above this window light bosons affect the distribution of masses and spins of BHs via the superradiance process~\cite{Arvanitaki:2010sy}, see Sec.~\ref{sec:GWs}.

\emph{Large scale clustering of DM:} 
Next generation CMB probes~\cite{Abazajian:2016yjj,Ade:2018sbj,Sehgal:2019ewc} will reach deeper into the non-linear regime of CMB lensing, probing the clustering of DM. The Sunyaev-Zeldovich effect also probes DM with precision bounds on the epoch of reionisation~\cite{Calabrese:2014gwa}. Even more powerful probes of DM clustering are offered by upcoming galaxy surveys measuring the power spectrum through galaxy clustering and weak lensing. These surveys, including, Euclid, the Vera C.~Rubin Observatory, and VLT, combined with the CMB, will reach sensitivity to measure the minimal neutrino mass, and possibly discern the neutrino hierarchy. Consequently, bounds on other mixed and non-cold DM models, and DM interactions with baryons, can also be significantly improved. If DM plays a role in the so-called ``discordances'' in $\Lambda$ cold DM (the $\sigma_8$ and $H_0$ tensions), surveys offer a significant opportunity to test this hypothesis. 

\emph{High-$z$ Universe}: The high redshift Universe is an un-tested environment in which DM could have manifested itself differently than today. 
 DM particles could have imprinted their signature on high redshift observables by affecting the growth of structure, and/or modifying the thermal and ionization histories of the intergalactic medium. Several classes of DM models, such as warm DM -- a prototype of which is a keV-scale mostly sterile neutrino -- and fuzzy DM result in suppressed small scale structure which could lead to a delay in star formation, cosmic heating and ionization~\cite{Schultz:2014eia,Bozek:2014uqa,Dayal2015}. Typically, in these theories the Universe has more time to cool down resulting in colder inter-galactic gas at the onset of star formation and prior to the phase of X-ray heating, and stars form in more massive DM halos than in cold DM imprinting larger ionized bubbles. The impact of DM physics  on the processes of heating and reionization can be probed by the  high-redshift 21-cm signal of neutral hydrogen, as done by EDGES~\cite{Bowman:2018yin}.
 Many DM models also predict non-trivial features in the power spectrum close to the cutoff, which could even survive down to redshifts relevant for Lyman alpha observations~\cite{Bose:2018juc}. Fully classifying the possible interplay of DM particle physics and structure formation in general is still an outstanding task (though the ETHOS framework~\cite{Cyr-Racine:2015ihg} 
offers first steps in this direction).
This epoch of the Universe is also synergistic to other constraints on energy injection by DM discussed below.

\emph{The UV Luminosity Function}: Observations of the first generations of galaxies provide a window into the early epoch of structure formation. Current data on the UV luminosity function of high-$z$ galaxies from HST allow us to probe the state of the Universe up to $z \approx 10$ and scales down to $k \approx 10\, \mathrm{Mpc}^{-1}$, with upcoming telescopes, such as JWST, being able to further extend these ranges and thus our handle on the evolution of DM. Besides space-based telescopes, ground-based facilities, such as the 39-m European Extremely Large Telescope, the Thirty Meter Telescope and the 25-m Giant Magellan Telescope will be valuable in performing spectroscopic follow-ups of sources discovered by JWST~\cite{finkelstein_2016}. This will paint a more complete picture of the astrophysical conditions around the time of reionisation, where various feedback processes (e.g. from supernovae) can mimic the suppression of structures at small scales caused by a non-CDM species~\cite{Wechsler:2018pic, Cuby2019Unveiling}.
It will also become important to combine these measurements with data from other probes, such as line intensity mapping surveys/21-cm~\cite{Kovetz:2019uss, furlanetto2019astro2020, Furlanetto:2019jzo}.

\emph{Intensity mapping}: Intensity mapping offers the possibility of a complete view of the Universe from low $z$ in galaxies, to high $z$ in the dark ages. Many surveys will begin producing data in the coming years. The signal itself is more complicated to model than e.g. weak lensing or the CMB, and requires more theoretical input in preparation for data beyond the next ten years. 
In the late Universe, the suppression of structure formation in warm DM and fuzzy DM-like theories leads to neutral hydrogen residing in more massive halos, consequently biasing and increasing the large scale intensity mapping signal. If measured by e.g.~SKA, HIRAX, CHIME, PUMA, this could detect deviations from cold DM in the form of a mixed DM component at the sub-percent level.
More specifically, constraints on decaying and annihilating DM can be obtained in the future by intensity mapping experiments such as the SKA by looking at the high-z energy injection in the gas and extra anisotropies induced by DM decays (see e.g.,~\cite{Furlanetto:2006wp}).
Regarding PBH of masses above 1000 $M_\odot$, for models in which PBHs could provide the seeds for super massive black holes (BHs), signatures would appear in intensity mapping observations at high-z (see e.g.~\cite{Dolgov:1992pu}). \footnote{Gravo-thermal collapse of self-interacting DM halos can also lead to the formation of BHs and potentially seed super massive BHs~\cite{Pollack:2014rja}.}

Although the above discussion has mostly focused on observables due to peculiar DM ``transfer function'' effects, surveys may also shed light  on features in the primordial power spectrum (such as extra Poisson clustering in PBH cosmologies) or anomalous structure growth, e.g.~(self-)interactions in the DM sector.
Especially important to highlight DM properties at small scales would be the measurements of strong and extreme (e.g.~caustic crossings) lensing events~\cite{LSSTDarkMatterGroup:2019mwo,Diego:2017drh}.
 
\subsubsection{Travelling messengers}
DM can inject high-energy particles into the environment through different processes such as decay or annihilation (particle DM), evaporation or accretion (PBH). The final stable particles produced can then travel towards Earth, further interacting with the environment (e.g.~in the interstellar medium of galaxies), and, eventually, be detected by telescopes on the ground and in space. 
\emph{Travelling messengers} can be of different nature: photons at different wavelengths, neutrinos, and charged cosmic rays, and can be used to discover/constrain the nature of DM, see also Sec.~\ref{\TM}. 
The challenge of traditional DM indirect detection is to disentangle a DM signal over the largely dominant emission coming from ``standard'' astrophysical processes, by 
maximising the signal over background, and  minimising systematic uncertainties of both theoretical (e.g.~the modelling of the DM distribution) and instrumental nature. For recent reviews on the subject, we refer the reader to~\cite{Gaskins:2016cha,Leane:2020liq,PerezdelosHeros:2020qyt}. 

DM particles below the electron mass scale can produce photons only through decay. 
In this case the energy spectrum of the signal is a narrow line at an energy corresponding to the mass scale of the DM particle. This line is then modulated by redshift evolution for the cosmological signal.
Axions and axion-like particles (ALPs), depending on their actual mass, can produce a signal from low radio frequencies (240 MhZ for $\mu$eV axions), to the IR (for meV to eV scale particles), to X rays (keV scale particles)~\cite{Irastorza:2018dyq,DiLuzio:2020wdo}.
Monochromatic radio emission (MHz - GHz) is also expected by DM axion/ALP-photon conversion 
through either resonant conversion from highly magnetised neutron stars (NSs), or white dwarf stars~\cite{Pshirkov:2007st,Huang:2018lxq,Hook:2018iia} or non-resonant transitions in the vicinity of the Galactic center and/or of discrete astrophysical objects~\cite{Sigl:2017sew}. Radio telescopes such as LOFAR and SKA have the potential to probe the 0.1 - 100 $\mu$eV QCD DM axion.
For appropriate ranges of parameters, light DM particles would contribute to  energy loss of stars, hence impinging  observable changes in their evolution~\cite{Raffelt:1990yz}, see also Sec.~\ref{\PS}.

Superradiance around PBHs can produce extremely dense axion clouds around the BH, resulting in stimulated axion decay producing millisecond-bursts at GHz frequencies~\cite{Rosa:2017ury}. 
Detecting such signals represents not only a data analysis challenge, but also entails new theoretical and numerical developments to realistically model the expected signal, as for example the full ray-tracing
computation of the conversion process of ALPs in realistic NS magnetosphere models~\cite{Leroy:2019ghm}.
Sterile neutrinos in the keV to MeV mass range are well probed by X rays, which provide stringent bounds on their mixing with ordinary neutrinos, see~\cite{Drewes:2016upu} for a review. 
Identifying new viable mechanisms to produce keV sterile neutrino DM without conflicting various cosmological and astrophysical bounds remain an important theoretical objective, see Sec.~\ref{sec:neutrinoBSM}. 
 
For DM particles with mass above the electron mass scale, annihilation (or decay) into charged products allows the production of a vast spectrum of final particles through prompt or radiative (i.e.~``secondary'') emission. 
Being electromagnetic signals, they can be searched for both in the Galaxy as well as in the cosmological radiation backgrounds, or can be used to investigate specific targets like dwarf galaxies, external galaxies, clusters or cosmological filaments.
More specifically: radio signals originate from synchrotron emission of DM leptonic products on ambient magnetic fields; X rays are produced through inverse Compton scattering on low-energy radiation fields; the CMB power spectrum is sensitive to injection of particles (especially leptons) which can alter the optical depth of the Universe or offer an additional mechanism to the spectral distortions of the CMB (see also below).
Pion and meson decay channels open up progressively for higher DM masses, leading to prompt $\gamma$-ray emission from GeV to TeV energies. For this mass range, complementary searches in the fluxes of cosmic rays, especially antimatter, and neutrinos are of relevance, see discussion in Sec.~\ref{\TM}. Combining them properly will require further developments of theoretical models and numerical tools. %
For particles heavier than 100~TeV, a testable scenario with current and up-coming $\gamma$ and $\nu$ telescopes is decaying DM, see e.g.~\cite{Chianese:2021htv,Esmaili:2021yaw, Maity:2021umk}, since WIMP-like annihilation processes are theoretically disfavored by the unitarity bound.

A sensible way to probe all these signals is to leverage on spectral and angular ``features'', optimising analysis strategies in this respect. As an example, the cross-correlation between DM emission spectrum and specific gravitational tracers of the DM distribution in the Universe represents a promising tool to probe the DM particle nature~\cite{Fornengo:2013rga}.
The GeV domain has been successfully covered by the superb performance of Fermi-LAT: a variety of DM searches have been performed, demonstrating all the richness an all-sky $\gamma$-ray instrument can offer and highlighting the complementarity of different targets which are affected by different systematics~\cite{Charles:2016pgz}. Prospects for improvement in the 10 - 100 GeV mass range will be dictated mostly by our ability to better describe astrophysical foreground/background emission towards the Galactic centre and also at higher latitudes,  
to be pursued through a fully multi-messenger and multi-wavelength investigation. This approach can also shed light onto the origin of yet mysterious ``anomalies'' in the $\gamma$-ray sky~\cite{Prantzos:2010wi,Murgia:2020dzu}, for which DM is often invoked as a possible explanation. 
Progresses are expected in the modelling of the DM distribution in the Galaxy, exploiting the output of large future optical surveys such as the Vera C.~Rubin Observatory~\cite{Drlica-Wagner:2019xan}, but also in the predictions of DM radiative emission which is largely affected by uncertainties on the Galactic magnetic field and confinement and diffusion of the electrons.
At TeV scale, further improvements are required for robust and more accurate calculations of the heavy DM spectra affected by electroweak radiative corrections~\cite{Bringmann:2017sko,Bauer:2020jay}. 
From an observational point of view, future MeV instruments have a great potential for discovery/constraining sub-GeV DM~\cite{DeAngelis:2017gra}.

Axions/ALPs can leave signatures on the cosmic electromagnetic radiation due to oscillations into photons, ranging from radio signals up to TeV $\gamma$ rays.
The search for below neV axions/ALPs produced by Primakoff effect in all past core collapse supernovae (see also Sec.~\ref{\PS}) can be uniquely constrained by future instruments covering the MeV gap~\cite{Calore:2020tjw}. Likewise, diffuse axion/ALPs emission at TeV energies from star forming galaxies is within the reach of HAWC, LHAASO, CTA~\cite{Vogel:2017fmc}.
Of relevance for neV -- peV axion/ALPs are also searches for oscillation patterns (induced by conversion into photons and back) in the spectra of Galactic and extragalactic sources from X to $\gamma$ rays, or for absorption features which alter the transparency of the Universe to photons. The search for former signal will benefit from improved energy resolution of X and $\gamma$-ray instruments, e.g. Athena, e-ASTROGAM, while the latter is among the CTA science cases~\cite{Abdalla:2020gea}. 
These searches do not assume axions/ALPs to be all the DM in the Universe, but are carving deeper 
into the allowed DM parameter space.

PBHs lighter than $10^{-16} \, M_\odot$ and heavier than $1 \, M_\odot$ are strongly constrained by emission of radiation during accretion of gas in the recombination history of the Universe and today, affecting CMB, radio and X rays observations, see~\cite{Carr:2020xqk,Green:2020jor} for recent reviews. 
Significant theoretical uncertainties affect the robustness of these bounds: on the astrophysical side, the accretion rate and also the ionizing effects of the radiation are little understood and phenomenological prescriptions are adopted, on the signal side, constraints are usually derived under the assumption of a delta-function (or monochromatic) mass function, while more realistic/complex mass functions can impact the limits in a non-trivial way.
Further studies on the dependence on mass function and clustering are required to assess the robustness of the bounds.  
Evaporation constraints from MeV -- GeV $\gamma$ rays set the lower bound of a surviving open PBH DM window of about $10^{-16} \, M_\odot$ (unless PBHs form relics). 
Future MeV and X-ray space observatories can allow more precise measurements of the high-latitude diffuse (soft) $\gamma$ and X-ray emission and improved constraints can be placed on PBH up to $10^{-15} \, M_\odot$.

CMB spectral features can also be used to investigate both axions/ALPs in the sub-eV range, besides heavier DM annihilating into non-thermal electrons.
Spectral distortions of the CMB energy spectrum offer a further avenue to constrain various DM models. 
A possible future detection of spectral distortions, which is a goal for cosmology in the coming decades \cite{Chluba:2019nxa,DiValentino:2020vhf,Kogut:2019vqh}, will allow us to explore otherwise inaccessible regions of DM parameter space and probe the power spectrum at small scales.
Spectral distortions may be sensitive to decaying/annihilating DM, ALPs, and $\mathcal{O}(10)M_\odot$ PBHs~\cite{Chluba:2019nxa,Kogut:2019vqh}.

There is no observational evidence for or against multicomponent DM (for example within extended beyond Standar Model structures), but candidates which are independently viable may not be mutually compatible. As an example, the high density halo of DM which is expected to form around a PBH (assuming PBHs are not all of the DM) would lead to a huge annihilation signal from fiducial WIMPs, which would be such a strong signal that a detection of PBHs (within most mass ranges) or of WIMPs would effectively rule out the existence of the other \cite{Lacki:2010zf,Eroshenko:2016yve,Adamek:2019gns,Bertone:2019vsk,Carr:2020mqm}. A challenge for theorists is to find more examples of mutually incompatible DM models which may be tested in the foreseeable future. 

\subsubsection{Gravitational waves and compact objects}
\label{sec:GWs}
Gravitational waves (GW) are a promising tool for investigating DM properties. 
GW signals are expected to be generated by local DM environments modifying the GW signal from a merger of two compact objects in a distinctive way (see \cite{Bertone:2019irm,Bertone:2018krk, Giudice:2016zpa,Barausse:2020rsu} for reviews).
As for BHs, the DM structures around them can reflect properties such as particle mass, spin and self interactions, and their presence will change the trajectory of any inspiralling object 
 due to the backreaction of their energy density on the gravitational metric, in addition to distinctive dynamical friction and accretion effects \cite{Barausse:2014tra}.
Well motivated possibilities exist that would result in sufficient enhancements to the DM density to generate an observable signature; in particular, the phenomena of superradiance in light bosonic DM candidates, the effect of self interactions, and the formation of DM ``mini-spikes" where seed BHs grow adiabatically in DM haloes. Simulations of the evolution of DM in strong gravity environments can confirm in which cases such structures are likely to form around isolated BHs, and/or subsequently survive the merger process.
Space-based detectors (e.g.~LISA) will probe supermassive BHs and extreme mass ratio inspirals, which offer several advantages for DM detection over the BH solar mass range targeted by ground based detectors (e.g.~LIGO/Virgo). Environmental effects are stronger in more massive BHs as the same DM density will give a larger curvature relative to that of the BH itself. %
Extreme mass ratio inspirals also offer longer timescales over which characteristic dephasing of the GW signal might be observed.
The next ten years will be critical in preparations for the LISA mission, quantifying the potential effects of DM environments for different models. Should sufficiently promising regimes be identified, more work will be needed to produce template waveforms incorporating DM effects that are accurate enough to be used for data analysis purposes, with advancements needed in both numerical relativity simulations and second-order self-force calculations~\cite{Barausse:2020rsu}.

GWs from a binary NS inspiral detected by ground-based detectors can also be used to constrain  
DM around NSs (see e.g.~\cite{Hook:2017psm}), but also DM cores inside NSs, see e.g.~\cite{Ellis:2017jgp}. 
Constraints on QCD axion from NS mergers are complementary to those expected from e.g.~the CASPEr experiment~\cite{Zhang:2021mks}. 
Studying DM effects on NSs is also particularly interesting with regard to the new NICER results, see also Sec.~\ref{\NA}.

There is also the possibility that DM itself may form compact objects that generate GWs via merger events, e.g.~boson stars, or that the DM are PBHs \cite{Sasaki:2018dmp,Carr:2020gox,Carr:2020xqk,Green:2020jor}. The PBH abundance can be constrained by their merger rate and therefore GW detections \cite{Nakamura:1997sm,Bird:2016dcv,Clesse:2016vqa,Sasaki:2016jop}. While future GW experiments will certainly improve our estimates of the overall BH-BH merger rate, progress on this also requires theoretical progress on PBH merger rates, which comes from a better understanding of PBH binary formation and potential disruption before merging. Since BHs have no hair, it is challenging to discriminate between astrophysical and primordial BHs but large statistics about the BH spin, redshift and mass distributions can provide signatures, whilst the cleanest signal would be the detection of a sub Chandrasekhar mass compact object. More work on distinguishing primordial and astrophysical BHs signatures is important, relating them back to the formation mechanism. If PBHs formed from the collapse of large amplitude density perturbations after horizon entry, then the required amplitude would be so large that a stochastic GW background would inevitably be generated, and this is a valuable indirect probe of PBHs, especially to close the ``asteroid'' mass range which is the primary remaining window where all of the DM could be PBHs. 
A stochastic GW background would be produced also in dark phase transitions and non-perturbative DM production~\cite{Bertone:2019irm,Giudice:2016zpa}.
Another future diagnostic tool in the search of PBH in the stellar mass interval will be offered by cross-correlation among GW events and large scale structures~\cite{Scelfo:2018sny,Maggiore:2019uih}.

Finally, DM capture in compact (and less compact) objects such as stars and other celestial bodies can offer an alternative way to direct detection experiments to probe the DM-nucleon scattering cross-section over a broad range of DM masses, see e.g.~\cite{Baryakhtar:2017dbj,Nisa:2019mpb,Leane:2021ihh}.
One theoretical goal is to make  capture rate calculations more and more refined and better define observational prospects.
This constitutes an alternative channel to obtain sub Chandrasekhar mass BH, with detailed signatures associated to the transmutation event that remain to be worked out via numerical simulations.
For more details about constraining DM with BHs, NSs, and, more generally, through stars we refer the interested reader to Sec.~\ref{\PS}.

\subsubsection{Galactic astrophysics and near-field cosmology}
Galactic astrophysics probes the local distribution of DM in a complementary manner to cosmic surveys.
Information about the distribution of DM within our Galaxy has been obtained either with ``global'' methods  -- based on the known Galactic rotation curve and bringing information about the spatial distribution of DM from few kpc up to tens of kpc from the center -- or more ``local'' methods -- instead based on the use of tracers of the gravitational potential of a local region around the Earth's position of order 100 pc. The two methods infer quantities
which are compatible within the uncertainties (see for instance Figure 5 in \cite{Benito:2020lgu} and references therein), whilst they leave big unknowns on the actual velocity distribution of DM, which direct detection experiments are quite sensitive to.

The above is based on the assumption that the entirety of DM in the Milky Way is in a steady, gravitationally relaxed state. The Milky Way, however, is expected to have experienced one or a few significant mergers in its recent past, and the question arises whether the amount of DM involved in the merger is dynamically relevant, and whether this can teach us anything about the nature of DM, complementing the information one can obtain from direct detection. Recently, GAIA data have allowed the identification of stellar streams, which are thought to be associated to DM streams (see \cite{Helmi:2020otr} and references therein). 
The ongoing work seems to indicate that the amount of DM from streams can be sizable with respect to the density in the ``steady state'' halo. This does not change the determination of the local DM density, but may dramatically alter the velocity distribution, and have profound implications for the interpretation of results of direct detection experiments~\cite{Necib:2019zka, Necib:2019zbk}. 

High density DM substructures can form from enhanced amplitude perturbations on smaller scales than are probed on the CMB. These are called ultracompact minihaloes (UCMHs). A similar mechanism of enhanced density perturbations due to a phase transition also seeds axion miniclusters~\cite{Hogan:1988mp}. If a significant portion of the local DM is locked up in such structures, this has severe implications for direct detection, and so determining constraints on the fraction locally and globally is an important goal of astroparticle physics. Recent numerical simulations of both UCMHs and miniclusters~\cite{Gosenca:2017ybi,Delos:2017thv,Eggemeier:2019khm} have sharpened predictions, but much is still to be done. Evidence for such objects will also have profound implications for our understanding of the early Universe. 

Gravitational microlensing is sensitive to DM substructures ranging from asteroid to solar masses. This can be used to constrain PBHs, axion miniclusters, UCMHs or even boson stars, and depends on the finite, extended, size of the lens~\cite{Croon:2020wpr}. The Roman Space Telescope, Euclid, and the Vera C.~Rubin Observatory will provide us with precise microlensing surveys of the Galaxy which can probe sub-Earth mass DM structures. 
Substructures inlcuding PBHs, UCMHs, and axion miniclusters are largely unconstrained in the ``asteroid mass gap'', and thus closing this window is a major goal to achieve in the near future and e.g.~microlensing of X-ray pulsars with large effective area, future, X-ray telescopes (e.g. AstroSat, LOFT) has the potential to do so~\cite{Bai:2018bej}.  

In the phase transition leading to minicluster production the only free parameter is the axion particle mass, which can thus be fixed by the DM relic density if the phase transition and subsequent evolution can be computed accurately enough. Computations have made significant progress (e.g.~\cite{Klaer:2017ond,Gorghetto:2020qws}) and further advancements could make a precise prediction in the near future. Such models predict miniclusters in the asteroid mass gap. 

\subsubsection{Terrestrial experiments}
Terrestrial DM experiments are highly complementary to astroparticle probes, in particular where degeneracies can be broken (e.g. local density), or fundamental properties such as spin, parity, gauge interactions etc can be identified. For an extensive review of experimental effort see an APPEC Report~\cite{Billard:2021uyg}, and references therein.

\emph{WIMP-like Particles ($m\gtrsim MeV$)}: New results are soon expected from liquid xenon detectors XENONnT, LZ, and PandaX-4T, to be followed within the  next few years by new-generation liquid argon experiment DarkSide-20k and a liquid xenon-based DARWIN. For cross sections below a certain threshold, direct detection experiments lose sensitivity due to background events from coherent neutrino-nucleus scattering, with experiments in the next decade approaching this ``neutrino floor''. Liquid noble gas detectors will also probe sub-GeV DM through the Migdal and Bremsstrahlung effects. This effort will be complemented by the search for sub-GeV DM via nuclear recoils by, e.g., the CRESST-III, EDELWEISS, and SuperCDMS low-threshold experiments. Another milestone for the next decade is the independent tests performed by the Anais, Cosine, SABRE, and Cosinus experiments of a DM interpretation of the DAMA annual modulation signal. Over the same time window, the search for sub-GeV DM particles in condensed matter systems is expected to become increasingly important.  The growing interest in sub-GeV DM should strengthen the R\&D of novel low-band gap detector materials that are optimised for this search~\cite{Geilhufe:2018gry}, such as graphene~\cite{Baracchini:2018wwj}. From the theoretical side detailed calculations of the response of operating and proposed detector materials are required~\cite{Catena:2019gfa,Catena:2021qsr,Kurinsky:2020dpb,Trickle:2020oki}. The theoretical and experimental studies outlined above are expected to foster new synergies between the astroparticle and condensed matter physics communities. 

\emph{Ultra Light Bosons}: The coming decade will see a large number of new searches for ultralight bosons moving beyond the prototype stage. In particular, searches for axions will reach sensitivity to the QCD axion across most of its natural and allowed parameter space, $10^{-11}\text{ eV}\lesssim m\lesssim 10^{-2}\text{ eV}$, in the coming decade, corresponding to frequencies of order kHz to THz. For example, MADMAX and CAPP extend to the upper frequencies not currently covered by ADMX, while NMR e.g.~CASPEr and lumped circuits e.g.~ABRACADABRA probe lower frequencies. Axion searches at different frequencies probe the early Universe history of the axion and its production mechanism. At low frequencies, this tests the ``anthropic window'' and the scale of inflation~\cite{Hertzberg:2008wr}, while at high frequencies this probes models of DM production from cosmic strings~\cite{Gorghetto:2018myk}, and the possible existence of miniclusters~\cite{Hogan:1988mp}. Similarly, precision searches for ultralight scalars will reach some maturity across wide frequency ranges. Ultralight boson searches can often be performed in tandem with gravitational wave detectors, offering further synergy (e.g.~\cite{Dimopoulos:2008sv,Badurina:2019hst}). Signal modeling should be essentially informed by astrophysical considerations and theoretical simulation of the wavelike behaviour~\cite{Foster:2017hbq,Hui:2020hbq}. If ultralight bosons are detected in the laboratory by a axion haloscope-like experiment the signal lineshape could be used to measure properties of the local DM distribution such as velocity dispersion and clumpiness~\cite{OHare:2017yze}. 

\emph{Detector Networks}: Spatially distributed networks of sensors allow probing of more complicated spatiotemporal signatures of bosonic DM. Bosonic DM can form stable field configurations, like boson stars~\cite{QBalls}, strings, and domain walls \cite{DomainWalls} as well as gravitationally bound halos around the Sun and Earth \cite{Banerjee2020}. 
Signals due Earth collisions with such objects can be identified with network detectors. Currently operating networks include GNOME, searching for magnetization induced by pseudoscalars, and the atomic clocks of the Global Positioning System (GPS) \cite{Roberts:2017hla}, which tests scalar induced variation of constants, and gravimeter networks to search for DM clumps \cite{Hu2020}. The theory community should continue to develop models for the signatures and rates of such exotic DM events to exploit this data.

Further experimental probes of the couplings between the dark and SM sectors come from beam-dump, collider, and neutrino-experiments searching for either missing energy/momentum signatures or direct scatterings of DM particles in the detectors, as well as for decays of mediator particles, see e.g.~\cite{Beacham:2019nyx} and Sec.~\ref{sec:neutrinoBSM}.

\subsection{Confirming the nature of dark matter}
Even if a new particle is discovered, a key challenge will be to confirm that it is \textit{the} DM in the Universe.
One possibility would be to measure the local DM density $\rho_\chi$ and compare with the density inferred from the local kinematics of stars or from global mass modeling of the Milky Way~\cite{Read:2014qva}. Unfortunately, such a measurement is only possible with terrestrial experiments when DM-Standard Model interactions are strong, in which case the degeneracy between $\rho_\chi$ and the scattering cross section can be broken by Earth-scattering effects~\cite{Bramante:2018qbc,Kavanagh:2020cvn}. Even with such an ultra-local measurement of $\rho_\chi$, comparisons with astronomical estimates may be confounded by the presence of substructure in the local DM phase space~\cite{Necib:2018iwb}.

A single experimental technique is insufficient to probe the vast landscape of viable DM candidates, which display distinct properties. The lack of a unique theoretical guideline calls for a combined search strategy \cite{Arrenberg:2013rzp} in which different experimental methods (indirect, direct and accelerator-based searches) complement each other, not only probing more DM models, but also providing independent confirmation of potential observations, helping to remove backgrounds, and, eventually, contributing to a better measurements of the DM properties.

In some DM scenarios only one of these complementary signals may be available, but there are models in which many measurements can be expected and complementarity can be fully exploited. This is e.g.~the case of WIMPs, for which the combination of direct and indirect searches~\cite{Bergstrom:2010gh} might be used to reduce the effect of astrophysical uncertainties and  to confirm that the detected particle would indeed have been produced with the correct relic abundance~\cite{Roszkowski:2016bhs}. This can also be achieved by comparing results of indirect searches from different astrophysical origins. 
A similar approach should also be possible for axion DM, by combining direct searches with experiments which aim to produce and then detect new light particles~\cite{Bahre:2013ywa}, or measure solar axions~\cite{Dafni:2018tvj}. Such a confirmation of the nature of DM would be made even more difficult in the presence of a non-standard cosmology~\cite{Arias:2019uol} or multi-component DM.

It will therefore be crucial in the near future to extend such studies to as wide a range of DM models as possible, as well as to map out which combinations of search strategies to yield the best chance of confirming the nature of DM.

\section{Dark Energy}
{\bf Coordinators:} Alessandra Silvestri and Julien Lesgourgues.\\
{\bf Contributors:}   Emilio Bellini, David Benisty, Chris Byrnes, C.G.~B\"ohmer, Clare Burrage,  Anne-Christine Davis, Claudia de Rham,  Recai Erdem, Mario Herrero Valea,  Ali Rida Khalifeh, Kazuya Koyama, Jackson Levi-Said, Lucas Lombriser, Matteo Martinelli, Carlos J.A.P. Martins, Peter Millington, Lorenzo Pizzuti, Marco Raveri, Javier Rubio,     Nikolina \v{S}ar\v{c}evi\'c, Martin Sahlen,  Konstantinos Tanidis, Cora Uhlemann, Filippo Vernizzi, Sebastian Zell. \\

Shedding light on the source of cosmic acceleration and the nature of gravity on large cosmological scales will remain one of the main scientific goals in the next decade. Since the discovery of cosmic acceleration in 1998 
\cite{SupernovaCosmologyProject:1998vns}, much progress has been made both on the observational and theoretical sides to tackle these questions. The standard model of cosmology, $\Lambda$CDM, has passed with flying color an increasing number of tests based on precise cosmological measurements. Yet, we do not have a satisfactory understanding of the cosmological constant from the theoretical point of view. Besides, we are just starting to collect highly precise cosmological data 
in order to test General Relativity on large scales with a precision comparable to that of Solar System tests \cite{Will:2014kxa}. Furthermore, some tensions among different cosmological datasets have emerged in the recent years, possibly hinting at new physics beyond $\Lambda$CDM. The most notable tensions to date are related to the amount of matter clustering, commonly encoded in the $S_8$ parameter \cite{Joudaki:2019pmv}, and to the Hubble parameter $H_0$ \cite{Riess:2020fzl}.
While in the past two decades the evidence in favor of $\Lambda$CDM has been building up, and while this model remains arguably the most economic fit to the data, small inconsistencies started to appear. This is not a complete surprise as we entered the era of precision cosmology. It is fair to notice that more recent Large Scale Structure (LSS) data from DES \cite{DES:2021wwk} show a reduction of the tension in $S_8$, and that unidentified systematics in local measurements still remain a possible explanation for the $H_0$ tension (e.g. \cite{Freedman:2019jwv,Freedman:2020dne,Efstathiou:2020wxn,Mortsell:2021tcx,Mortsell:2021nzg}).


The long-standing evidence for accelerated expansion and the recent tension hints have triggered interest for various models of dynamical Dark Energy (DE), modifications of Einstein gravity on large scales (Modified Gravity, MG), or departures from the basic assumptions of the FLRW model (small-scale averaging problem, strong local inhomogeneity, large-scale deviation from homogeneity). A plain cosmological constant $\Lambda$ provides the simplest possible  explanation of accelerated expansion, but suffers from fine-tuning problems (since radiative corrections should bring it to considerably larger values \cite{Weinberg:2000yb}), and also most likely from instability problems at the quantum gravity level. Indeed, a stable $\Lambda$ with the required order of magnitude seems to be incompatible with low-energy quantum gravity \cite{dvali2014qcg,dvali2014quantum_exclus,Dvali_2017_quantum_brreak} or string theory \cite{Obied_2018_Sitter,Ooguri_2019_Distance}. Next-level solutions are also contrived: several scalar field (quintessence) models also face fine-tuning issues (like the cosmic coincidence problem), many MG models are unstable \cite{Clifton:2011jh}, and the backreaction of small-scale perturbations seems way too small to trigger acceleration \cite{Green:2014aga} (see however \cite{Buchert:2015iva}). 

Note that some authors are still questioning the very evidence for accelerated expansion in the framework of inhomogeneous cosmology \cite{Mohayaee:2021jzi}, but such a discussion goes beyond the scope of this section. Note also that DE and MG models have been recently proposed for modifying the {\it early} cosmological evolution in such a way to reduce the Hubble tension (e.g. \cite{Karwal:2016vyq,Poulin:2018cxd,Braglia:2020auw}), but here we will rather focus on the {\it late} (post-recombination) evolution, and on upcoming opportunities to test the nature of gravity and possibly shed light on late-time DE. 

\subsection{Theory}
\subsubsection{Theoretical Landscape}
On the theoretical side, much effort has gone into the exploration of the gravitational landscape, including dynamical DE and MG (see e.g.~\cite{Clifton:2011jh} for a review). Significant progress has been made in  constraining and, in certain cases, ruling out specific classes of models and in developing unifying frameworks to efficiently chart the space of alternatives to $\Lambda$CDM. In this Section we provide an overview, with emphasis on future directions on the theory side.\\

\noindent{\bf Old cosmological constant problem.} Given that a cosmological constant or vacuum energy associated to the energy scale of $10^{-3}$~eV faces huge problems of stability against radiative corrections and against decay, typical DE/MG models focus on sourcing cosmic acceleration while assuming that the vacuum energy exactly vanishes (an exception would be models with the degravitation mechanism). We are then  back to the old cosmological constant problem. While this problem could be solved at the level of high-energy physics by imposing symmetries (e.g. minimal supergravity), or even more fundamentally at the level of quantum gravity, the problem can also be investigated from the point of view of a cosmologist, invoking for instance extra dimensions, the sequestering mechanism \cite{Kaloper:2013zca} or, perhaps, unimodular gravity \cite{alvarez_2015qc}. This old cosmological constant problem is still crucial and should receive high priority. \\  

\noindent{\bf Gravitational landscape.}  General relativity (GR) has been exceedingly successful in meeting observational challenges over the century since its inception, despite several theoretical shortcomings ranging from its renormalizability to the difficulty of reformulating it as a gauge theory. There are related issues concerning the presence of singularities in the theory. These problems, added to the discovery of cosmic acceleration and the rise of tensions within $\Lambda$CDM, have prompted through the years a consideration of several alternative gravity theories. A thorough review of the latter is not in the scope of this paper. We refer the reader to the extensive literature on this (see e.g.\cite{Clifton:2011jh}), while here we provide only a brief recap, and then focus on future directions. 

A general classification of alternative theories can be based on the  modifications to the Einstein-Hilbert action for gravity and the action for matter fields. They range from the explicit addition of scalar, vector or tensor fields (quintessence is an example),  the  inclusion of  geometric invariants (e.g. functions of Lovelock scalars, $f(R)$ being the most prominent example), higher-dimensions \cite{Maartens:2010ar} (offering a possible explanation for the weakness of the gravity force, as well as a scenario for realizing the degravitation mechanism \cite{Dvali:2007kt}) or non-local theories (possibly alleviating the fine-tuning problem for some fundamental constants of nature \cite{Belgacem:2017cqo,Calcagni:2018gke}) all the way to Lorentz-violating theories. Models of Lorentz-violating gravity can provide a solution to the renormalizability problem of GR, with the most notabe example being Ho\v rava gravity  \cite{Horava:2009uw}. 
Furthermore, geometric gravity can  appear not only through curvature deformations but also torsion and non-metricity. These so-called teleparallel theories of gravity have been poorly studied in comparison to curvature-based modifications of general relativity, but they shown promise in producing a gauge theory of gravity \cite{Bahamonde:2021gfp}.

 Models involving additional fields and geometric scalars can be organized in broader classes, such as Horndeski gravity \cite{Horndeski:1974wa} (later rediscovered as Generalized Galileon gravity \cite{Deffayet:2011gz,Kobayashi:2011nu}), beyond Horndeski gravity and DHOST \cite{Zumalacarregui:2013pma,Gleyzes:2014dya,Langlois:2015cwa}. They can further be classified depending on the so-called screening mechanism, as we discuss in the next subsection. 
 
 Through constraints from local tests of gravity and the propagation of  gravitational waves, Horndeski, DHOST and beyond Horndeski gravity have been severely constrained in recent years \cite{Creminelli:2019kjy,Ezquiaga_2018darkenergy}. The surviving sector can still produce an interesting phenomenology at the level of LSS \cite{Peirone:2017ywi}. It shall also be noted that some of the constraints on, e.g., the speed of sound from GWs cannot be directly applied to some of these theories since, as DE theories,  they have a low-energy cut-off which lies below the frequency range characteristic of LIGO-Virgo.

\subsubsection{Screening mechanisms} 

Many MG models effectively add new light scalar degrees of freedom coupled to gravity. To satisfy solar system and laboratory tests (e.g., Lunar Laser Ranging tests or E\"{o}t-Wash experiments \cite{Will:2014kxa}), these models require a screening mechanism that suppresses any fifth force locally. Screening models can be divided into three major categories (e.g.,~\cite{Ishak:2018his} and references therein). Chameleon models, including $f(R)$ models or scalar-tensor theories of Brans-Dicke type, rely on a conformally coupled scalar field with an environment-dependent effective mass. In symmetron or dilaton models, screening is controlled via an environment-dependent matter coupling. Alternatively, self-interactions with higher-order derivatives produce screening via the Vainshtein or k-mouflage mechanisms, which are characteristic of some Horndeski, but mostly beyond Horndeski and DHOST theories. In all cases, GR is recovered locally, but the fifth force leaves a characteristic imprint in the formation 
of cosmological structures that 
depends on the 
environment (e.g.,~\cite{Pogosian:2007sw,Burrage:2018dvt}). Upcoming LSS surveys hold great promise in constraining such features, especially in combination with CMB observations \cite{Amendola:2016saw, Ishak:2018his}. However, in this section, we shall focus on smaller scales and local  experiments \cite{Hui:2009kc, Koyama:2015oma,Baker:2019gxo}, like constraints from galaxy clusters (as explored e.g. in~\cite{Pizzuti:2016ouw,Barreira:2014zza, Cataneo:2018mil,Terukina:2013eqa, Sakstein:2016oel}) or galaxy morphology \cite{Desmond:2020gzn}. Different works have shown that the background mass of the scalar in chameleon gravity should be $\mathcal{O}$(Mpc) to pass solar system constraints, while~\cite{Saltas:2018mxc} put stringent bounds on beyond Horndeski/DHOST models  by using white dwarfs.

 It has been shown that, at least for certain classes of models, theories can be ‘un-screened’ in carefully designed laboratory experiments. In recent years, a number of laboratory experiments, including atom-interferometry \cite{Jaffe:2016fsh, Hamilton:2015zga, Sabulsky:2018jma}, neutron bouncing \cite{ Brax:2011hb, Cronenberg:2015bol, Jenke:2014yel, Ivanov:2012cb}, neutron interferometry \cite{Li:2016tux, Lemmel:2015kwa, Brax:2014gja, Brax:2013cfa, Pokotilovski:2012xuk}, torsion balance \cite{Upadhye:2012rc, Brax:2014zta} and Casimir force experiments \cite{ Brax:2007vm, Almasi:2015zpa, Elder:2019yyp} have proved particularly powerful in constraining theories that screen through a chameleon or symmetron mechanism. The fact that screening relies on non-linearities means that connections between laboratory and cosmological constraints are model-dependent \cite{ Burrage:2017qrf}. Similarly, the design of optimal experiments (providing the most interesting constraints on the parameters relevant for cosmology) may differ from model to model. This is certainly a direction to be further explored in the near future; interestingly, laboratory experiments can typically be performed by small groups on short timescales. Thus, they could contribute to rapid improvements in our understanding of cosmological theories.

In addition, the behaviour of fifth forces cannot be disentangled from the
extension of the Standard Model (SM) of particle physics into which they are embedded. If this model is scale invariant, the existence of a conserved dilaton current for certain MG theories implies the absence of fifth forces~\cite{Garcia-Bellido:2011kqb, Ferreira:2016kxi}, as can be verified by an explicit calculation of scalar exchanges~\cite{Burrage:2018dvt}. Conversely, in the case of the SM, certain classes of MG theories are equivalent to Higgs-portal theories in the Effective Field Theory (EFT) sense~\cite{Burrage:2018dvt}. Quantum corrections can also be important, e.g., for matter sources with a spatial extent much smaller than the Compton wavelength of the fifth-force mediator (see, e.g.,~\cite{Burrage:2021nys}); or in the absence of scale or shift symmetries, which leaves scalar models unprotected from radiative corrections. There is a need to go beyond existing analyses of screened fifth-force models based on the classical equations of motion, and to fully account for quantum corrections.

\subsubsection{Effective and parametrized approaches}
Shortly after the discovery of cosmic acceleration, the main efforts focused on DE models and their impact on the background dynamics, which is typically parametrized by the equation of state $w(z)$, with $w=-1$ in $\Lambda$CDM.  Building on a decade-long exploration of the phenomenology of perturbations in DE/MG theories, this parametrization has been extended to three functions $(w,\mu,\Sigma)$, where $\mu$ and $\Sigma$ respectively encode MG effects on matter and light~\cite{Amendola:2007rr,Bertschinger:2008zb,Pogosian:2010tj}. This complete set describes the phenomenology of linear scalar perturbations, in a very general way which is quite agnostic about the underlying theory. $\mu$ and $\Sigma$ are functions of time and scale for which  analytical, model-specific expressions can be found after restricting to the so-called quasi-static regime. This \emph{phenomenological} framework has been adopted by several collaborations (see e.g.~\cite{Planck2020results}). It will still represent one of the most efficient avenues to interface with upcoming observables. Analogous frameworks are being developed for constraining DE/MG with the propagation of gravitational waves \cite{LISACosmologyWorkingGroup:2019mwx}.
As we are nearing the release of relevant data, it becomes necessary to construct more efficient frameworks that maximally leverage on the power of cosmological observations to shed light on DE and gravity.  In more recent years, much effort has gone into the development of unifying frameworks that have a closer link to theory. The most relevant example is the EFT of DE, which represents a natural generalization of the EFT of inflation to late times \cite{Creminelli:2008wc,Gubitosi:2012hu,Bloomfield:2012ff,Gleyzes:2013ooa,Gleyzes:2014rba,Bellini:2014fua,Frusciante:2019xia}. Building on symmetry arguments, these methods generally provide a unifying action for studying linear perturbations in models with an additional scalar degree of freedom, most notably Horndeski gravity. They improve over the phenomenological framework in that they allow to treat both scalar and tensor perturbations at once, and to study constraints from LSS and GWs all together. On the other hand, they include several unknwon functions of time (e.g.  five  for Horndeski gravity), making it very challenging to obtain meaningful constraints from data. A very promising avenue is the combination the phenomenological and EFT frameworks into a theory-guided non-parametric reconstruction of $(w,\mu,\Sigma)$ \cite{Pogosian:2016pwr,Espejo:2018hxa}. This capitalizes on a very powerful aspect of the EFT of DE, i.e.,  that of naturally endowing us with \emph{theoretical priors} that restrict the viable theory space during the agnostic exploration of models, as we review in the following. 
 
All these frameworks have been mostly developed in the linear regime of cosmological perturbations. However, most of the data that will be collected by future LSS surveys will concern scales where the linear approximation is insufficient.  An obvious direction for the next future is that of building a parametrized approach valid beyond the linear regime. As a first step, one can study scales where fluctuations are small but higher-order corrections are nevertheless important. One can capture these scales by using perturbation theory \cite{Bernardeau:2001qr} and introducing a  finite number of unknown coefficients (or counterterms), whose scale-dependence is dictated by symmetries. These coefficients parameterize the effect of highly nonlinear scales \cite{Baumann:2010tm,Carrasco:2012cv,Porto:2013qua}. This approach has the disadvantage of introducing new parameters, but it allows for a controlled accuracy. An obvious direction of future research, initiated in \cite{Cusin:2017mzw,Cusin:2017wjg}, is to combine it with the EFT of DE. 

To describe scales beyond the reach of perturbation theory, one needs to resort to $N$-body simulations, emulators or other numerical approaches. These approaches are technically more involving, but they allow to compute the effect of DE/MG in terms of the EFT of DE parameters only.  

\subsubsection{Theoretical Priors}

When dealing with EFTs of DE, one typically has in mind at least one additional degree of freedom, often a scalar,  in addition to gravity. Kinetic interactions of the DE field and potential non-trivial mixing between gravity and the DE field can lead to pathologies that are only manifest at the level of fluctuations. The consistency of the background solution requires all the physical d.o.f present in the EFT to be stable. 
The stability of the background solution against unbounded fluctuation growth demands that both the kinetic and gradient terms of the physical d.o.f be positive definite. Their respective squared mass term could in principle be negative (signaling a tachyonic instability) as long as it is of the same order as the Hubble parameter (or smaller, such that the time scale of the associated instability is at least as long as that of the background DE evolution). Such types of theoretical priors are now commonly imposed on any type of DE theory  and have been shown to be particularly powerful in cutting the parameter space \cite{Raveri:2014cka,Peirone:2017lgi,Traykova:2021hbr}. They can be applied either before fitting any model to the data, or in an integrated way when the data and theory prior covariances are combined at the analysis level \cite{Espejo:2018hxa}. On can consider further classes of theoretical priors that rely on additional requirements beyond the strict stability of the background solution. This represents a natural direction for future works. Such requirements include: \emph{Preserving perturbative unitarity -} The absence of strong coupling issues further requires that both the kinetic and gradient terms ought to be finite and non-vanishing \cite{deRham:2017aoj};
 \emph{Preserving Causality -} In a standard non-gravitational theory, the requirement of causality is linked with that that of subluminality; however the condition $(0<)c_s^2<1$  only makes sense when gravity can be fully decoupled. In theories of DE, the notion of subluminality is not frame-invariant and the presence of a gravitational exchange allows for a small amount of superluminality to be present without being linked with causality \cite{deRham:2019ctd,deRham:2020zyh};
 \emph{Embedding in a Standard High Energy completion - } The requirement that a DE model can in principle be embedded in a ``standard" high energy completion have motivated the derivation of so-called positivity bounds. Those bounds can be imposed directly on the low-energy EFT like DE, but they feed on the requirement that the high-energy completion is unitary, causal, local and fundamentally Lorentz invariant (even though the cosmological background we may be interested in may spontaneously break Lorentz invariance)~\cite{Bellazzini:2019xts,Tokuda:2020mlf,Herrero-Valea:2020wxz}. Preliminary studies have  shown that positivity bounds can  significantly reduce the allowed region of parameter space \cite{Melville:2019wyy,deRham:2021fpu}. The derivation of these bounds has reached maturity in the framework of a  Minkowski background, but their direct application to the realm of DE is still an ongoing matter 
 (see \cite{Grall:2021xxm} for initial progress in this direction); \emph{Technical Naturalness -} Finally, in DE models, the presence of a small parameter that either manifests itself as a small cosmological constant or an equivalently small DE mass is inevitable. The consistency of such DE theories typically demands that the quantum corrections remain under control and that, even if this parameters is tuned, it remains technically natural (see e.g. \cite{deRham:2012ew,deRham:2013qqa,Heisenberg:2014raa,Heisenberg:2020cyi}).

\subsection{Observational Outlook}

\subsubsection{What should we try to constrain with observations?}

In order to constrain models of DE and MG, we should of course aim at better measurements of the expansion of the universe, which test the Einstein equation at the background level (that is, in the standard paradigm, the Friedmann equation), and the dilution law of a possible DE component. We should simultaneously use all probes of cosmological perturbations to test the clustering and lensing properties of matter (and, therefore, the perturbed Einstein equation) and to constrain possible DE perturbations. To this extent, the combination and cross-correlation of different probes can play a key role to test characteristic imprints of DE/MG. It is also important to notice that the clustering properties of matter can be unveiled by a number of probes which are highly complementary to simple 2-point correlation functions, like density-weighted probes with special DE/MG sensitivity (see e.g. \cite{Hellwing:2017pmj,Boyle:2020bqn,Sahlen:2018cku,White:2016yhs,Spolyar:2013maa}). 

Concerning the possibility that the apparent acceleration relates to departures from the Friedmann model, we can use the angular dependence of several observables (related e.g. to luminosity distances, angular distances, time delays, redshift drifts, the Sunyaev-Zel'dovitch effect or bulk flows) to improve the determination of our local velocity with respect to the cosmological frame, and at the same time, to exclude a possible unexpected level of local inhomogeneities. Within this list, redshift drift measurements will be particularly interesting, because they will provide a direct comparison between different past-light cones. Compared to traditional methods limited to one past-light cone, for which one needs a model to compare different redshifts, they will bring additional model-independent information and break degeneracies \cite{Corasaniti:2007bg,Alves:2019hrg}. The comparison of all these observables with CMB maps can be used to better test the statistical homogeneity of our universe on the largest observable scales.

The previous list of observables relates to photon geodesics. Gravitational wave observations rely instead on graviton geodesics. In the upcoming decade, they will offer complementary probes of the homogeneous expansion and the dynamics of perturbations~\cite{Schutz:1986gp,Mukherjee:2020hyn,Scelfo:2018sny}, while  differences between photon and graviton geodesics could provide direct evidence for a departure from Einstein gravity~\cite{Mukherjee:2020mha}. There is currently no evidence for such differences, but the tests can be pushed to higher precision. Modified gravity can affect the propagation of gravitational waves in different ways. This can be used as a test for alternative theories to general relativity. For instance, in the case of scalar-tensor gravity, the scalar field acts as a Lorentz breaking medium, possibly inducing refraction, dispersion, absorption and polarization mixing (see e.g.~\cite{Ezquiaga:2018btd,Tambalo:2020nvg}). These effects can be tested on a wide range of wavelengths,  from the sub-kilometer (with ground-based interferometry) up to the parsec scale (with pulsar timing array). In the future it will be very important to develop predictions for different types of models, sources and detectors.

Beyond measuring the gravitational clustering properties of DM, and possibly of DE, LSS data can be used to test the fundamental symmetries assumed by General Relativity (like translational and rotational invariance or the equivalence principle), which lead to specific properties of cosmological correlators \cite{Peloso:2013zw,Kehagias:2013yd,Creminelli:2013mca}. The violation of one of these properties can be tested in simulations or in future LSS data (see e.g.~\cite{Creminelli:2013nua,Esposito:2019jkb,Crisostomi:2019vhj}). 

On smaller scales, beyond the traditional tests of Einstein gravity on solar system scales and terrestrial scales, several astrophysical tests can be used to probe specific MG paradigms, like screening mechanisms or violations of Lorentz invariance.

\subsubsection{What are the actual observational prospects?}

In order to measure the background expansion with very high accuracy (and test the level of isotropy of this expansion), there are prospects to improve on current techniques (e.g Baryon Acoustic Oscillations and/or Supernovae luminosity from the Dark Energy Spectroscopic Instrument (DESI), Euclid, the Rubin Observatory, the Roman Space Telescope and the Square Kilometer  Array Observatory (SKAO)). New techniques are also emerging. The Extremely Large telescope (ELT) will measure quasar absorption systems precisely enough for directly detecting their drift over about $10$ years between $z\sim1.9$ and 4.5 \cite{Liske:2008ph,Alves:2019hrg,2020NatAs...4..603P,2020MNRAS.493.3997M}. Future radioastronomy observations (from e.g. the SKAO, CHIME, HIRAX) can also infer the drifts indirectly from intensity mapping \cite{Alves:2019hrg}. If the $\Lambda$CDM paradigm is correct, these instruments should be able to detect the switch between a positive and negative drift around $z\simeq 2.1$. GW telescopes like LIGO-Virgo-KAGRA, LISA or the Einstein Telescope (ET) will provide a wealth of information relevant for cosmology, either together with the detection of electromagnetic counterparts, or in combination with other cosmological data. Indeed, they will provide a completely independent Hubble diagram based on the luminosity distance to standard sirens, provide a comparison between the propagation of photons and GWs, and constrain the gravitational landscape \cite{LISACosmologyWorkingGroup:2019mwx,Mukherjee:2020mha}.

The clustering properties of matter, the possible existence of DE fluctuations and the consistency conditions implied by fundamental symmetries will be tested at a high precision level by future galaxy surveys and cosmic shear surveys such as DESI, Euclid, the Rubin Observatory, the Roman Space Telescope or SKAO. Future CMB experiments (Simons Observatory, CMB Stage 4, Litebird) will also improve the measurement of CMB lensing and help to break degeneracies in the analysis of LSS surveys. Thanks to multi-messenger cosmology, cross-correlations of these surveys with data from GW telescopes will also probe dark matter fluctuations and relativistic effects from inhomogeneities along the line of sight \cite{Garoffolo:2020vtd,Scelfo:2021fqe}. GW telescopes will also offer a new  window on cosmological perturbations through the observation of anisotropies in the stochastic GW background, which could play the role of a new CMB~\cite{Ricciardone:2021kel,Cusin:2018rsq}. All these studies are at their infancy, and still at the forecast level, but the probes bear great promise and represent one obvious focus for the next decade.

Some classes of modified gravity models can be discriminated using specific tests. We have already mentioned promising laboratory tests. Additionally, evidence for screening mechanisms could arise from the observation of an asymmetry between the leading and trailing streams of tidally disrupted dwarf galaxies in the Milky Way halo, accessible to the GAIA satellite (see e.g. \cite{Naik:2020uby}). Screening mechanisms would also affect the splashback radius of DM halos, which will be measured with high accuracy by future galaxy surveys around cluster halos (see e.g. \cite{Adhikari:2018izo,Contigiani:2018hbn}). MG models violating Lorentz invariance also call for specific tests like, for instance, strong gravity tests around black holes to be carried by the ELT~\cite{Johannsen:2015mdd}. 

\subsection{Numerical and Statistical Aspects}

Having reviewed where we stand from the theory point of view and what the most promising observables are, we should discuss the tools needed to exploit them.

\begin{figure}[t!]
\centering
\includegraphics[width=0.9\textwidth]{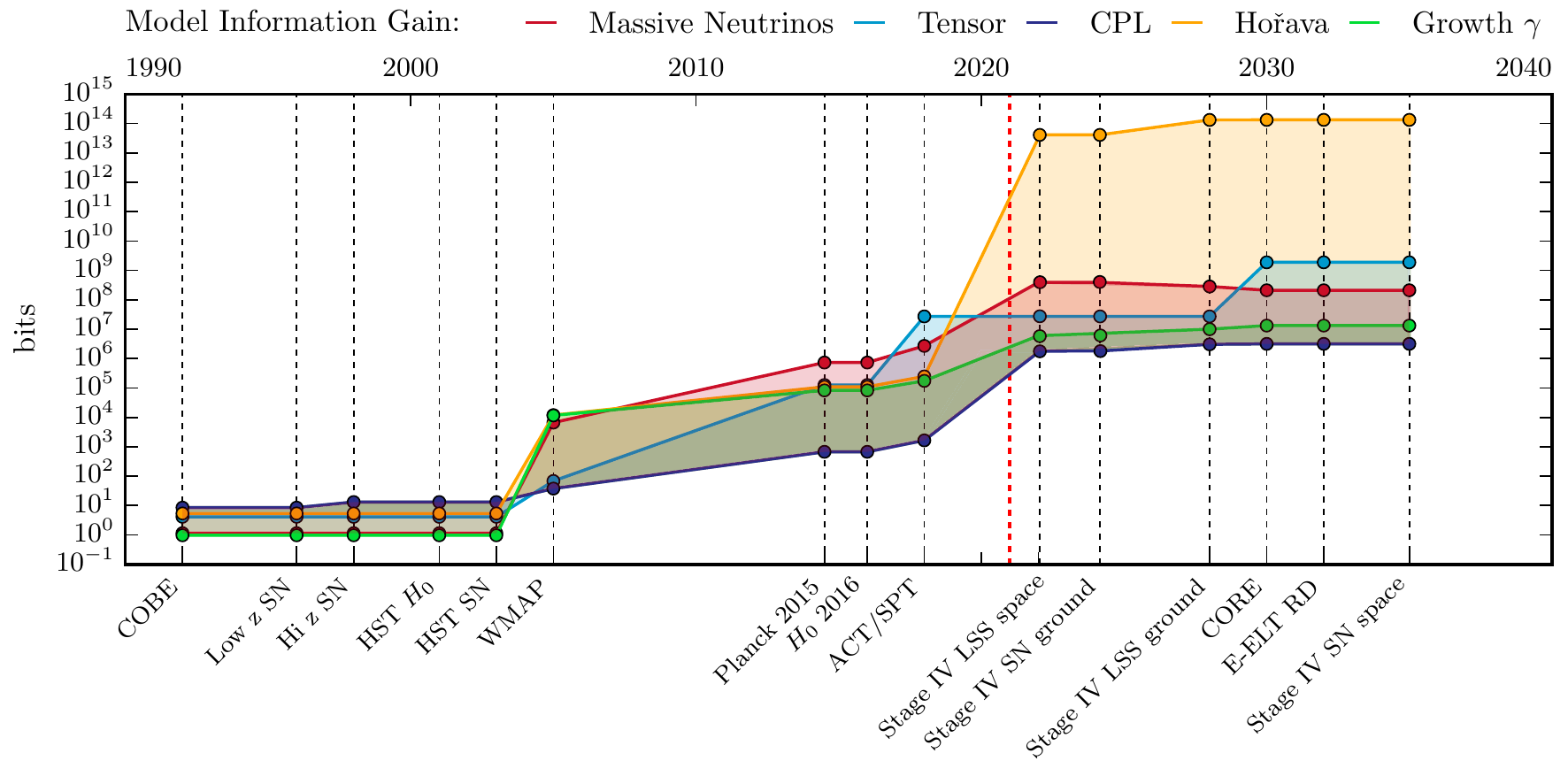}
\vspace*{-0.5cm}
\caption{Model cumulative Information Gain. Different lines and colors show the cumulative information gained on the parameters defining extensions of the $\Lambda$CDM model, as explained in the legend. The red dashed curve indicates the present time. Adapted from \cite{Raveri:2016xof}.}
\label{fig_information_gain}
\end{figure}

\noindent{\bf Linear regime -} In the last five years or so, some very important progress has been made in the development of numerical tools that allow a precise and efficient calculation of observables related to the background cosmology and linear perturbations for large classes of DE/MG theories: fluid-based description of DE in \texttt{CAMB}~\cite{Lewis:1999bs} and \texttt{Class}~\cite{Lesgourgues:2011re,Blas:2011rf}, scalar-field-based description of DE in \texttt{Class}, phenomenological parametrisation $(w,\mu, \Sigma)$ in \texttt{MGCAMB} \cite{Zhao:2008bn}, EFT of DE in \texttt{EFTCAMB}~\cite{Hu:2013twa} and \texttt{HiClass}~\cite{Zumalacarregui:2016pph}. These tools are now mature, have been validated to sub-percent level and will allow a significant step forward in the ability to constrain gravity and DE with new data in the next decade \cite{Bellini:2017avd}. Nevertheless, important developments should be carried over upcoming years, e.g. extending the codes to fully exploit GW probes (including theory-dependent relativistic corrections and the cross-correlation of these probes with LSS and CMB lensing), and including further conditions of  stability in the form of theoretical priors.\\
\noindent{\bf Non-linear regime -} N-body codes for the nonlinear structure formation have been developed for a range of alternative gravity models (see \cite{Winther:2015wla,Valogiannis_2017efficient,Li2018simulating} and references therein). These simulations have focused on specific modified gravity models. There are a number of  N-body codes that have been developed  for parametrized modifications of gravity, e.g. MG-evolution~\cite{hassani2020nbody} and more \cite{brax2012mg_screened,srinivasan2021mg_nbodysim,Lombriser_2016parametrisation}, or for parametrized DE perturbations including relativistic corrections, like in \cite{Hassani:2019lmy,Hassani:2019wed}. 
For parameter estimation analyses with observational data, emulator approaches for the nonlinear matter power spectrum of the Hu-Sawicki f(R) (n=1) gravity model and Jordan-Brans-Dicke gravity have been developed in~\cite{Winther2019emulators,ramachandra2020matter,joudaki2020testing}. References~\cite{Cataneo_2019ontheroad,Bose_2020ontheroad} developed the ReACT code for generic MG/DE models, modeling the reaction of a $\Lambda$CDM power spectrum to new physics using higher-order perturbation theory, the spherical collapse model, and the halo model. The community should maintain its efforts in trying to incorporate more and more DE/MG models in N-body codes, either explicitly like in previous references, or, whenever possible, with generic and economic approaches based on a post-processing of the results of standard model simulations~\cite{Brando:2020ouk,Brando:2021jga}.\\ 
\noindent{\bf Statistical aspects -} Upcoming surveys will deliver a wealth of high-precision data. Making sense of these data requires advanced statistical techniques. Moreover, as these cosmological experiments will hit the limit imposed by the
fact that we can only observe one Universe, it is 
more important than ever to maximize the physical information that can be extracted from the data. In the context of DE/MG, particular attention needs to be paid to possible tensions, while model selection is complicated by the large number of theories and parameters. To this extent, further efforts should go into the development of advanced statistical tools to assess the concordance among data sets, detect and quantify tensions, identify residual systematics and, eventually,  single out physical interpretations.\\

In a summary, while the past decade has been about constraining with ever increasing precision the parameters of $\Lambda$CDM, and incidentally unveiling some tensions, the next decade will be all about discerning among various extensions of the minimal $\Lambda$CDM model, while finally tackling head-on long-standing questions about the nature of DE and gravity on cosmological scales, as Fig.\ref{fig_information_gain} highlights.

\section{Astrostatistics}
{\bf Coordinators:} Christoph Weniger and Roberto Trotta.\\
 {\bf Contributors:} David Benisty, Alex Cole, Adam Coogan, Thomas D. P. Edwards, Stephen M. Feeney, Natalie B. Hogg, Konstantin Karchev, Bradley J. Kavanagh, Michael Korsmeier, Silvia Manconi, Jurgen Mifsud, Kathrin Nippel, Elena Sellentin, Roberto Trotta, Cora Uhlemann, Christoph Weniger.
 \\
\label{sec:astrostatistics}

\subsection{New data will bring new challenges}

Scientific progress is based on comparing theoretical models with data.
A myriad of statistical methods are available for this task, and although seldom in the limelight they provide the bridge between theory and observation, and ultimately determine what we consider scientifically established truth.
An inefficient or incorrect use of statistics and data analysis may lead to weaker or entirely incorrect conclusions.
Given the ever-increasing complexity of astroparticle physics observations, the scientific return of many upcoming observations and experiments is expected to be limited by the efficiency and sophistication of our statistical inference tools~\cite{Trotta2017}.
We will here review recent developments with focus on astroparticle physics applications (for detailed reviews see~\cite{Algeri:2018zph, 2021AnRSA...8..493F}).
The selected topics cover only a small part of the current statistical landscape, and reflect what we believe to be most relevant to the astroparticle physics and cosmology communities.

\begin{figure}[h]
	\centering
	\includegraphics[width=1\textwidth]{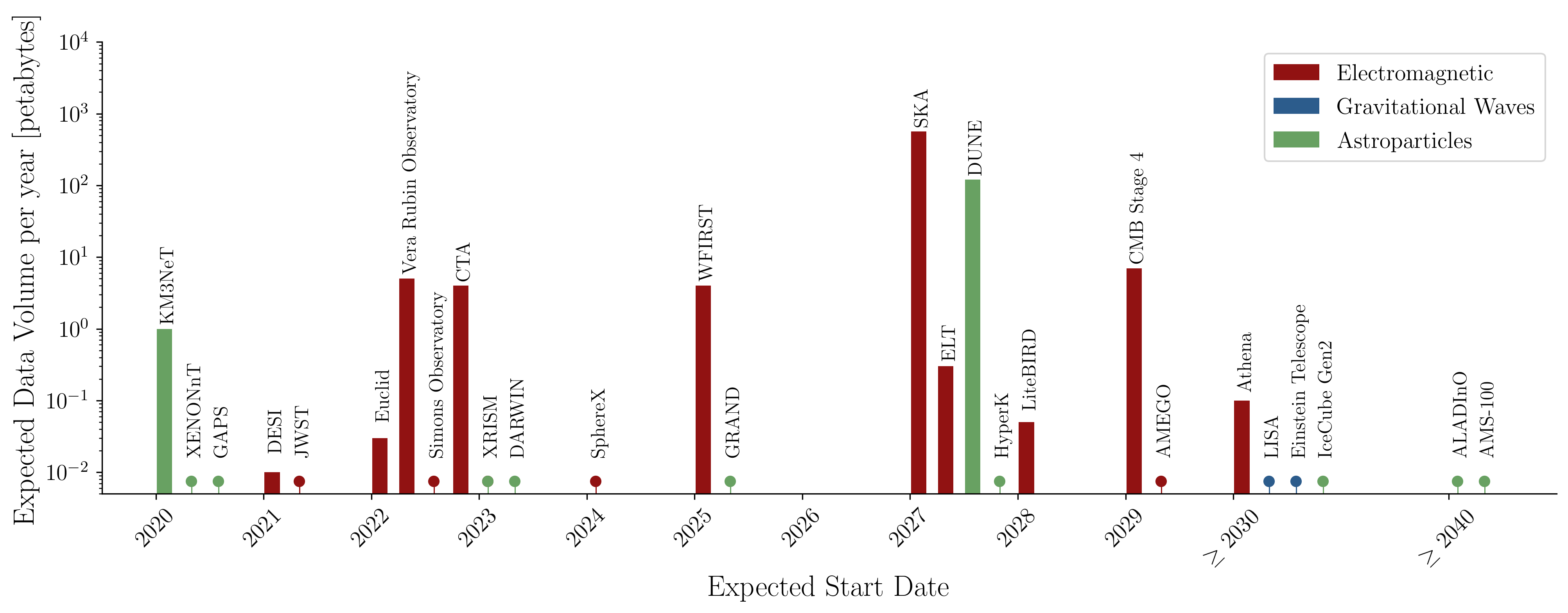}
	\caption{
	This plot shows rough estimates for the minimum expected volume of data per year in petabytes for a range of current and upcoming experiments and surveys in astroparticle physics and cosmology. 
	Note that both the data volume and start dates are indicative; the plot is intended to give a general overview of the current and future data analysis challenges of the field.}
	\label{fig:data_volume}
\end{figure}

Astroparticle physics data is rapidly increasing both in volume and precision, as we can see from Fig.~\ref{fig:data_volume}, showing the minimum volume of data per year expected to be produced by a range of upcoming surveys and experiments. 
In order to infer underlying physical processes and to draw scientific conclusions, an abundance of statistical tools and physical simulation codes have been developed within the community.\footnote{\label{softwarelist}An extensive list of the tools used in astrophysics, cosmology and high energy physics can be found here: \url{https://github.com/nikosarcevic/HEP-ASTRO-COSMO}. }
The huge volumes and high complexity of these data sets demand higher realism and fidelity of physical models and simulators.   This leads to an increase of both the computational cost of simulators and the number of model parameters.
For example, simulating the propagation of cosmic rays for a model with $\sim$ 10 parameters can take from $\mathcal{O}$(minutes) in 2D and up to $\mathcal{O}$(hours) in a more detailed, 3D setup.
In many problems, well-established algorithms like likelihood-based inference with MCMC or nested sampling~\cite{Skilling2004}
or likelihood-free methods like Approximate Bayesian Computation (ABC) would require a very large number of simulator runs, which is computationally impractical or even unfeasible.
This development takes established inference algorithms of the kind described in Sec.~\ref{sec:LikelihoodBase} -- which generally do not scale favorably with model complexity -- to their limits, and is often the most constraining factor for progress.

Statistical inference tools that employ a much smaller number of simulations, or can work with faster, approximated simulations are therefore becoming of paramount importance.  In recent years, new classes of scalable and simulator-efficient inference and search algorithms have been enabled by breakthroughs in deep learning, which could play a major role in the successful analysis of upcoming astroparticle physics data.
A major remaining hurdle for scientific applications is a robust uncertainty quantification for deep neural networks (NN) (see Ref.~\cite{2020arXiv200410710C}), which however becomes feasible for a subset of emerging techniques (see \fex Sec.~\ref{sec:NeuralSimulatorBasedInference}).
Improved computational efficiency is also imperative from an environmental point of view: slow and wasteful simulations running on high performance computing clusters can be as damaging to the environment as commercial aviation \cite{Avgerinou2017,Stevens:2019ntb, PortegiesZwart:2020pdu}.
This impact can be mitigated by improving computer hardware, cluster cooling mechanisms and the re-use of generated heat \cite{Auweter2011}, as well as the calculation and offsetting of the carbon footprint (as mentioned in e.g. \cite{Lepori:2021lck}).
Additionally, faster and more efficient statistical tools, and methods for recycling and re-using expensive computer simulations, must here play a central role. 

The rest of this section is structured as follows:  Sec.~\ref{sec:common_task} summarizes common statistical analysis tasks, and Sec.~\ref{sec:current_developments} current developments; Sec.~\ref{sec:best_practices} discusses best practices, and in Sec.~\ref{sec:conclusions} we conclude. Due to constraints on the number of citations in this white paper we aimed at citing a representative sample of papers.

\subsection{Common statistical analysis tasks}\label{sec:common_task}

\subsubsection{Inference \& the inverse problem}
A common situation is to have a parametric physical simulator that maps physical input parameters $\btheta$ onto simulated observations $\bx$.
Simulators can be deterministic (\fex~cosmic-ray propagation in the Milky Way), or feature stochastic states $\bz$ (\fex initial conditions of N-body simulations, or in general a layer of latent variables in a hierarchical model context).
A common problem (often referred to as the ``inverse problem''~\citep[\fex][]{Trotta2017, Innes2019}) is to ask ``which range of physical parameters $\btheta$ is the most probable given my data''? 
This is answered in a Bayesian context by determining the posterior probability distribution (pdf) $p(\btheta|\bx)$, which encapsulates our final state of knowledge about parameters $\btheta$ starting from the prior $p(\btheta)$ and including information from the data $\bx$ via the likelihood. The posterior is
obtained via Bayes' theorem as
\begin{equation}
\label{eqn:post}
    p(\btheta | \bx) = \frac{\int d\bz\, p(\bx|\btheta, \bz) p(\bz|\btheta)  p(\btheta)}{p(\bx)}\;.
\end{equation}
Here, $p(\bx|\btheta, \bz)$ is the data likelihood which can also depend on stochastic states $\bz$, $p(\bz|\btheta)$ is the prior on the stochastic states, and $p(\bx)$ the Bayesian evidence or model likelihood.
We refer to $p(\bx, \btheta, \bz) = p(\bx|\btheta, \bz) p(\bz|\btheta) p(\btheta)$ as joint distribution. 
Challenges in evaluating Eq.~\eqref{eqn:post} are related to the large dimensionality of $\bz$ (\fex for image data) or computationally expensive simulator runs for likelihood evaluation (\fex N-body simulations).
While Bayesian and frequentist inference agrees in a limited subset of cases (notably, Gaussian likelihood and uniform priors), there is no generally agreed solution to reconciling inferences based on marginal posteriors with those based on profile likelihood intervals in the cases where they disagree~(e.g., see \cite{RN128}). 

\subsubsection{Searches \& robust detection}
Searches for new physics or astrophysical events and objects often proceed by constructing a background and signal model, followed by performing a likelihood ratio hypothesis test between the null (background only) and alternative (background + signal) hypotheses.
Detection criteria vary between domains.
For instance, particle physics experiments typically take a frequentist approach and require the $p$-value (the probability under the null hypothesis of obtaining data as extreme or more extreme than observed through chance alone) to cross a predetermined threshold (see Sec.~\ref{sec:best_practices}  for common pitfalls, and Ref.~\cite{Cowan:2010js} for commonly used asymptotic distributions).  Upper limits are often based on the frequentist-inspired CLs method~\cite{Read:2002hq}.
On the other hand, gravitational wave searches use a combination of the false positive rate (frequentist) and the probability of being of astrophysical origin (Bayesian) to determine an event's significance~\cite{LIGOScientific:2018mvr}.
Signal searches become computationally very expensive when the search space becomes large. 
In the case of gravitational wave searches, waveform models typically contain at least four parameters, and robust searches requires a grid of $\mathcal{O}(10^5-10^6)$ template points~\cite{2021PhRvD.103h4047M}.

\subsubsection{Combination of data \& global scans}
Drawing conclusions in astroparticle physics  often involves a combined analysis of multiple data sets: observations at multiple wavelengths, scales, redshifts, different messengers, and/or multiple laboratory results.
Typical examples are here cosmological parameter fits~\cite[\fex][]{Planck:2018vyg} or global scans for new physics scenarios~\cite[\fex][]{GAMBIT:2017yxo}.
This brings multiple challenges:
Often no likelihood is published, which makes it impossible to perform a joint analysis in a statistically principled way~\cite{RN120}.
Published experimental results typically rest on underlying model assumptions for the signal, thus introducing difficulties in recasting them for different physical models.
Furthermore, overlapping signal regions can complicate a sound statistical combination of measurements.
Lastly, combining observations featuring each many nuisance parameters increases the overall parameter count of a model, leading to challenges for established inference algorithms.
Many of these problems can be addressed by end-to-end forward modeling, which we discuss in sections~\ref{sec:VariationalInference} and~\ref{sec:NeuralSimulatorBasedInference}.

\subsection{Current developments}
\label{sec:current_developments}

\begin{figure}[t]
    \centering
    \includegraphics[width=0.72\linewidth]{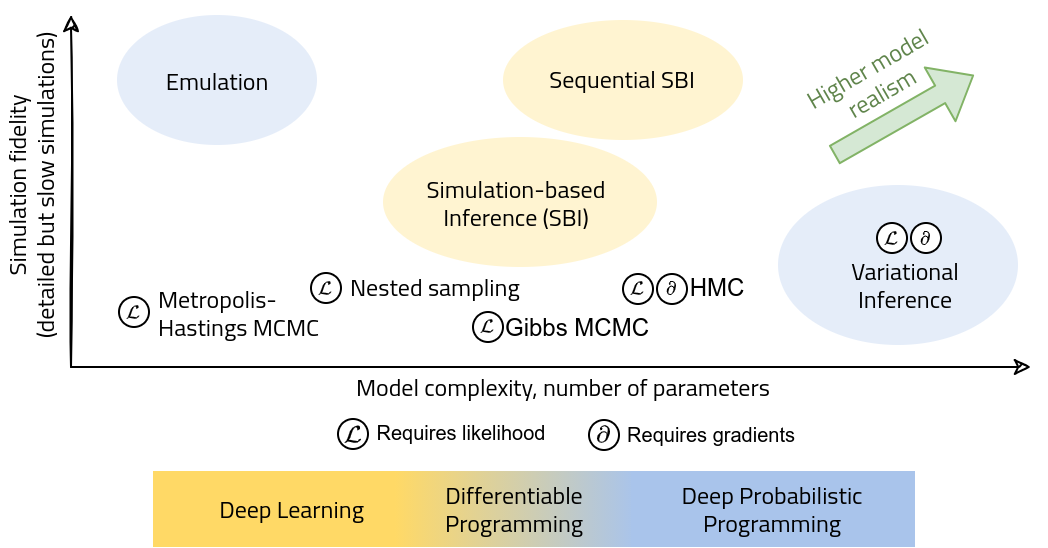}
    \caption{Illustration of modern statistical techniques discussed in this white paper section. Variational inference (VI, Sec.~\ref{sec:VariationalInference}) allows inference of very large numbers of model parameters, and requires gradient calculations and explicit likelihoods. Simulation-based inference (SBI, Sec.~\ref{sec:NeuralSimulatorBasedInference}) works in a likelihood-free setting, and in particular sequential versions are very simulation efficient.  Emulation (Sec.~\ref{sec:emulation}) can efficiently substitute highly realistic but slow simulators. Driving computational technologies for these new methods are deep learning, differentiable and deep probabilistic programming.}
    \label{fig:overview}
\end{figure}

In this section, we highlight a range of new developments in statistics, machine learning and computer science that we consider promising for solving the most pressing data analysis challenges in the field of astroparticle physics and cosmology, see Fig.~\ref{fig:overview}.

\subsubsection{Established likelihood-based methods}
\label{sec:LikelihoodBase}
In a Bayesian context, the posterior Eq.~\eqref{eqn:post} is typically approximated with samples drawn via Markov Chain Monte Carlo (MCMC) methods or nested sampling. 
The main challenge is to ensure convergence of the sampling to the posterior distribution, especially in high dimensional parameter spaces, for multi-modal posteriors and/or curving degeneracies. 
The simplest MCMC algorithms (such as Metropolis-Hastings) have now largely been superseded by more powerful approaches (Hamiltonian Monte Carlo~\cite{RN555},  affine-invariant ensembles,
particle filters, Gibbs sampling~\cite{RN534}). 
Nested sampling is particularly effective for multi-modal distributions, with MultiNest~\cite{RN595} having become the {\em de facto} standard approach for parameter spaces of up to 30 dimensions, past which its efficiency plummets due to the curse of dimensionality.
Other nested sampling approaches exist that scale more favourably with the dimensionality of the parameter space (such as PolyChord~\cite{2015MNRAS.450L..61H}). 
In a frequentist setting, inference is based on the maximum likelihood value. Maximising a function, especially in high dimensions, is a difficult task.  Posterior samplers can be adapted to it, but are not designed for this task. Optimizers exist that use many strategies to try and locate the global maximum 
(genetic algorithms, local optimizers, annealing schemes~\cite[\fex][]{DarkMachinesHighDimensionalSamplingGroup:2021wkt}) and to map the likelihood around it to obtain confidence regions.

\subsubsection{Variational inference and differentiable simulators}
\label{sec:VariationalInference}
Stochastic variational inference (VI) \citep{svi,zhang2018advances} transcends established sampling methods (see Sec.~\ref{sec:LikelihoodBase}) and recasts inference of the posterior $p(\btheta | \bx)$ as an optimization problem that scales to high-dimensional models, for which exact inference is usually intractable. The target of optimization is the evidence lower bound (ELBO),
\begin{equation}\label{eqn:elbo}
    \operatorname{ELBO}[q_\phi] \equiv \mathbb{E}_{\btheta \sim q_\phi(\btheta)} \left[ \log p_\eta(\bx, \btheta) - \log q_\phi(\btheta) \right] \, ,
\end{equation}
where the model may depend on hyperparameters $\eta$. The ELBO can also be written as the sum of the model evidence $p_\eta(\bx)$ and the reverse Kullback-Leibler (KL) divergence (a non-negative quasi-distance measure) between the true posterior $p_\eta(\btheta | \bx)$ and a \emph{variational proposal} $q_\phi(\btheta)$, parametrized by $\phi$. ELBO maximization simultaneously recovers the best approximate posterior (within the assumed parametrization) and approximately optimizes the model hyperparameters.
The proposal distribution is usually of a simple analytic type, often assumed to fully factorize across the components of $\btheta$ (the so-called mean-field approximation).
In combination with the mode-seeking nature of the reverse KL divergence \citep{DBLP:journals/corr/Goodfellow17}, this leads to a tendency of VI to underestimate the posterior's variance,
and the optimal hyperparameters may be strongly biased (see \fex \cite{karchev2021stronglensing}). More flexible proposals
\citep{vi_cascading_flows}, and/or a modified objective function based on the R\'enyi $\alpha$ divergence \citep{renyivi} can help mitigate these issues.

The ELBO is most easily optimized with respect to the parameters $\phi$ by using stochastic gradient descent and the reparametrization trick \citep{kingma2014autoencoding,rezende2014stochastic} to differentiate the expectation in Eq.~\eqref{eqn:elbo}.
Critically, this requires the simulator
to be differentiable with respect to $\btheta$. The automatic-differentiation capabilities of machine learning frameworks enable full automation of VI \citep{kucukelbir2016automatic}, making it possible to obtain realistic approximate posteriors even for models with hundreds of millions of parameters and tera-scale datasets \citep{vi_astro_catalogs} at comparable computational costs to maximum {\em a posteriori} estimation. A less efficient but more flexible technique called black-box variational inference \citep{bbvi} can be applied when the variational proposal is non-reparametrizable.
VI has so far been employed in only a few astrophysics works, including deblending starfields \citep{vi_starfields}, mapping gas, dust and diffuse gamma-ray emission \citep{vi_dust,vi_gc_gammarays}, strong gravitational lensing \citep{coogan2020targeted,karchev2021stronglensing}, and cataloging and generating astronomical images \citep{vi_astro_catalogs}. Examples of differentiable astrophysical simulators are catalogued below in our recommendations to theorists (Sec.~\ref{sec:conclusions}).  Differentiable and deep probabilistic programming languages (see~\cite[\fex][]{Innes2019}) significantly ease the implementation of physics models in a VI framework.

\subsubsection{Neural simulator-based inference}
\label{sec:NeuralSimulatorBasedInference}
Often the likelihood factor in Eq.~\eqref{eqn:post} is extremely hard to compute, \fex due to the need to marginalize numerous unobserved parameters or non-trivial instrumental effects (examples are N-body simulations, or observations affected by selection effects). 
Simulation-based inference (SBI) circumvents evaluating the likelihood 
by instead relying on a complete simulation of the relevant physical, statistical, and instrumental effects.
The prevalent approach is Approximate Bayesian Calculation (ABC) 
\citep{Sisson_Fan_Beaumont_2020},
which however requires a large number of examples and often domain-motivated summarizing and comparison procedures.

\textit{Neural SBI methods} (see \cite{Cranmer2019frontiers,benchmarking_sbi} for details and a deeper literature overview)
improve upon ABC on two fronts: by automating feature extraction and accelerating inference via neural networks (NN). 
Provided with pairs of data and parameters of interest, a NN can be trained to marginalize nuisance parameters and produce either directly the posterior, the likelihood, or a ratio of probability densities: the likelihoods of different parameters  or the likelihood and the evidence.
Likelihood and posterior estimation are unsupervised tasks, which benefit from advances in neural density estimation \citep{Papamakarios2021normalizingflows,Bond-Taylor2021nde}
and allow subsequent sampling from the respective distributions.
In contrast, ratio estimation solves a supervised classification task, which enables directly evaluating marginal posteriors, while providing principled diagnostics to judge the convergence of training.
All three approaches allow using the full spectrum of NN architectures (\fex convolutional NNs) for pre-processing the data and can thus be thought of as automatically extracting flexible relevant summaries. Moreover, the learning is \emph{amortized}: after an up-front cost of training, posteriors can be obtained for arbitrary new data without re-training, allowing efficient application to multiple i.i.d.\ observations (\fex relevant for GW parameter estimation \citep{Delaunoy2020lightning}). Furthermore, the number of simulations required for training \textit{all} posteriors (through amortization) can be significantly smaller than the required simulations for established sampling methods to obtain \textit{one} posterior estimate \citep{Alsing2019delfi, Miller2020swyft}. Neural SBI methods have found application in DM substructure inference (see \citep{Hermans2020streams,coogan2020targeted} and references therein),
gravitational wave and standard siren analysis (see \citep{Delaunoy2020lightning,Gerardi2021unbiased} and references therein),
and cosmology and related fields \citep{Huppenkothen2021xray,Rouhiainen2021fields,Zhang2021microlensing}.
Further work is needed to estimate (and mitigate) the impact of mis-specifying the data model in SBI analyses and to verify SBI-derived posteriors (though note that diagnostics are available for some variants \citep[\fex][]{nre_Hermans2020}). Use in combination with likelihood-based techniques (\fex VI: see Sec.~\ref{sec:VariationalInference} and \cite{coogan2020targeted}) may address this issue.

\textit{Active learning} (AL)
is a strategy to optimize the use of a simulation budget and perform inference from limited data more efficiently by discovering and targeting informative regions of parameter space at the expense of reusability of training. Sequential versions of the neural SBI algorithms (refer to \cite{benchmarking_sbi})
achieve this by using the current estimate of the posterior (either alone or in combination with the prior \citep{alsing2018optimal}) to target regions in agreement with the data.
An alternative is the Bayesian-optimization approach that takes into account the \emph{uncertainty} of the current posterior estimate (evaluated \fex via a Bayesian NN or from an ensemble of estimators) and targets parameters that are expected to improve it most \citep{
jarvenpaa2019acquisition,lueckerman2018}.
Examples of AL can be found in \cite{Alsing2019delfi, LSST_AL2020}.

\subsubsection{Emulation}
\label{sec:emulation}
Likelihood- and simulation-based inference require repeated, accurate modeling of the data or its statistics. As the computational cost of modeling increases, {\em emulation} -- training flexible models capable of rapidly but accurately approximating the necessary inputs -- becomes increasingly attractive. The most obvious use case for emulation is when the forward model is too slow to calculate more than a handful of times. It is, however, also impactful when the cost per model evaluation is low, but the number of evaluations required is high, for example, in population analyses or when augmenting training sets.

At its heart, emulation is a problem of interpolation. The varying complexity and number of physical simulations available in different settings motivate a range of emulation algorithms, including Gaussian Processes~\cite[\fex][]{Moore_2014, Rogers_2019, McClintock_2019, Ho:2021tem}, 
deep neural networks~\cite[\fex][]{chen2020learning, Khan_2021}, 
variational autoencoders~\cite[\fex][]{Chianese_2020,Lanusse_2021},
and generative adversarial networks~\cite[\fex][]{Mishra2019cmbgan, Mustafa_2019}. 
Astroparticle applications abound, including a variety of nonlinear power spectra~\cite[\fex][]{Rogers_2019, Ho:2021tem}, 
large-scale structure maps~\cite[\fex][]{Mustafa_2019, chen2020learning} 
and their summaries~\cite[\fex][]{McClintock_2019},
galaxy images~\cite[\fex][]{Chianese_2020,Lanusse_2021},
and gravitational waveforms~\cite[\fex][]{Moore_2014, Khan_2021}. 
Optimizing these emulators, through, \fex Bayesian optimization~\cite{Rogers_2019} or coupling of slow and fast simulations~\cite{Ho:2021tem, Chartier_2021}, is a key ongoing avenue of research. Understanding how to assign uncertainties on the emulated output is also crucial: see, \fex recent developments in the LHC context \cite{bellagente2021understanding}.

\subsection{Best practices}\label{sec:best_practices}

A sound statistical methodology requires following a principled approach to inference, and for example avoids looking exclusively for confirming evidence to a hypothesis; using an incorrect or over-simplified sampling distribution; ignoring important observational effects affecting the data (\fex selection effects) or unrepresentative training data or simulations in machine learning. In all of these circumstances and many others, best practice helps achieve sounds and robust results, that are more likely to survive scrutiny and be corroborated by larger or more precise future data. We give a few examples in the following.    

\subsubsection{Searches}
From a frequentist perspective, statistical significance is often quantified in terms of a $p$-value (see Sec.~\ref{sec:common_task}). 
However, the $p$-value is often mis-interpreted as reflecting the probability that a positive result is a false positive (the false positive risk)~\cite{Colquhoun2017}, a question which can only be addressed in a Bayesian framework. 
A more technical complication is in actually calculating a $p$-value. This is often done using a likelihood-ratio test  statistic (LRTS), for which the sampling distribution (and therefore $p$-value) can be determined from Wilks' Theorem~\cite{Wilks:1938}, under certain regularity conditions. However, there are a number of scenarios where these regularity conditions break down~\cite{Algeri:2019arh}. For example, if the parameter of interest lies on the boundary of the parameter space (\fex positive signal strength parameters, $\mu \geq 0$), then the LRTS 
typically leads to an overestimate of the $p$-value if one incorrectly uses Wilks' theorem~\cite{Chernoff:1954eli}. 
Wilks' Theorem also requires that the hypotheses under comparison should be nested (i.e.~that one model can be obtained by restricting to a subset of parameters in the other),
though a number of techniques for calculating $p$-values for non-nested scenarios have been developed~\cite[\fex][]{2017arXiv170106820A}.
A related issue is the ``look-elsewhere effect'' (LEE). In a situation where multiple hypothesis tests are carried out, there is a high probability that spurious detections with seemingly high significance will occur. Quoting the `local' $p$-value (which ignores multiple testing) may lead to a much-increased false positive rate. 
Strategies for dealing with the LEE~\cite[\fex][]{Bayer:2020pva}
are receiving increased attention in the context of astrophysics~\cite{Algeri:2016gtj}.

\subsubsection{Bayesian model comparison}
Many of the above problems can be addressed by taking a Bayesian approach, which favours model comparison over hypothesis testing~\cite{RN169}. The central quantity in Bayesian model comparison is the Bayes factor, the ratio of the evidences of the two models being compared. The Bayes factor updates prior odds between the models to their posterior odds in light of the data. It has a more straightforward interpretation than $p$-values, it automatically accounts for multiple testing, and is free from e.g. stopping rule pathologies~\cite{RN280}. However, Bayes factors suffer from an asymptotic dependence on the model parameters' prior, which controls the strength of the Occam's razor factor penalizing needlessly `complex' models~\cite{RN176}. 

\subsubsection{Unrepresentative data}
Astroparticle data often suffer from selection effects: a common example is when the probability of an object being observed is a function of its apparent magnitude (called `Malmquist bias' in astronomy). In statistical terms, the observed sample is non-representative of the population of objects whose parameters one might wish to infer. In a supervised learning setting, this translates into poor generalization onto the test from a non-representative training set. So-called `covariate shift' occurs when the distribution of labels conditional on the covariates is the same between source and target domain, but the unconditional distribution of covariates differs. Ignoring covariate shift leads to incorrect inferences on physical properties, and affects the robustness of the results (\fex for population studies for SNIa classification~\cite{2019NatAs...3..680I}, photon sources \cite{Luo:2020bbk}, or gravitational waves \cite{Veske:2021qis}).
Possible solutions include correcting for selection effects using simulations \cite[\fex][]{RN336},  data augmentation  \cite[\fex][]{2020arXiv201212122C}, and unsupervised learning~ \cite[\fex][]{Finke:2020nrx}. Recently, a powerful and general solution has been proposed, which uses propensity scores stratification to obtain approximate balanced groups across covariates~\cite{2021arXiv210611211A}. 

\subsubsection{Data compression}
The use of data compression and summary statistics is prevalent both in cosmology~\cite[\fex][]{Gualdi:2017iey} and particle physics~\cite[\fex][]{Brehmer:2016nyr}. Cosmology in particular traditionally only analyses low-order correlation functions~\cite{Bernardeau:2001qr}, because often these carry most of the information, but also because model predictions are either unavailable or unreliable on small scales (where both non-linear gravity and baryonic effects become important). 
Computational advances and a simulation-based approach allow for the construction of sophisticated nonperturbative summary statistics both by hand \cite[\fex][]{Biagetti:2020skr} and via neural compression schemes \cite{2018PhRvD..97h3004C}. Similar methods, along with more accurate physical modeling, will be essential in the coming years in order to extract the full scientific information from data sets whose volumes and complexity will increase dramatically.

\subsection{Recommendations}
\label{sec:conclusions}

\subsubsection{For theorists}
Modern statistical methods are very promising for addressing many analysis challenges in current and upcoming data. 
In particular, they enable the use of realistic but slow physical simulators, and can handle very large numbers of nuisance parameters and uncertainties. This paves the way for confronting upcoming data with much more realistic simulations.  
However, in order to realize fully the potential of many modern methods, a shift towards simulation-based inference, and the development of the necessary computational and educational infrastructure are necessary.
Furthermore, some modern methods (variational inference, Hamiltonian Monte Carlo) are only applicable for end-to-end differentiable physical simulators. The immense advantages that this can bring remains underappreciated, although pioneering examples for various different applications do exist (stellar evolution \cite{Jermyn_2021}, gravitational lensing \cite[\fex][]{bohm2020madlens}, light curves and spectra for exoplanet research \cite{Morvan_2021}, $N$-body simulations and modeling dark matter halo assembly \cite[\fex][]{hearin2021differentiable}, and improving performance of telescopes \cite{Pope_2021}).
Often, this can be achieved with little extra effort through auto-differentiation libraries that are standard in deep learning packages.  
Furthermore, we encourage the community to explore ways to share simulation results in a way that allows simulation reuse; for example, by using random sampling techniques rather than grids, and providing accurate metadata.

\subsubsection{For experimentalists}
As experiments reduce statistical uncertainties while data becomes increasingly detailed and complex, a more refined understanding of the systematic uncertainties of both theoretical models and measurements is essential to achieve robust scientific conclusions.  New inference algorithms that can handle slow, detailed physical simulations with large numbers of model parameters will very likely play a critical role in future breakthroughs in astroparticle physics.
Those algorithms are typically based on forward simulations rather than likelihood evaluations.
In order to fully exploit the high quality of upcoming data, we strongly encourage experimental collaborations to release their instrumental forward simulations together with their data. These \emph{official} simulations should be used for the analysis strategies outlined in Sec.~\ref{sec:NeuralSimulatorBasedInference}.  
We also encourage experimental collaborations to provide more detailed information about systematic uncertainties and all relevant correlations. 
Lastly, simulation-based inference generalizes best when trained on \emph{randomized} parameters rather than on parameter grids. This is especially relevant in the case of cosmological measurements, which have to date focused on producing multiple simulations for a given fiducial cosmology for the purposes of covariance estimation. 

\subsubsection{Developments of standard tools and benchmarks}
Standard tools developed following the recommendations outlined in the previous sections should be open source and community-based efforts.
Standardised tools for solving common problems will enable the definition of a benchmark figure of merit of good statistical properties, see Sec.~\ref{sec:best_practices},  
thereby facilitating the comparisons between different analysis algorithms. 
Data challenges have been demonstrated to drive the rapid development of advanced machine learning and statistical techniques for current and upcoming datasets \cite[\fex][]{plasticc,Metcalf_2019,SKA_data_challenge_1}
(see also\footnote{\url{https://lisa-ldc.lal.in2p3.fr/challenge2}} and LHC/particle physics \cite{darkmachines_lhc}). 
However, these challenges are built around fixed training data and hence focus on an improvement of the inference strategies alone.  In many astrophysics and cosmology applications, higher scientific return can be obtained only by jointly improving physical simulators, models and inference pipelines.
We hence encourage future data \emph{and simulation challenges} to be conducted using the techniques outlined in Sec.~\ref{sec:current_developments}, guided by well-posed questions specific to the physical scenario being studied.
Summer schools and workshops would be ideal settings to host future challenges, as well as bringing other benefits to the field.

\subsubsection{Education and training}
Education and training in astrostatistics would be beneficial for researchers in both cosmology and astroparticle physics. We recommend that, given the prevalence of Bayesian methods in these fields,  courses on introductory Bayesian statistics are offered as standard to undergraduate physics students. 
Summer and winter schools and workshops, such as the \href{http://ada.cosmostat.org/}{ADA series}, the \href{https://astronomy.outreach.psu.edu/}{SCMA series} and the \href{https://cosmostatistics-initiative.org/}{COIN initiative}, which focus solely on astrostatistics, are an excellent way to guarantee that early career researchers are abreast of both the best practices and the newest methods in the field.
We recommend the organization of more such schools, particularly with targeted financial support for participants from low and middle income countries and for those with additional needs such as childcare, and paying attention to ensuring diversity of both lecturers and participants. 
We also recognise that astrostatistics is a field with rich opportunities for collaboration with researchers in computing, statistics and machine learning. We therefore propose that funding bodies should create both new small grants which explicitly support cross-disciplinary collaboration in all of these fields, as well as new fellowships in astrostatistics.

\newpage

\section{Outlook}

Astroparticle Physics is in undergoing a phase of profound transformation. The discoveries made in the past 50 years have radically changed our understanding of the universe, and the next decade promises to be equally exciting, notably thanks to an avalanche of new data from a variety of observational probes. 
As this document demonstrates, a vigorous theoretical effort is essential in order to fulfil the true discovery potential of upcoming experiments. We have in particular assessed here the key opportunities and challenges for 10 sub-fields of Astroparticle physics, and argued that while each sub-field has its own, interdisciplinary and collaborative work is crucial to address the most fundamental questions. 

\begin{itemize}
    \item {\bf Early Universe.} The physics of the very early universe remains an important frontier in cosmology. Key challenges include a systematic classification of inflationary predictions, and the refinement of calculations of non-Gaussian correlations in view of upcoming galaxy surveys. Furthermore it is essential to refine calculations of other probes of the early universe, including reheating, thermal relics, baryogenesis and phase transitions, in order to  understand what information new discoveries can provide on the underlying new physics. 
    \item {\bf Dynamical spacetime.} The direct detection of GWs has forever transformed  astroparticle physics. Future interferometers promise to address and hopefully solve long-standing problems in cosmology, astrophysics, and particle physics. High-precision theoretical predictions yielding more accurate waveform models -- with numerical, perturbative, EFT-based and amplitude-based methodologies -- are crucial to enable new discoveries in astroparticle physics with next-generation GW observatories such as LISA~and~ET.
    \item {\bf Nuclear astrophysics.}  Moving forward the frontier of nuclear astrophysics requires implementing gravitational, strong and electroweak interactions, as well as diverse collective phenomena in unexplored regimes, far from first-principles theory and established knowledge. A strong interdisciplinary effort is crucial to develop sufficiently accurate theoretical models, interpretations, and analysis strategies.
     \item {\bf Cosmic accelerators.} The physics of particle acceleration is often nonlinear, self-regulated and complicated, and covers a vast range of length and energy scales, requiring a variety of numerical and other theoretical methods. Interdisciplinary work will be needed to unlock the potential of multi-messenger observations, as well as close collaboration between scientists and institutions with different expertise.
     \item {\bf Traveling and Interacting Messengers.} Understanding the micro-physics of cosmic rays is essential to understand the long-standing problem of their origin, to interpret upcoming observational data, and to maximize the potential for the discovery of new physics. Global CR-transport simulations are ultimately required, as well as developing new theoretical methods for a first-principle derivation of cosmic ray diffusion.
     \item {\bf Neutrino properties.} Cosmological surveys, neutrinoless double-beta decay experiments, and beta decay endpoint measurements aim to precisely measure the neutrino physics scale. We must be ready to interpret these future results. At the same time, we must prepare to interpret new observations that may help us connect CP violation and lepton number violation with the origin of matter in the Universe, and to explore neutrinos as a window to new physics.
     \item{\bf Particles from Stars.} Stellar interiors are excellent laboratories for particle physics. 
     Major theoretical progress is needed across all scales, from macroscopic to microscopic processes, as well as new techniques to combine the information coming from heterogeneous data-sets and different messengers. We need in particular to improve stellar evolution models throughout evolutionary phases and stellar masses and consistently model the impact of particles on stellar physics. 
     \item{\bf Dark Matter.} The physical nature of dark matter remains a mystery. Continued development of theoretical tools is needed in order to keep up with the ever increasing precision of data. It is crucial to extend current searches to as wide a range of DM models as possible, as well as to map out which combinations of search strategies  yield the best chance of confirming the nature of DM.
     \item{\bf Dark Energy.}  In the past decade we have constrained with ever increasing precision the parameters of the standard cosmological model, and unveiled some tensions. The next decade will be all about discerning among various extensions of standard cosmology, while finally tackling head on long-standing questions about the nature of dark energy and gravity on cosmological scales.
     \item{\bf Astrostatistics.} Given the ever-increasing complexity of astroparticle physics observations, the scientific return of many upcoming observations and experiments is expected to be limited by the efficiency and sophistication of our statistical inference tools. In order to fully embrace the potential of many modern methods, we emphasize the importance to shift towards simulation- and/or gradient-based inference techniques, and to develop adequate computational and educational infrastructures.
\end{itemize}

The organisation of this white paper around research themes, rather than around messengers and experimental collaborations, underlines the importance of tackling upcoming challenges with a broad multi-messenger perspective, and to encourage interdisciplinary research. 
We hope that this document will increase the awareness of theoretical challenges and opportunities in the next decade, and help the community to prepare for the interpretation of upcoming data and to inform the design of future experimental probes. 
Addressing the most fundamental questions of astroparticle physics ultimately will require the collaboration of theorists with different backgrounds and skills, as well as the collaboration with experimentalists, observers, data scientists, and computer scientists. 

EuCAPT's mission is precisely to enable this collaboration, and to coordinate the community-wide effort to push the frontiers of astroparticle physics. To achieve this goal, we recommend the following:

\begin{itemize}
    \item National and international funding bodies should provide adequate support for theoretical efforts. In order to promote truly interdisciplinary work, we recommend in particular the funding of {\bf independent short and long-term positions} that go beyond geographic, thematic, or experimental boundaries.
    \item Adequate {\bf computational resources} should be made available to the international theory community in Europe, so they don't have to rely solely on infrastructure overseas. 
    \item A concerted effort should be made by scientists and institutions to build extensive {\bf open-access repositories} to share scientific software and services to the science community and enable open science (see e.g. intiatives such as the {\it European Science Cluster of Astronomy and Particle physics ESFRI research infrastructures}). \item We encourage {\bf interdisciplinary initiatives} that explore potential synergies in technology, physics, organization and/or applications, such as the JENAS call recently launched by APPEC, ECFA and NuPECC.
    
    \item We recommend that institutions and funding agencies provide adequate support for {\bf education and training in machine learning methods and astrostatistics}, in order to maximize the scientific return of future observations and experiments.
    \item Particular attention should be paid to ensure {\bf diversity} in speakers, panelists, lecturers, and participants at all EuCAPT initiatives, as well as {\bf equal opportunities} and access to all scientific resources and funding instruments. 
    \end{itemize}
\vspace{1cm}
\begin{center}
    {\bf \Large Acknowledgements}
    \end{center}
    This White Paper is the result of a very large group effort. We thank all of the scientists who contributed to the Mattermost channels, google docs, and EuCAPT symposium -- too many to thank them all individually -- for any and all contributions. 
  \newpage  
\section*{List of endorsers}
\label{sec:endorse}
\addcontentsline{toc}{section}{\nameref{sec:endorse}}
\begin{longtable}{l l}
Asmaa Abada & Theory Pole, IJCLAB\\
Amine Ahriche & University of Sharjah (AE)
\\
Zhuk Alexander & Odessa National University\\
Roberto Aloisio & GSSI,  L'Aquila, Italy\\
Loris Amalberti & Padua University \\
Elena Amato & INAF  Arcetri\\
Antonio Ambrosone &  Naple University\\
Shin'ichiro Ando & GRAPPA Institute, University of Amsterdam\\
Stefan Antusch & University of Basel\\
Alexandre Arbey & Univ Lyon, Lyon\\
Maria Archidiacono & Milano University\\
Pedro Avelino & Universidade do Porto, Portugal\\
Anastasios Avgoustidis & University of Nottingham\\
Santiago Avila &  Universidad Aut\'onoma de Madrid, \\
A.B. Balantekin & University of Wisconsin, Madison\\
Csaba Balazs & School of Physics and Astronomy,  Melbourne\\
Iason Baldes & Universit\`e Libre de Bruxelles\\
Guillermo Ballesteros & IFT UAM-CSIC\\
Mihaela-Andreea Baloi & West University of Timisora\\
Enrico Barausse & SISSA\\
Suzan Basegmez du Pree & Nikhef \\
Sebastian Baum & Stanford Institute for Theoretical Physics\\
Andreas Bauswein & GSI Helmholtzzentrum fur Schwerionenforschung\\
Nicola Bellomo &  University of Texas\\
Ido Ben-Dayan & Ariel University\\
David Benisty & University of Cambridge\\
Maria Benito &  NICPB Tallinn\\
Marcus Berg & Karlstad University\\
Atri Bhattacharya & STAR Liege\\
Neven Bilic & Rudjer Boskovic Institute\\
Diego Blas & IFAE Barcelona\\
Pasquale Blasi & Gran Sasso Science Institute\\
Josao Luis Blazquez-Salcedo & Universidad Complutense de Madrid\\
Artem Bohdan &  DESY\\
Beatrice Bonga &  Radboud University\\
Alexander Bonilla Rivera &  Universidade Federal de Juiz de Fora\\
Alexey Boyarsky & Leiden University\\
Gustavo  C. Branco & CFTP/IST, U. Lisboa\\
Philippe Brax & Universit\`e Paris-Saclay, CNRS\\
Jarle Brinchmann & Instituto de Astrofisica e Ciencias do Espaco\\
Richard Brito & CENTRA, Lisboa\\
Robert Brose & Dublin Institute of Advanced Physics\\
Roberta Calabrese & Naples UNiversity\\
Guadalupe Casas-Herrera &  Leiden University\\
Francesco Capozzi &  Virginia Tech\\
Silvia Celli & La Sapienza Universit\`a di Roma and INFN\\
Jose A. R. Cembranos &Universidad Complutense de Madrid\\
David G. Cerdeno &  IFT-UAM/CSIC Madrid, Spain   \\
Marina Cermelo & Universit\`e catholique de Louvain\\
Prasanta Char & University of Liege\\
Andrew Cheek & Nicolaus Copernicus, Polish Academy of Sciences\\
Mu-Chun Chen & University of California, Irvine\\
Maria Chernyakova & DCU\\
Caprini Chiara &  Geneva University\\
Gihyuk Cho & DESY\\
Talal Ahmed Chowdhury &  University of Dhaka\\
Xiaoyong Chu & Institute of High Energy Physics Vienna\\
Marco Cirelli & LPTHE CNRS\\
Johann Cohen-Tanugi & LUPM, Universit\`e de Montpellier
\\
Philippa Cole & GRAPPA  Amsterdam\\
Geoffrey Comp\`ere & Universit\`e Libre de Bruxelles\\
Omar Contigiani & CITA\\
Adam Coogan & GRAPPA Amsterdam\\
Claudio Corian\`o & University of Salento \\
Javier Coronado-Blazquez &  UAM-CSIC Madrid\\
Josao Cortes & Universidad de Zaragoza\\
Jean-Reneux Cudell & Universit\`e de Li\`ege\\
Giulia Cusin & Geneva University\\
Virgile Dandoy &  KIT Karlsruhe\\
Anirban Das & SLAC \\
Antonio da Silva & Universidade de Lisboa\\
Anne-Christine Davis & DAMTP,  Cambridge University\\
M.Patrick Decowski &  Nikhef\\
Nathalie Degenaar & University of Amsterdam\\
Paul De Jong &  Nikhef \\
Luis Del Peral &  University of Alcala\\
Valerio De Luca & Geneva University\\
Peter Denton & Brookhaven National Laboratory\\
Francesco D'Eramo & University of Padua\\
Claudia de Rham & Imperial College London\\
Valentina De Romeri & CSIC-Universitat de Valencia\\
Kyriakos Destounis &  IAAT Tubingen\\
Konstantinos Dialektopoulos & Aristotle University of Thessaloniki\\
Jose M. Diego & IFCA Santander\\
Mattia di Mauro & INFN Torino\\
Eleonora Di Valentino & University of Sheffield\\
Arache Djannati-Atai & APC-CNRS\\
Goran S. Djordjevic & University of Nis\\
Christoph Dlapa &  DESY\\
Caterina Doglioni & University of Manchester\\
Fiorenza Donato & Torino University\\
Daniela Doneva &  University of Tuebingen\\
Amelia Drew & DAMTP Cambridge, Wilberforce Road\\
Ippocratis D. Saltas & CEICO, Czech Academy of Sciences\\
Christopher Eckner & CNRS Annecy, France\\
Timon Emken & Stockholm University\\
Rikard Enberg & Uppsala University\\
Torsten En$\beta$lin & Max Planck Institute for Astrophysics\\
Celia Escamilla-Rivera & Universidad Nacional Autonoma de Mexico\\
Domenec Espriu & Universitat de Barcelona\\
Giuseppe Fanizza &  Universidade de Lisboa\\
Tobias Felkl & UNSW Sydney\\
Pedro G. Ferreira & University of Oxford\\
Pavel Fileviez Perez & Case Western Reserve University, USA\\
Damiano F. G. Fiorillo & Naples University\\
Jernej F. Kamenik &  University of Ljubljana\\
Ottavio Fornieri & Gran Sasso Science Institute\\
Roberto Franceschini & Rome 3 University\\
Sebastian Franchino &  Universitat Heidelberg\\
Gabriele Franciolini & Geneva University\\
Kare Fridell & TUM\\
Carlos F. Sopuerta &  CSIC Barcelona\\
Javier Galan Lacarra & University of Zaragoza\\
Viviana Gammaldi & Universidad Autonoma de Madrid\\
Juan Garcia-Bellido & Universidad Autonoma de Madrid\\
Marcos A. Garcia Garcia &  Padova University\\
Sebastian Garcia-Saenz & Southern University Shenzhen\\
Mathias Garny & TUM\\
Alice Garoffolo & Leiden University\\
Shao-Feng Ge & Tsung-Dao  Shanghai Jiao Tong University\\
Laszlo Gergely &  University of Szeged\\
Jacopo Ghiglieri & Subatech  Nantes\\
Oindrila Ghosh &  Universitat Hamburg\\
Maurizio Giannotti & Barry University\\
Gian Giudice & CERN\\
Carlo Giunti & INFN Torino\\
Florian Goertz & Max-Planck-Institut  Heidelberg\\
Mario E. Gomez & Universidad de Huelva\\
    Rebeca Gonzalez Suarez & Uppsala University\\
Alex Gough & Newcastle University\\
Alessandro Granelli & SISSA Trieste\\
Jonathan Granot & The Open University of Israel\\
Anne Green & University of Nottingham\\
Christophe Grojean & DESY Hamburg\\
Jon Gudmundsson & Stockholm University\\
Claire Guepin & University of Maryland\\
Tao Han & University of Pittsburgh\\
Rasmus S. L. Hansen & Niels Bohr Institute Copenhagen\\
Martin Hardcastle &  University of Hertfordshire\\
Chandan Hati & TUM  \\
Andreas Haungs & Karlsruhe Institute of Technology\\
Lavinia Heisenberg &  ETH Zurich\\
Jan Heisig &  RWTH Aachen University\\
Andi Hektor & KBFI, Tallinn\\
Carlos Herdeiro &  CIDMA \\
Bohdan Hnatyk & University of Kyiv\\
Shunsaku Horiuchi & Virginia Tech\\
Patrick Huber & Virginia Tech\\
Katri Huitu & Helsinki Institute of Physics\\
Gert Hutsi &  NICPB Tallinn\\
Oksana Iarygina & Leiden University\\
Jose Ignacio Illana & University of Granada\\
Jordi Isern & CSIC\\
Sudip Jana & Max-Planck-Institut  Heidelberg\\
Alexander C. Jenkins &  University College London\\
 Yu Seon Jeong & Chung-Ang University\\
Philippe Jetzer & University of Zurich\\
Cristian Joana &  CURL Louvain\\
Adil Jueid &  Korea Institute for Advanced Study, Seoul\\
Felix Kahlhoefer &  RWTH Aachen University\\
Alba Kalaja & University of Groningen\\
Gregor Kaelin & DESY\\
Kristjan Kannike & NICPB Tallinn, Estonia\\
Dimitrios Kantzas &GRAPPA  Amsterdam\\
Alexander Kappes & University Muenster\\
Alexandros Karam & NICPB Tallinn, Estonia\\
Konstantin Karchev & SISSA/GRAPPA\\
Timo Karkkanen & Lorand University\\
Uli Katz & FAU Erlangen\\
Joern Kersten & University of Bergen\\
Venus Keus & University of Helsinki\\
Ali Rida Khalifeh &  University of Barcelona\\
Claus Kiefer & University of Cologne,\\
Doojin Kim & Texas University\\
Jinsu Kim &  CERN\\
Stefan Kis & Universit\`e Catholique de Louvain\\
Michael Klasen & WWU Muenster\\
Pyungwon Ko & KIAS\\
Kostas Kokkotas & University of Tuebingen\\
Thomas Konstandin & DESY\\
Joachim Kopp & CERN\\
Matthias Koschnitzke & University of Hamburg/DESY\\
Dmitriy Kostunin & DESY\\
Ely Kovetz & Ben-Gurion University of the Negev\\
Michael Kramer & RWTH Aachen University\\
Christian Krager & Universitat Tobingen\\
Jutta Kunz &  University of Oldenburg\\
Martin Kunz & Geneva University\\
Kerstin E. Kunze & Universidad de Salamanca\\
Jui-Lin Kuo &  Austrian Academy of Sciences\\
David Langlois &  CNRS\\
Rebecca K. Leane &  Stanford University\\
Hyun Min Lee & Chung-Ang University\\
Marek Lewicki &  University of Warsaw\\
Sarah Libanore &  Padova University\\
Stefano Liberati & SISSA\\
Michele Liguori &  University of Padova\\
Eugene Lim & King's College London\\
Axel Lindner & DESY\\
Manfred Lindner & Max-Planck-Institut  Heidelberg\\
Zhengwen Liu & DESY\\
Francisco Lobo &  Universidade de Lisboa,\\
Francesco Longo & University of Trieste \\
Michele Lucente & RWTH Aachen University\\
Yang Ma &University of Pittsburgh\\
Oscar Macias & GRAPPA\\
Jonathan Mackey & Dublin Institute for Advanced Studies\\
Elisa Maggio &La Sapienza Roma \\
Michele Maggiore & University of Geneva\\
Farvah Mahmoudi &  CNRS Lyon\\
Yann Mambrini & Universit\'e Paris-Saclay\\
Fabio Marchesoni & Universita' di Camerino\\
Alexandre Marcowith & Laboratoire Univers Particle Montpellier\\
David Marsh & Stockholm University\\
Jorge Martin Camalich & Instituto de Astrof\'isica de Canarias\\
Luca Marzola &NICPB\\
Sabino Matarrese & Padova University\\
Anupam Mazumdar & Van Swinderen Institute\\
Davide Meloni & University of Roma Tre\\
Scott Melville &  University of Cambridge\\
Olga Mena & IFIC \\
Luca Merlo & Universidad Autonoma de Madrid\\
Philipp Mertsch & RWTH Aachen University\\
Benjamin Kurt Miller & University of Amsterdam\\
Josao Pedro Mimoso &  Universidade de Lisboa \\
Jordi Miralda-Escud\'e & ICREA Barcelona\\
Rukmani Mohanta &  University of Hyderabad\\
Paulo Moniz & UBI - Universidade da Beira Interior\\
Azadeh Moradinezhad Dizgah & Geneva University\\
Giovanni Morlino & INAF Arcetri\\
Emmanuel Moulin & CEA Saclay\\
Suvodip Mukherjee & Perimeter Institute\\
Wanga Mulaudzi & University of Cape Town\\
Eike Moeller & Stockholm University\\
Victor Munoz & University of Valencia-CSIC\\
Ilia Musco &  La Sapienza Roma \\
Kristjan Muursepp & NICPB Tallinn\\
Enrico Nardi & INFN Frascati\\
Pavel Naselsky & Niels Bohr Institute\\
Kenny Chun Yu Ng & The Chinese University of Hong Kong\\
Argyris Nicolaidis &  Aristotle University of Thessaloniki\\
Kenichi Nishikawa & Alabama  University\\
Samaya Nissanke & GRAPPA Amsterdam\\
Emil Nissimov & Bulgarian Academy of Sciences\\
Jose Miguel No & UAM/CSICMadrid\\
Nelson J. Nunes &  University of Lisbon\\
Martin Obergaulinger & Universitat de Valencia\\
Giorgio Orlando &  University of Groningen\\
Per Osland & University of Bergen\\
Sydney Otten & University of Amsterdam \\
Rami Oueslati & STAR Institute  Liege\\
Francesco Pace &  Bologna University\\
Ian Padilla-Gay & Niels Bohr Institute\\
Sergio Palomares-Ruiz & IFIC  Valencia \\
Paolo PANCI & University of Pisa\\
Sohyun Park & CERN\\
Roman Pasechnik &  Lund University\\
Sergio Pastor & IFIC Valencia\\
Chris Pattison &  University of Portsmouth\\
Vasiliki Pavlidou & University of Crete \\
Carlos Perez de los Heros &  Uppsala University\\
Mihael Petac & University of Nova Gorica\\ 
 Serguey Petcov & SISSA Trieste\\
Kallia Petraki & LPTHE\\
Oleh Petruk &  NASU\\
Valeria Pettorino &  CNRS\\
Ivica Picek &  University of Zagreb\\
Wodzimierz Piechocki & Warsaw National Centre for Nuclear Research\\
Mathias Pierre & DESY Hamburg\\
Massimo Pietroni & Universit\`a di Parma \\
Luigi Pilo & University of l'Aquila\\
Elena Pinetti & University of Turin\\
Lorenzo Pizzuti & Astronomical Observatory Aosta Valley\\
Pavlo Plotko & DESY Zeuthen\\
Martin Pohl & University of Potsdam\\
Vivian Poulin & CNRS Montpellier\\
Julien Poyatos &  Universidade do Porto\\
Gianfranco Pradisi &  University of Rome Tor Vergata\\
Josef Pradler & Austrian Academy of Sciences\\
Geraint Pratten & University of Birmingham\\
Giovanna Pugliese & University of Amsterdam\\
Pablo Quielez & DESY\\
Antonio Racioppi & NICPB Tallinn\\
Johann Rafelski & The University of Arizona\\
Martti Raidal &  NICPB Tallinn\\
Justin Read & University of Surrey\\
Margarida Nesbitt Rebelo &  Universidade de Lisboa \\
Marco Regis &  Universit\`a di Torino\\
Patrick Reichherzer & Universitat Bochum\\
Fabrizio Renzi & Leiden University\\
Gerasimos Rigopoulos &  Newcastle University\\
Andreas Ringwald &  DESY\\
Tania Natalie Robens & Rudjer Boskovic Institute \\
Maria  Rodriguez Frias &  UAH Madrid\\
Marek Rogatko & Maria Curie Sklodowska University\\
Rogerio Rosenfeld &  ICTP-SAIFR\\
Leszek Roszkowski & Astrocent Nicolaus Copernicus \\
Alberto Rozas-Fernandez &  University of Lisbon\\
Diego Rubiera-Garcia & Madrid Complutense University \\
Roberto Ruiz de Austri & IFIC-UV/CSIC Valencia\\
Ester Ruiz Morales & Universidad Politecnica de Madrid \\
Kari Rummukainen & University of Helsinki\\
Vishu Saini & Indian Institute of Technology Bombay\\
Mairi Sakellariadou & King's College London\\
Filippo Sala & LPTHE CNRS \\
Ennio Salvioni & University of Padua\\
Miguel Sanchez-Conde & UAM Madrid  \\
Elena Santopinto & INFN Genova\\
Emmanuel Saridakis & National Observatory of Athens\\
 Ninetta Saviano &  Naple University\\
Joern Schaffran & DESY\\
Joop Schaye &  Leiden University\\
Michael A. Schmidt & The University of New South Wales\\
Kai Schmidt-Hoberg & DESY\\
Fabian Schussler &  CEA Saclay\\
Dominik Schwarz & Bielefeld University\\
David Seery & University of Sussex\\
Dmitri Semikoz & APC, Paris\\
Manibrata Sen & Max-Planck-Institut Heidelberg\\
Olga Sergijenko &  National University of Kyiv\\
Bibhushan Shakya & DESY\\
Hua-Sheng Shao & LPTHE\\
Chia-Hsien Shen & UCSD\\
Graham Shore & Swansea University\\
Vitalii Sliusar & University of Geneva\\
Josao Sobrinho &  Universidade de Lisboa\\
Philip Soerensen & DESY  Hamburg\\
Anastasia Sokolenko & University of Chicago\\
Thomas Sotiriou &  University of Nottingham\\
Lara Sousa & Universidade do Porto\\
Thomas F.M. Spieksma & GRAPPA Amsterdam\\
Antonio Stamerra & INAF-OAR and SNS-Pisa\\
Danielle Steer & APC, University of Paris\\
Nikolaos Stergioulas & Aristotle University of Thessaloniki\\
Patrick Stocker & KIT Karlsruhe\\
Sabin Stoica &  University of Chinese Academy of Sciences\\
Konstantinos Tanidis & CEICO\\
Nial Tanvir & University of Leicester\\
Marco Taoso &INFN  Torino\\
Gianmassimo Tasinato & Swansea University, UK\\
Andrew Taylor & DESY\\
Nikolaos Tetradis &  University of Athens\\
Luigi Tibaldo & IRAP  Toulouse\\
Anna Tokareva & University of Jyvaskyla\\
Andrew Tolley & Imperial College London\\
Giovanni Maria Tomaselli & University of Amsterdam\\
Eemeli Tomberg &  NICPB  Tallinn\\
Emilio Torrente-Lujan &  Murcia University \\
Mariam T\`ortola & IFIC Valencia\\
Andreas Trautner & Max-Planck-Institut Heidelberg\\
Sebastian Trojanowski &Nicolaus Copernicus Astronomical Center\\
Michel Tytgat & Universit\`e Libre de Bruxelles\\
Piero Ullio & SISSA \& IFPU\\
Michael Unger & Karlsruhe Institute for Technology\\
Patrick Valageas &  CNRS\\
Jos\`e W. F. Valle & IFIC/CSIC-Univ Valencia\\
Matthijs van der Wild & Durham University\\
Valeri Vardanyan & The University of Tokyo\\
Ville Vaskonen & IFAE Barcelona \\
Matyas Vasuth & Wigner RCP\\
Hardi Veermae &  NICPB  Tallinn\\
Sudhir Vempati & Indian Institute of Science\\
Licia Verde & ICREA and ICC-UB\\
Rodrigo Vicente &  IFAE\\
Matteo Viel & SISSA\\
Patricio Vielva & CSIC  Santander\\
Vincenzo Vitale & INFN Roma Tor Vergata\\
Stefan Vogl & Freiburg University\\
Benedict von Harling & IFAE Barcelona\\
Roland Walter &  University of Geneva\\
David Wands & University of Portsmouth\\
Dong-Gang Wang & DAMTP, University of Cambridge\\
Jin-Wei Wang & SISSA \\
Barry Wardell & University College Dublin\\
Meng-Ru Wu & Institute of Physics, Academia Sinica\\
Yongcheng Wu & Oklahoma State University\\
Lili Yang & School of Physics and Astronomy Zhuhai\\
Zixin Yang & DESY Hamburg\\
Stoytcho Yazadjiev & Sofia University\\
Doosoo Yoon &  University of Amsterdam\\
Gabrijela Zaharijas &  University of Nova Gorica\\
Robert Ziegler & KIT\\
Kai Zuber & TU Dresden\\

\end{longtable}

 \newpage
 \section*{List of experiments and missions}
\label{sec:exp}
\addcontentsline{toc}{section}{\nameref{sec:exp}}
    Acronyms and references of experiments and missions mentioned in the text

\begingroup
\setlength{\tabcolsep}{8pt}
\begin{center}

\resizebox{\textwidth}{!}{\begin{tabular}{cccc}
\hline\hline
  {\bf Experiment}   &  {\bf Field}  & {\bf Acronym} & {\bf Refs.} \\
Athena X-ray Observatory & X-ray astronomy & Athena & \cite{nandra_hot_2013} \\  
Borexino & Neutrino physics and astrophysics & Borexino & \cite{Borexino:2008gab,Borexino:2017rsf} \\ 
CERN Axion Solar Telescope & Particle physics & CAST & \cite{CAST:2004gzq} \\
Cherenkov Telescope Array & $\gamma$-ray astronomy & CTA & \cite{CTAConsortium:2018tzg} \\
Deep Underground Neutrino Experiment & Neutrino physics and astrophysics & DUNE & \cite{DUNE:2020lwj}\\
Einstein Telescope & Gravitational Waves & ET & \cite{Punturo:2010zz,Maggiore:2019uih} \\
Event Horizon Telescope & Astronomy and astrophysics & EHT & \cite{EHTPaperI} \\
enhanced X-ray Timing and Polarimetry mission & X-ray astronomy & eXTP & \cite{eXTP:2018anb} \\ 
Global Astrometric Interferometer for Astrophysics & Astronomy and astrophysics & Gaia & \cite{Gaia:2016zol,gaia_edr3} \\
High Altitude Water Cherenkov Detector & $\gamma$-ray astronomy & HAWC & \cite{tepe_hawc_2012} \\
Hyper-Kamiokande & Neutrino physics and astrophysics & Hyper-K & \cite{Hyper-Kamiokande:2018ofw} \\
IceCube Neutrino Observatory &  Neutrino physics and astrophysics & IceCube & \cite{IceCube_2013science} \\
International Axion Observatory & Particle physics & IAXO & \cite{Armengaud:2014gea} \\  
James Webb Space Telescope & Astronomy and astrophysics & JWST & \cite{Gardner:2006ky} \\
Jiangmen Underground Neutrino Observatory & Neutrino physics and astrophysics & JUNO & \cite{JUNO:2015zny} \\
Kepler & Exoplanetary and stellar physics & Kepler & \cite{Kepler:2010xwo} \\
Large High Altitude Air Shower Observatory & $\gamma$-ray and cosmic ray astronomy & LHAASO & \cite{bai_lhaaso_wp_2019} \\
Laser Interferometer Space Antenna & Gravitational Waves & LISA & \cite{Audley:2017drz}\\
Low-frequency Array & Radio astronomy & LOFAR & \cite{van_haarlem_lofar_2013} \\ 
Neutron Star Interior Composition Explorer Mission & X-ray astronomy & NICER & \cite{2016SPIE.9905E..1HG} \\
PLAnetary Transits and Oscillations of stars & Exoplanetary and stellar physics & PLATO & \cite{Plato:2014} \\
Square Kilometre Array & Radio astronomy & SKA & \cite{dewdney_square_2009} \\
Sudbury Neutrino Observatory+ & Neutrino physics and astrophysics & SNO+ & \cite{SNO:2021xpa}\\
Super-Kamiokande & Neutrino physics and astrophysics & Super-K & \cite{Super-Kamiokande:2001ljr} \\
Transiting Exoplanet Survey Satellite & Exoplanetary and stellar physics & TESS & \cite{TESS:2015} \\ 
THEIA & Neutrino physics and astrophysics & THEIA & \cite{Theia:2019non} \\ 
Tibet AS-gamma & $\gamma$-ray and cosmic ray astronomy & Tibet AS-$\gamma$ & \cite{huang_tibet_2011} \\
Euclid & Astronomy and astrophysics & --- & \cite{EUCLID:2011zbd}\\
Vera C. Rubin Observatory & Astronomy and astrophysics & --- & \cite{LSST:2019}\\
XENONnT & Dark matter direct detection & & \cite{Aprile:2020vtw}\\ 
LZ  & Dark matter direct detection  & ---& \cite{LZ:2019sgr}\\
PandaX-4T & Dark matter direct detection  & ---& \cite{PandaX:2018wtu}\\ 
DarkSide-20k &  Dark matter direct detection &--- & \\
DARWIN  & Dark matter direct detection  & ---&\cite{Aalbers:2016jon}\\ 
CRESST-III & Dark matter direct detection  &--- &\cite{Abdelhameed:2019hmk} \\ 
EDELWEISS & Dark matter direct detection  &--- &\cite{Arnaud:2017usi} \\ 
SuperCDMS  &Dark matter direct detection   &--- &\cite{Agnese:2016cpb}\\
Anais & Dark matter direct detection & ---& \cite{Amare:2019ncj} \\
Cosine & Dark matter direct detection & ---&\cite{Adhikari:2017esn} \\
SABRE  &Dark matter direct detection   & ---&\cite{Antonello:2018fvx} \\ 
Cosinus  &Dark matter direct detection   & &\cite{Angloher:2016ooq}\\
DAMA & Dark matter direct detection  & ---&\cite{Battaglieri:2017aum} \\
GNOME  & Dark matter direct detection  & ---&\cite{GNOME1}\\ 
MADMAX & Dark matter direct detection  & ---& \cite{TheMADMAXWorkingGroup:2016hpc}\\ 
CAPP &  Dark matter direct detection & ---&\cite{Semertzidis:2019gkj} \\
ADMX  & Dark matter direct detection  & ---&\cite{2010PhRvL.104d1301A}\\ 
ABRACADABRA & Dark matter direct detection &--- & \cite{Ouellet:2019tlz}\\
CASPEr & Dark matter direct detection & ---&\cite{Budker:2013hfa} \\
AstroSAT & X-ray astrophysics & ---&\cite{2006AdSpR..38.2989A} \\
Large Observatory For X-ray Timing & X-ray astrophysics & LOFT& 
\\
(Nancy Grace) Roman Space Telescope & Astronomy \& Astrophysics & --- & \cite{2012arXiv1208.4012G}\\ 
(Advanced) Laser Interferometer Gravitational-Wave Observatory & Gravitational waves & LIGO/AdLIGO & \cite{LIGOScientific:2014pky}\\
(Advanced) Virgo & Gravitational waves & Virgo/AdV & \cite{VIRGO:2014yos} \\
Fermi Large Area Telescope & $\gamma$-ray astronomy & Fermi-LAT & \cite{2009ApJ...697.1071A}\\ 
Cubic Kilometre Neutrino Telescope & Neutrino physics and astrophysics & KM3NET & \cite{2016JPhG...43h4001A}\\
Very Large Telescope & Astronomy and astrophysics & VLT & ---\\
Experiment to Detect the Global EoR Signature & Radio astronomy & EDGES & \cite{Bowman_2008}\\
Hubble Space telescope & Astronomy and astrophysics  & HST & ---\\
European Extremely Large Telescope & Astronomy and astrophysics  & E-ELT & \url{https://www.eso.org/sci/facilities/eelt/}\\
Thirty Meter Telescope & Astronomy and astrophysics  & TMT & \url{https://www.tmt.org}/ \\
Giant Magellan Telescope & Astronomy and astrophysics  & --- & \url{https://www.gmto.org/}\\
Hydrogen Intensity and Real-time Analysis eXperiment & Radio astronomy & HIRAX & \cite{Newburgh:2016mwi}\\
Canadian Hydrogen Intensity Mapping Experiment  & Radio astronomy & CHIME & \cite{CHIMEFRB:2018mlh}\\
Packed Ultra-wideband Mapping Array & Radio astronomy & PUMA & \cite{PUMA:2019jwd}\\
enhanced ASTROGAM & $\gamma$-ray astronomy & e-ASTROGAM & \cite{e-ASTROGAM:2017pxr}\\
     \hline
\end{tabular}
}
\end{center} 
\endgroup

\clearpage
\addcontentsline{toc}{section}{Bibliography}
\bibliographystyle{utphys}
\scriptsize
\bibliography{Main.bib,DE.bib,Astrostatistics.bib,references_particle_stars.bib,neutrino-bibliography.bib,DM_bib.bib,cosmicaccel.bib,references-early1.bib,references-early2.bib, NuclearAstro.bib}

\end{document}